\documentclass[conference]{IEEEtran}

\usepackage{cite}
\usepackage{graphicx}
\usepackage{amsmath,amssymb}
\usepackage{hyperref}
\usepackage{xurl}
\usepackage{subcaption}
\usepackage{wrapfig}
\usepackage{tikz}
\usetikzlibrary{arrows.meta,shapes,shapes.geometric,positioning,fit,backgrounds}

\title{Engineering Mythology: A Digital-Physical Framework for Culturally-Inspired Public Art}

\author{
    \IEEEauthorblockN{Jnaneshwar Das, Christopher Filkins, Rajesh Moharana, Ekadashi Barik, Bishweshwar Das, \\ David Ayers, Christopher Skiba, Rodney Staggers Jr, Mark Dill, 
    Swig Miller, \\ Daniel Tulberg, Patrick Smith, Seth Brink, Kyle Breen, 
     Harish Anand, Ramon Arrowsmith}
}

\begin{document}

\maketitle

\begin{abstract}
Navagunjara Reborn: The Phoenix of Odisha was built for Burning Man 2025 as both a sculpture and an experiment—a fusion of myth, craft, and computation. This paper describes the digital–physical workflow developed for the project: a pipeline that linked digital sculpting, distributed fabrication by artisans in Odisha (India), modular structural optimization in the U.S., iterative feedback through photogrammetry and digital twins, and finally, one-shot full assembly at the art site in Black Rock Desert, Nevada. The desert installation tested not just materials, but also systems of collaboration: between artisans and engineers, between myth and technology, between cultural specificity and global experimentation. We share the lessons learned in design, fabrication, and deployment and offer a framework for future interdisciplinary projects at the intersection of cultural heritage, STEAM education, and public art. In retrospect, this workflow can be read as a convergence of many knowledge systems—artisan practice, structural engineering, mythic narrative, and environmental constraint—rather than as execution of a single fixed blueprint.
\end{abstract}


\section{Introduction}

\begin{wrapfigure}{r}{0.23\textwidth}
  \centering
  \includegraphics[width=0.23\textwidth]{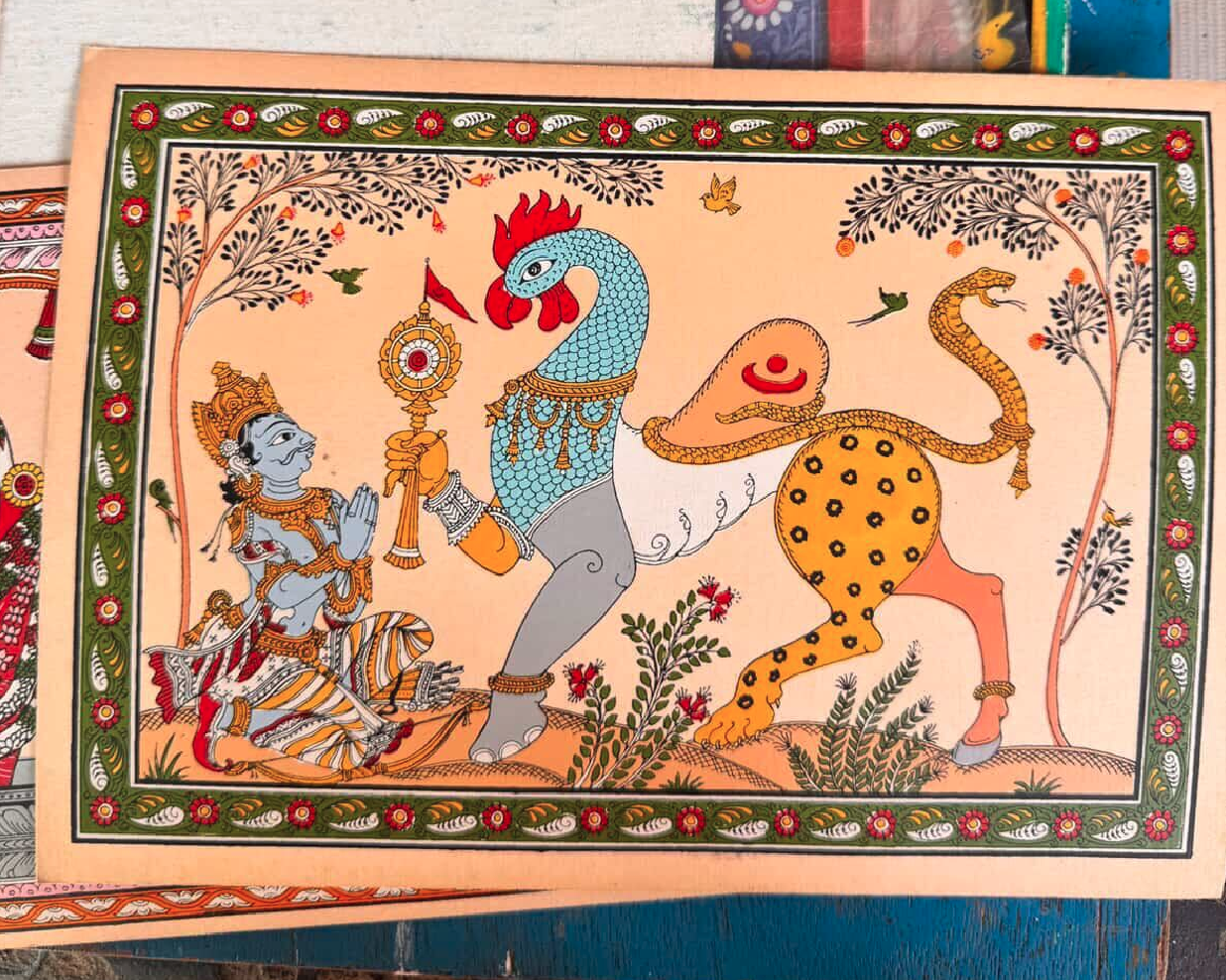}
  \caption{\small{Traditional `Pattachitra' painting of Navagunjara in Raghurajpur, Odisha, India. Photo courtesy: Lipikka Sahoo}}
  \label{fig:myfig}
\end{wrapfigure}

In recent decades, large-scale public art has become a stage where engineering and mythology meet. Burning Man~\cite{burningman2025}, through its Honoraria Arts Grant program, has nurtured this trend, supporting monumental installations that combine community-driven technical innovations with symbolic storytelling. Navagunjara Reborn was realized on a \$14,870 Honoraria grant against a total project budget of approximately \$25,000, with the balance supported by crowdfunding and the lead and co-artists, demonstrating that internationally distributed, craft-integrated public art at monumental scale is achievable with modest resources when digital-physical workflows are designed for resilience and adaptability.

\begin{figure}[htpb]
    \centering
    \includegraphics[width=3.5in]{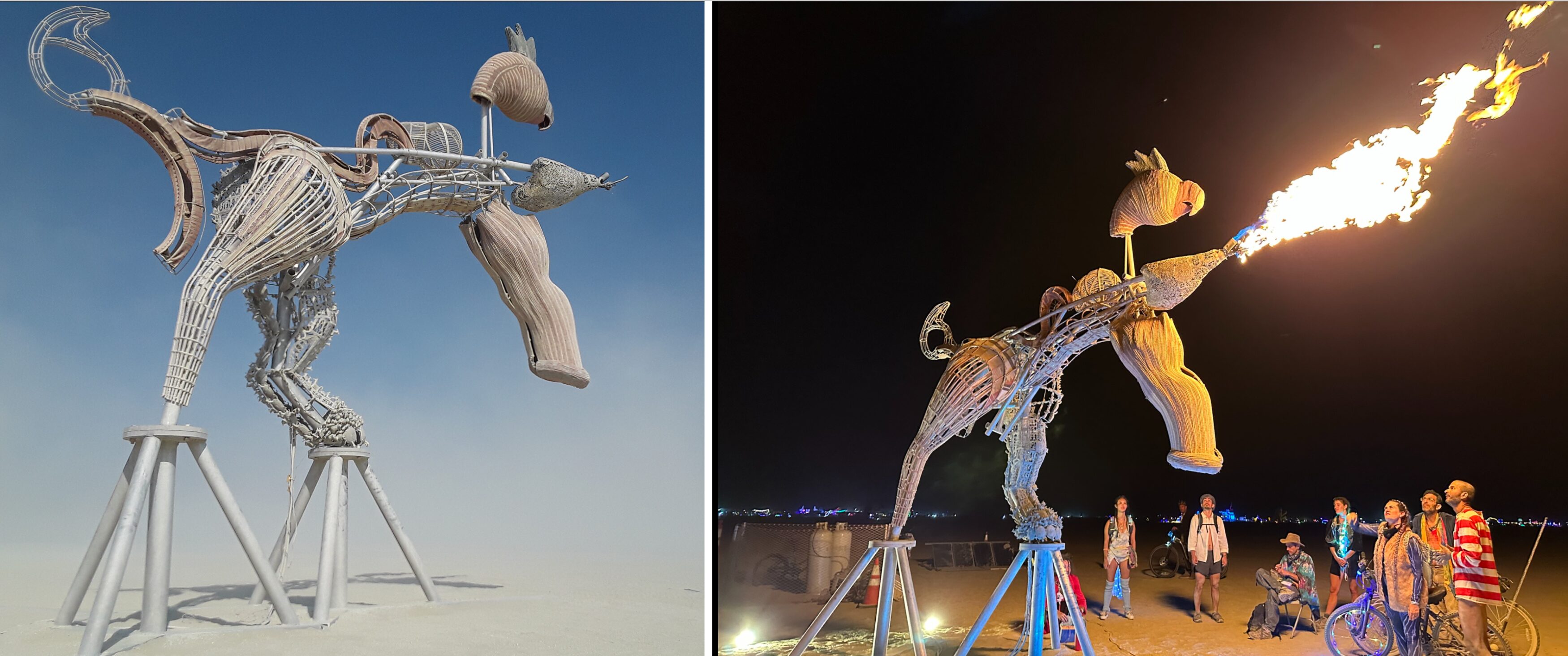}
    \caption{\small{Navagunjara Reborn: An 18-foot sculptural fusion of the Navagunjara and Phoenix myths, is illuminated by interactive flame effects at Burning Man 2025. Each animal section showcases traditional Odia crafts—including dhokra metalwork, sabai grass weaving, pattachitra painting, and indigenous textiles—brought together in a monumental installation that celebrates resilience, unity, and the transformative power of mythology through participatory public art.}}
    \label{fig:main-fig-sculptures}
\end{figure}

Navagunjara Reborn (Figure~\ref{fig:main-fig-sculptures}) was conceived in this spirit. The work fuses the Phoenix~\cite{hill1984phoenix,ungermann1999phoenix}—a universal emblem of transformation—with the Navagunjara~\cite{senapati2001myths}, a hybrid figure from Odia~\footnote{Odia refers to the culture and language of Odisha, a coastal state in southeastern India.} mythology composed of nine animals, together embodying multiplicity and renewal.

\begin{figure*}[!t]
    \centering
    \includegraphics[width=\textwidth]{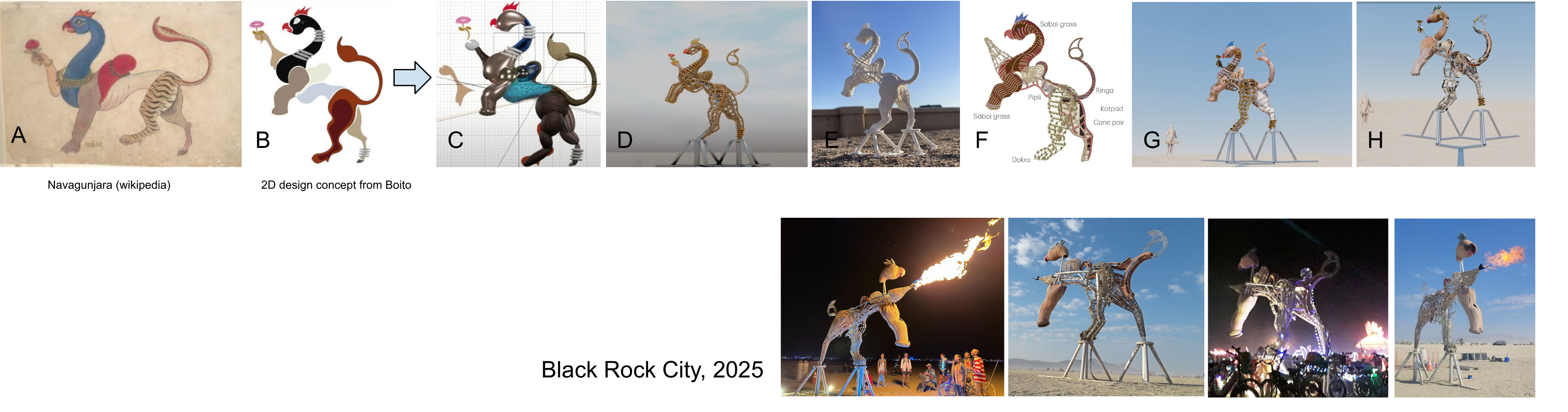}
    \caption{\small Conceptual synthesis and design-to-build progression of Navagunjara Reborn. \textbf{(A)}~Traditional painting of the Navagunjara (source: Wikipedia), the mythological nine-formed composite from Odisha's Mahabharata retelling that anchors the cultural concept. \textbf{(B)}~2D design concept from Boito. \textbf{(C)}~Refined Blender digital twin for India Art Fair 2025 (8~ft). \textbf{(D)}~3D render from the Burning Man 2025 full proposal (15--18~ft), submitted for the Honoraria Arts Grant. \textbf{(E)}~3D-printed physical twin of the sculpture, produced in Phoenix, AZ, for structural validation and artisan communication. \textbf{(F)}~Crafts atlas mapping traditional Odia craft techniques to each animal section, guiding the fabrication build in Odisha. \textbf{(G)}~Full digital twin during India crafts build phase. \textbf{(H)}~Completed aluminum CAD superstructure digital twin prior to transport to playa, with scanned crafts staged in Phoenix, AZ. Bottom row: Navagunjara Reborn at Black Rock City, August 2025, with active poofer flame effects.}
    \label{fig:workflow-synthesis}
\end{figure*}
The project was realized through an unusual global pipeline: 3D design in Arizona, traditional crafts sculpted in Odisha, structural elements optimized and fabricated in Arizona based on sculpted crafts, and final assembly in Black Rock Desert, Nevada, spanning a year-long journey. The outcome was an 18-foot hybrid sculpture shaped by both myths and algorithms, hands and machines. The design did not proceed from a fixed blueprint; rather, it emerged from a \emph{product kernel}: each contributor brought inductive priors encoding their domain—the lead artist's knowledge of Burning Man and mythology, the dhokra artisan's decades of metalwork intuition, the cane craftsperson's sense of tensile form, the structural engineer's material yield calculations, and the desert's own 75~mph wind requirement. The final form is the configuration that scored high under all these kernels simultaneously, arriving through constraint-satisfying iteration as craft realities, structural data, and field conditions progressively narrowed the reachable design space.

At its core, the project explores convergence—of artisans, designers, and builders—in the creation of a monumental public artwork. The digital–physical workflow documented here integrates diverse methods and original ideas, aiming to leave a practical, reproducible, and open framework for future works that connect cultural heritage with computational design and enable distributed creation of monumental art.

\section{Cultural Context}

\begin{table*}[h!]
\centering
\renewcommand{\arraystretch}{1.4}
\begin{tabular}{|p{3.3cm}|p{3.3cm}|p{5.5cm}|p{3.3cm}|}
\hline
\textbf{Sculptural Components} & \textbf{Crafts Techniques} & \textbf{Skillsets Required} & \textbf{Primary Artisans} \\
\hline
Elephant Leg, Peacock Neck, Rooster head    & Metal Wireframing + Sabai Grass Weaving    & Precision steel forming, structured weaving, dye work, large-format framework construction   & Rajesh Moharana, Malli Mani Nayak, Purnima Nayak, Pankajini Nayak, Arsu Marandi, Geeta Nayak, Jayanti Nayak, Gowri Mahapatra, Minu Nayak, Bina Pani Patra , Basamati Nayak  \\
\hline
Human Arm \& Tigress Leg        & Dhokra/brass metalwork/filigree        & Lost-wax investment casting with bronze, motifs with brass filigree                    & Rajesh Moharana \\
\hline
Deer Leg, Lioness Torso         & Cane Wood Mesh/exo-skeleton, Pattachitra Painting, Kotpad Weaving, Appliqué     & Cane curvature, tensioned meshwork, traditional painting                 & Ekadasi Barik, Akshaya Bariki, Jagabandhu Panika, Purnachandra Ghose, Kotuli Sitaka  \\
\hline 
\end{tabular}
\caption{Mapping Odisha artisan skills and techniques to sculpture components.}
\label{tab:artisan-skills}
\end{table*}

The Navagunjara~\cite{misra1995travel, senapati2001myths}, drawn from Odisha’s regional retellings of the Mahabharata~\cite{mitra1992mahabharat}, is a composite being of nine forms—peacock, serpent, lioness, elephant, tigress, deer, bull, human, and rooster. Each element represents a different dimension of experience, together expressing the multiplicity of paths toward truth and enlightenment~\cite{senapati2001myths}. In contrast, the Phoenix, rooted in global traditions, embodies a cycle of destruction and renewal through fire, becoming a universal emblem of transformation~\cite{hill1984phoenix}.

For Burning Man 2025, these archetypes were deliberately brought into dialogue. The Phoenix provided a universally legible icon for a global audience, while the Navagunjara anchored the work in the cultural specificity of Odisha~\cite{sahu2015traditional}. Their fusion generated a transcultural emblem of resilience—at once particular and universal—capable of speaking to both regional heritage and the broader mythic imagination. The final form represented different crafts from different regions of Odisha amalgamated to form the whole. These include dhokra~\cite{sinha2015dhokra, sethi2016dhokra, sahu2015traditional}, canewood, pattachitra (painted wood storytelling)~\cite{mohanty1984pata}, sabai grass weaving, and different forms of Odisha textiles such as pipli and kotpad~\cite{kapoor2019case, kapoor2025patchwork, dutta2025odisha, pandey2024fabric}. 

The craft selection itself underwent significant adaptation between the original proposal and the final build—an evolution driven by scale, logistics, material weight, artisan availability, and safety constraints. Table~\ref{tab:craft-evolution} documents this arc. The original full proposal envisioned terracotta (rooster), pattachitra painting (peacock), papier-mâché (human arm), ikat textiles (elephant), wood carving (cow), grass weaving (lioness), stone carving (tiger), silver filigree (deer), and dhokra metal casting (snake). The final build replaced nearly all of these: stone carving was eliminated due to shipping weight; terracotta and papier-mâché were replaced by metal wireframing and sabai grass which could survive desert conditions; ikat textiles gave way to sabai grass weaving on the elephant leg; and the tiger and human arm became dhokra metalwork. New craft forms not in the original proposal—kotpad weaving, pipli appliqué, and ringa textiles—entered as artisan relationships deepened in Odisha. The result was a craft vocabulary shaped as much by what artisans could best offer at scale as by the original design vision.

\begin{table}[htpb]
\centering
\renewcommand{\arraystretch}{1.3}
\begin{tabular}{|p{1.6cm}|p{2.8cm}|p{2.8cm}|}
\hline
\textbf{Animal} & \textbf{Proposed Craft} & \textbf{Actual Craft} \\
\hline
Rooster (head)     & Terracotta            & Sabai grass + metal wireframe \\
\hline
Peacock (neck)     & Pattachitra painting  & Sabai grass + metal wireframe \\
\hline
Human (arm)        & Papier-mâché          & Dhokra brass metalwork \\
\hline
Elephant (leg)     & Ikat/bandha textiles  & Metal wireframe + sabai grass \\
\hline
Lioness (torso)    & Grass weaving         & Cane mesh, pattachitra, kotpad \\
\hline
Tiger (leg)        & Stone carving         & Dhokra brass metalwork \\
\hline
Deer (leg)         & Silver filigree       & Cane mesh + pattachitra \\
\hline
Snake (tail)       & Dhokra metal casting  & Integrated into metalwork \\
\hline
\end{tabular}
\caption{Craft adaptation arc: proposed vs.\ realized material assignments per animal section. Scale, shipping weight, artisan availability, and LNT constraints drove the substitutions.}
\label{tab:craft-evolution}
\end{table}

Each of these traditions carries its own \emph{inductive kernel}: a function, learned through years of apprenticeship and practice, that evaluates which forms are possible, which joints are sound, and which motifs are true to tradition. Rajesh Moharana's dhokra kernel lives in his hands and his knowledge of lost-wax bronze topology; the sabai weavers' kernel is encoded in the tensile memory of grass and the geometry of wrapping; Ekadasi Barik's cane kernel expresses curvature constraints no CAD tool was asked to specify. The Navagunjara itself—a composite of nine distinct animal forms, no one of which can stand for the whole—is a mythological anticipation of this principle: multiplicity of perspective as the only adequate representation of truth. In retrospect, the Navagunjara can be read as a maximum-entropy configuration~\cite{jaynes1957} over animal morphologies subject to the constraint of expressing a unified being—the most diverse form that still coheres. That the myth arrived at what amounts to a MaxEnt solution millennia before the mathematics is a striking convergence, one the authors recognized only after the framework developed in this paper took shape. The final sculpture is precisely the configuration that scored high under all these kernels simultaneously: mythologically coherent, structurally sound, craftable by human hands, and resilient to desert winds.

This synthesis shaped three design imperatives:
\begin{itemize}
    \item Scaling artistic sketches into an aesthetically pleasing and stable 18-foot sculptural design.
     \item Translating traditional Indian craft techniques into computationally optimized engineering workflows.
    \item Ensuring resilience of the final installation against desert winds, dust, and storms.
\end{itemize}

\begin{figure*}[t]
\centering
\begin{tikzpicture}[
  x=0.55cm, y=0.65cm,
  >=Stealth,
  every node/.style={transform shape},
  usbox/.style={draw, rounded corners=1.5pt, fill=blue!8,
                text width=1.2cm, minimum height=0.45cm,
                align=center, font=\fontsize{4}{5}\selectfont, inner sep=1.5pt},
  inbox/.style={draw, rounded corners=1.5pt, fill=yellow!22,
                text width=1.2cm, minimum height=0.45cm,
                align=center, font=\fontsize{4}{5}\selectfont, inner sep=1.5pt},
  pribox/.style={draw, rounded corners=2pt, fill=orange!18,
                 text width=1.3cm, minimum height=0.7cm,
                 align=center, font=\fontsize{4}{5}\selectfont, inner sep=2pt},
  dtbox/.style={draw, ellipse, fill=teal!14,
                text width=1.1cm, minimum height=0.4cm,
                align=center, font=\fontsize{4}{5}\selectfont},
  fabbox/.style={draw, ellipse, fill=teal!14,
                 text width=1.1cm, minimum height=0.4cm,
                 align=center, font=\fontsize{4}{5}\selectfont},
  shipbox/.style={draw, rounded corners=1.5pt, fill=purple!10,
                  text width=1.1cm, minimum height=0.4cm,
                  align=center, font=\fontsize{4}{5}\selectfont, inner sep=1.5pt},
  mergebox/.style={draw, rounded corners=2pt, fill=red!12,
                   text width=1.1cm, minimum height=0.7cm,
                   align=center, font=\fontsize{4}{5}\selectfont\bfseries, inner sep=2pt},
  arr/.style={->, >={Stealth[length=2pt,width=1.5pt]}, thin, line width=0.4pt},
  darr/.style={->, >={Stealth[length=2pt,width=1.5pt]}, dashed, blue!55, line width=0.35pt},
  feedarr/.style={->, >={Stealth[length=2pt,width=1.5pt]}, dashed, teal!65, line width=0.35pt},
  looparr/.style={->, >={Stealth[length=1.5pt,width=1pt]}, gray!55, line width=0.3pt},
]

\draw[gray!22, line width=1pt] (2.4, 2.8) -- (29.0, 2.8);
\draw[gray!22, line width=1pt] (2.4,-2.8) -- (18.5,-2.8);

\node[anchor=east, font=\fontsize{4}{5}\selectfont\bfseries, text=gray!55] at (2.3, 2.8)
  {\textsc{US / Digital}};
\node[anchor=east, font=\fontsize{4}{5}\selectfont\bfseries, text=gray!55] at (2.3,-2.8)
  {\textsc{India / Craft}};

\node[pribox] (priors) at (1.2, 0) {
  \textbf{Inductive Priors}\\[2pt]
  {\tiny BM experience,}\\[-1pt]
  {\tiny mythology, motifs}
};
\draw[arr] (priors.north) to[out=90,in=180] (3.0, 2.8);
\draw[arr] (priors.south) to[out=270,in=180] (3.0,-2.8);

\node[dtbox]  (dt)       at ( 5.2,  2.8) {Digital\\Twin};
\node[usbox]  (loi)      at ( 8.2,  2.8) {BM 2025\\LOI (12--13')};
\node[usbox]  (postplan) at (10.7,  2.8) {Post-Accept.\\(15--18')};
\node[usbox]  (kickoff)  at (13.2,  2.8) {BM\\Kickoff};
\node[usbox]  (struct)   at (16.2,  2.8) {FEA\\(bounds est.)};
\node[shipbox](stage)    at (19.0,  2.8) {Phoenix\\Staging};
\node[usbox]  (feafull)  at (21.5,  2.8) {FEA Adj.\\(full SfM)};
\node[usbox]  (weld)     at (24.0,  2.8) {Superstructure\\Fabrication};
\node[usbox]  (transport)at (26.5,  2.8) {Load \&\\Transport};

\begin{scope}[on background layer]
  \node[draw=gray!40, dashed, rounded corners=5pt, fill=gray!6,
        fit=(struct)(weld), inner xsep=4pt, inner ysep=10pt,
        label={[font=\fontsize{3}{4}\selectfont,gray!60]above:Structural Design + FEA}] {};
\end{scope}

\node[inbox]  (iac)  at ( 4.5, -2.8) {IAF\\Constraints};
\node[inbox]  (iaf)  at ( 7.0, -2.8) {IAF 2025\\Proposal\\(8')};
\node[fabbox] (fab)  at (13.5, -2.8) {Craft Fab\\Odisha};
\node[shipbox](ship) at (16.5, -2.8) {Air Freight\\Shipping};

\node[mergebox] (bmc) at (28.5, 0) {BM\\Construct.};
\draw[arr] (feafull) -- (weld);
\draw[arr] (weld) -- (transport);
\draw[arr] (transport.east) to[out=0,in=90] (bmc.north);

\draw[arr] (dt)       -- (loi);
\draw[arr] (loi)      -- (postplan);
\draw[arr] (postplan) -- (kickoff);
\draw[arr] (kickoff)  -- (struct);
\draw[arr] (struct)   -- (stage);
\draw[arr] (stage)    -- (feafull);

\draw[arr] (iac) -- (iaf);
\draw[arr] (fab) -- (ship);
\draw[arr] (ship.north) to[out=90,in=270] (stage.south)
  node[pos=0.5, right, font=\fontsize{3}{4}\selectfont]{crates arrive};
\node[draw, rounded corners=2pt, fill=gray!15, text width=1.5cm, minimum height=0.55cm,
      align=center, font=\tiny, inner sep=2pt] (iafexec) at (9.5, -2.8) {IAF Exec.\\Feb 2025 (8')};
\draw[arr] (iaf) -- (iafexec);
\draw[dotted, gray!45, line width=0.7pt] (iafexec.east) -- (fab.west)
  node[midway, below, font=\fontsize{3}{4}\selectfont, text=gray!55]{\textit{indep.\ timelines}};

\draw[looparr]
  (dt.north west) to[out=145,in=35,looseness=5]
  node[above, font=\fontsize{3}{4}\selectfont, gray!60]{\textit{design iters.}}
  (dt.north east);
\draw[looparr]
  (fab.south west) to[out=215,in=325,looseness=5]
  node[below, font=\fontsize{3}{4}\selectfont, gray!60]{\textit{craft iters.}}
  (fab.south east);
\draw[looparr]
  (struct.north west) to[out=145,in=35,looseness=5]
  node[above, font=\fontsize{3}{4}\selectfont, gray!60]{\textit{FEA loop}}
  (struct.north east);

\draw[darr] (iaf.north) -- ++(0,1.2)
  -| (dt.south)
  node[pos=0.25, right, font=\fontsize{3}{4}\selectfont]{seeds DT};

\draw[darr] (kickoff.south) -- ++(0,-1.2)
  -| (fab.north)
  node[pos=0.25, left, font=\fontsize{3}{4}\selectfont]{authorises fab};

\draw[feedarr]
  (dt.south) to[out=280,in=160]
  node[pos=0.5, below, font=\fontsize{3}{4}\selectfont, text=teal!70, sloped]{templates $\rightarrow$}
  (fab.north west);

\draw[feedarr]
  (fab.north east) to[out=20,in=200]
  node[pos=0.5, above, font=\fontsize{3}{4}\selectfont, text=teal!70, sloped]{$\leftarrow$ SfM}
  (struct.south);

\draw[darr, red!45]
  (ship.north west) to[out=120,in=300]
  node[pos=0.5, left, font=\fontsize{3}{4}\selectfont, text=red!50]{delay}
  (struct.south);

\def\thumbsz{0.7cm}
\node[above=0.15cm of loi, inner sep=0pt]
  {\includegraphics[height=\thumbsz]{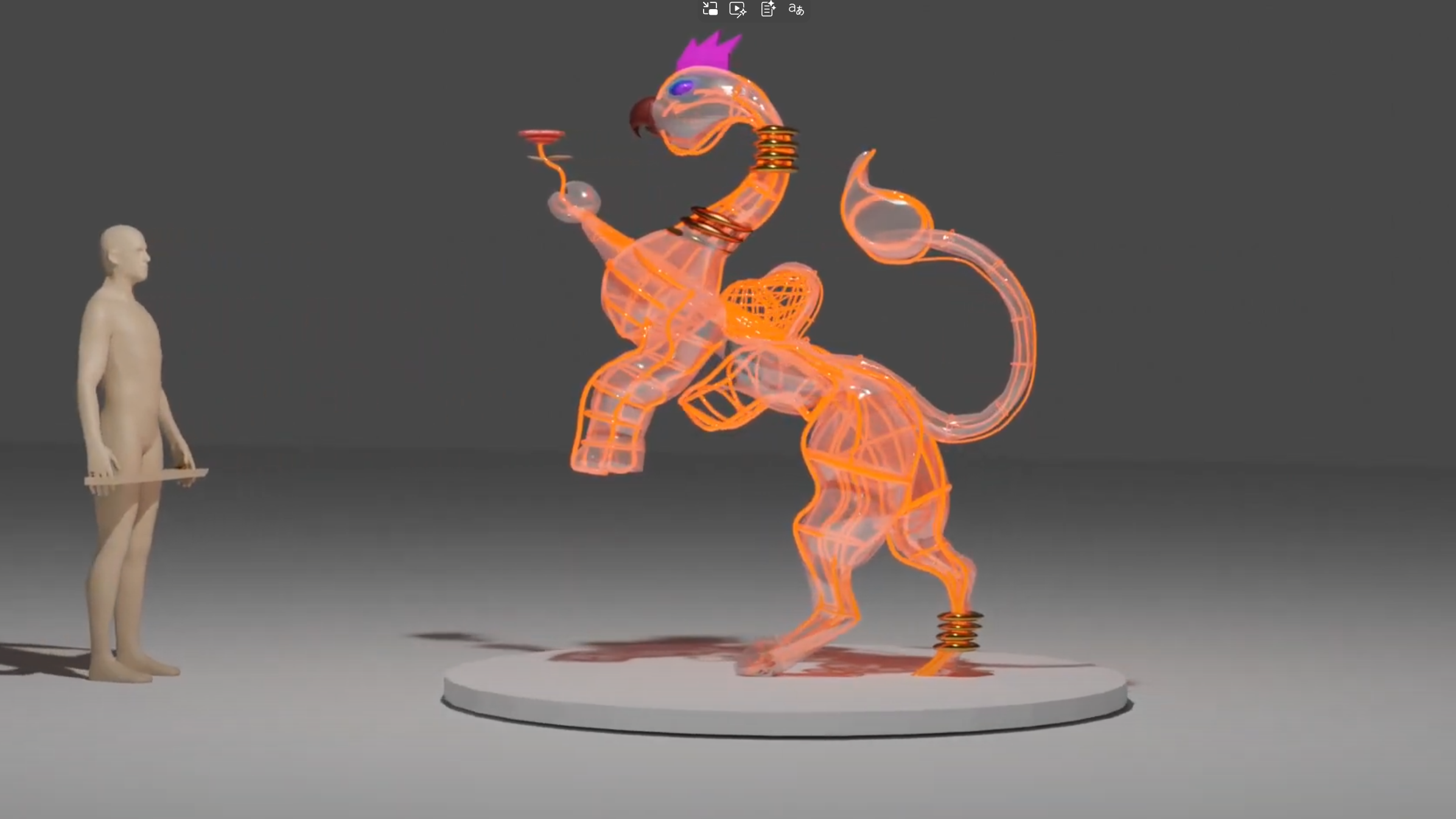}};
\node[above=0.15cm of postplan, inner sep=0pt]
  {\includegraphics[height=\thumbsz]{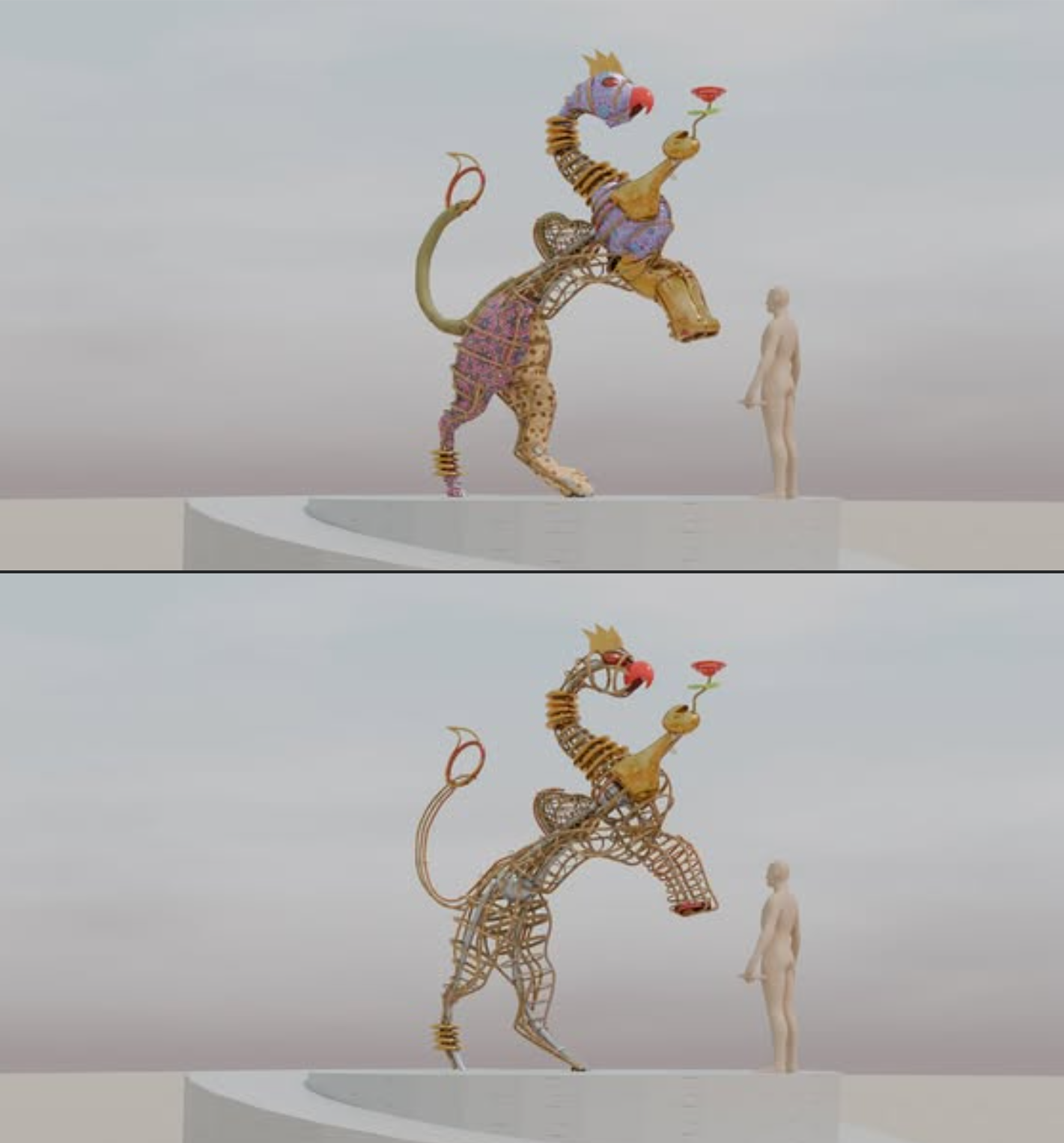}};
\node[above=0.15cm of kickoff, inner sep=0pt]
  {\includegraphics[height=\thumbsz]{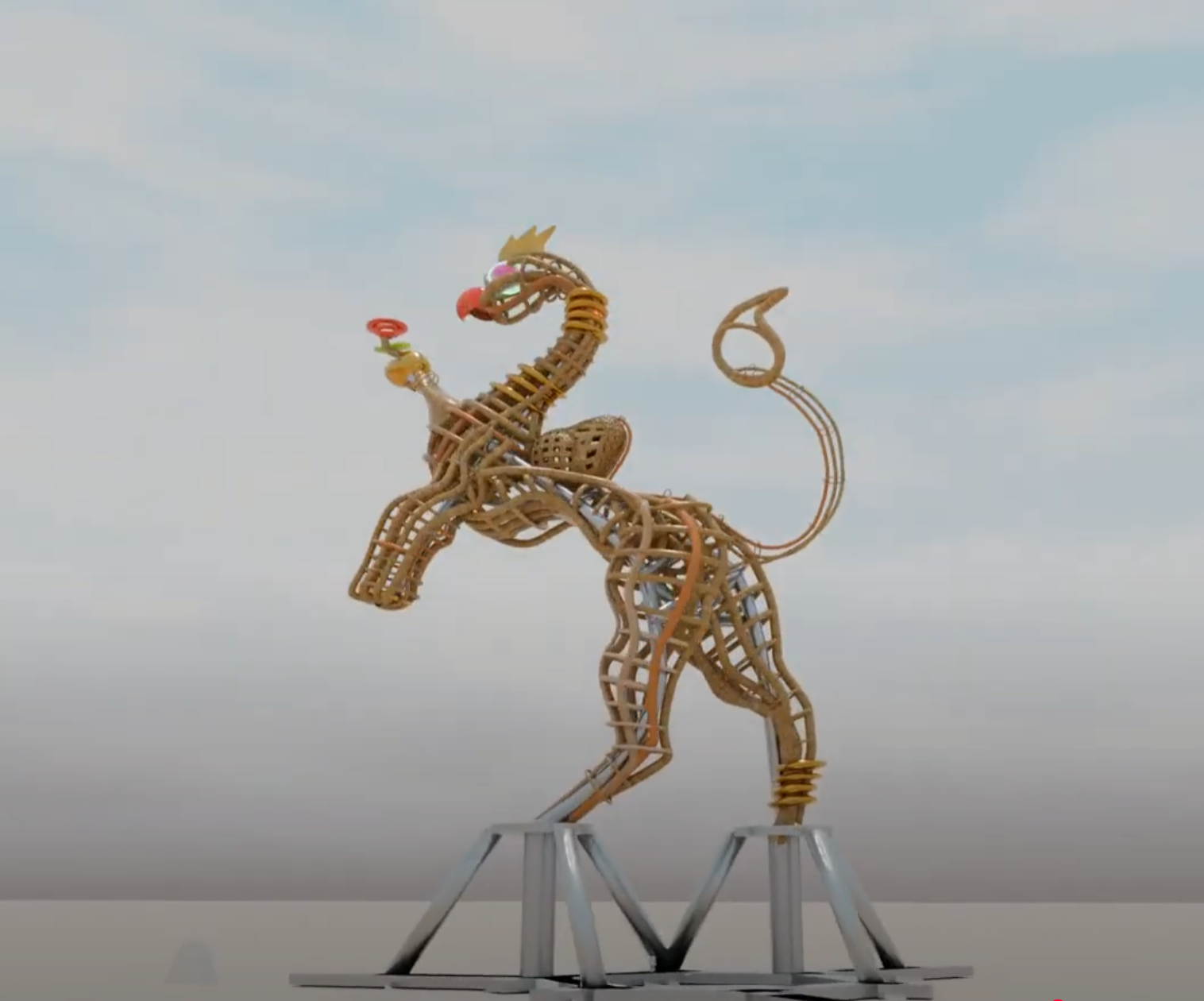}};
\node[above=0.15cm of struct, inner sep=0pt]
  {\includegraphics[height=\thumbsz]{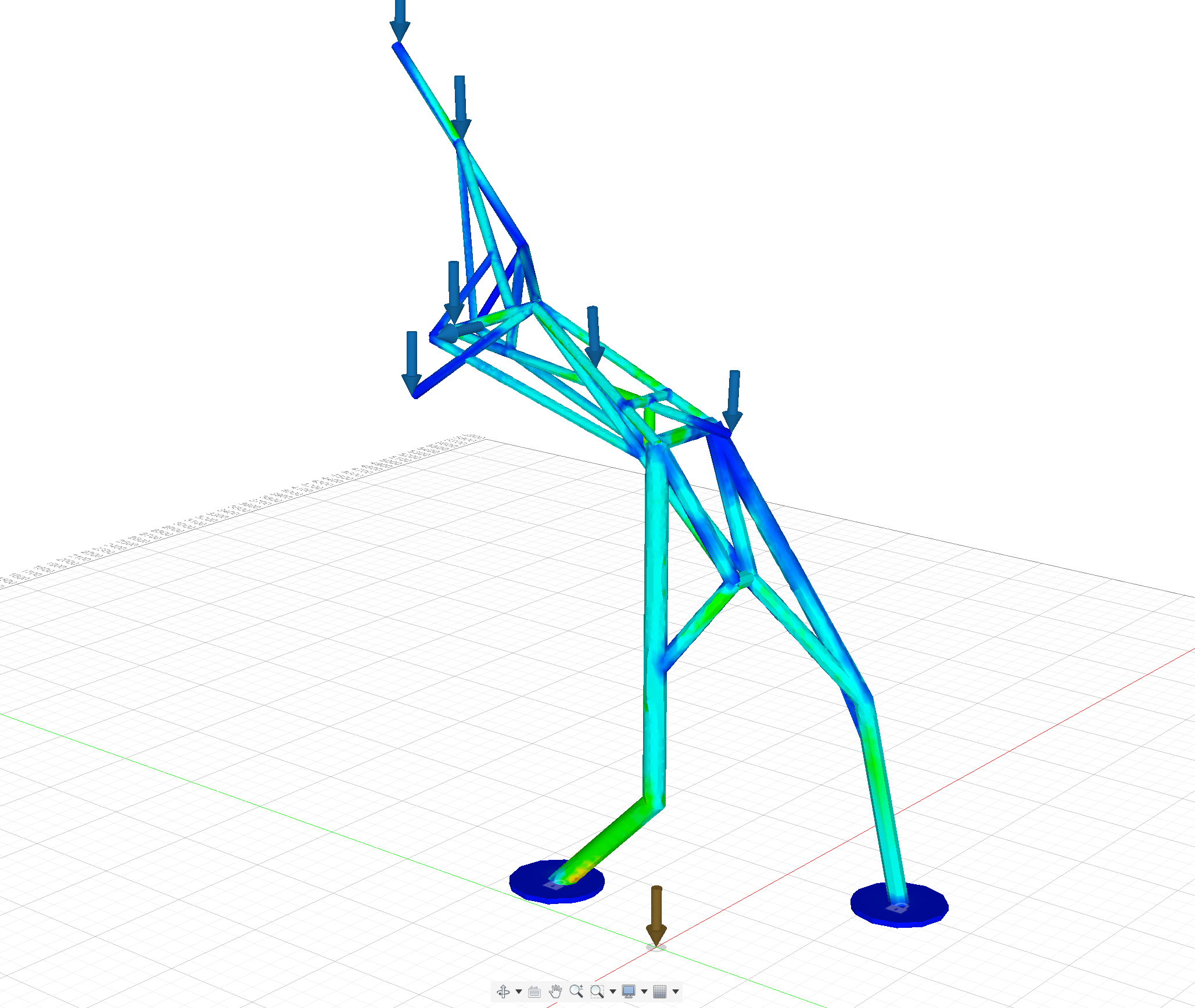}};
\node[below=0.15cm of iac, inner sep=0pt]
  {\includegraphics[height=\thumbsz]{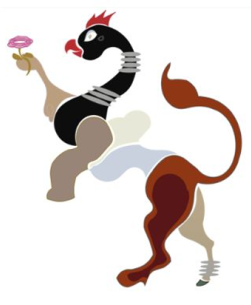}};
\node[below=0.15cm of iaf, inner sep=0pt]
  {\includegraphics[height=\thumbsz]{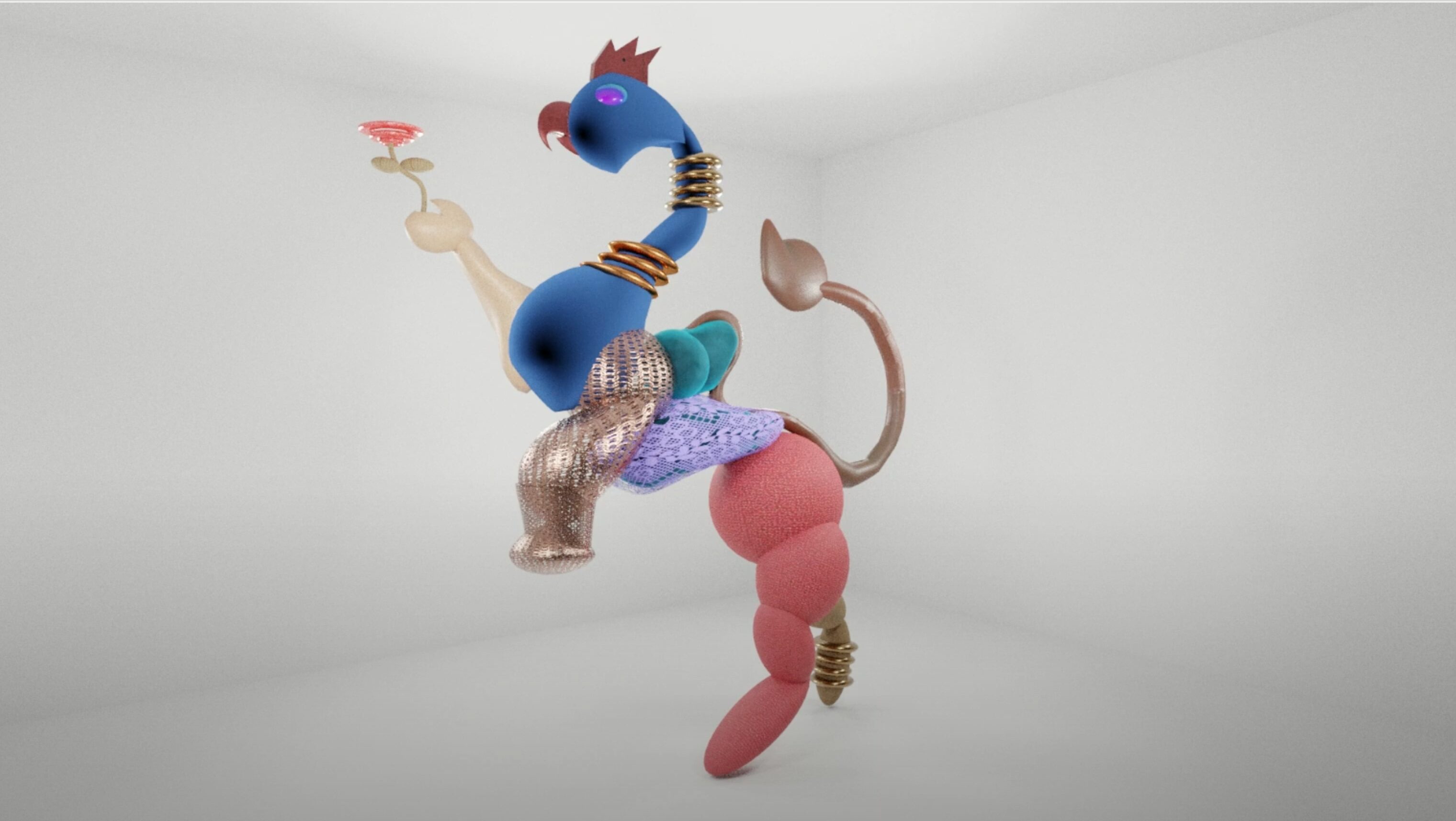}};
\node[below=0.15cm of fab, inner sep=0pt]
  {\includegraphics[height=\thumbsz]{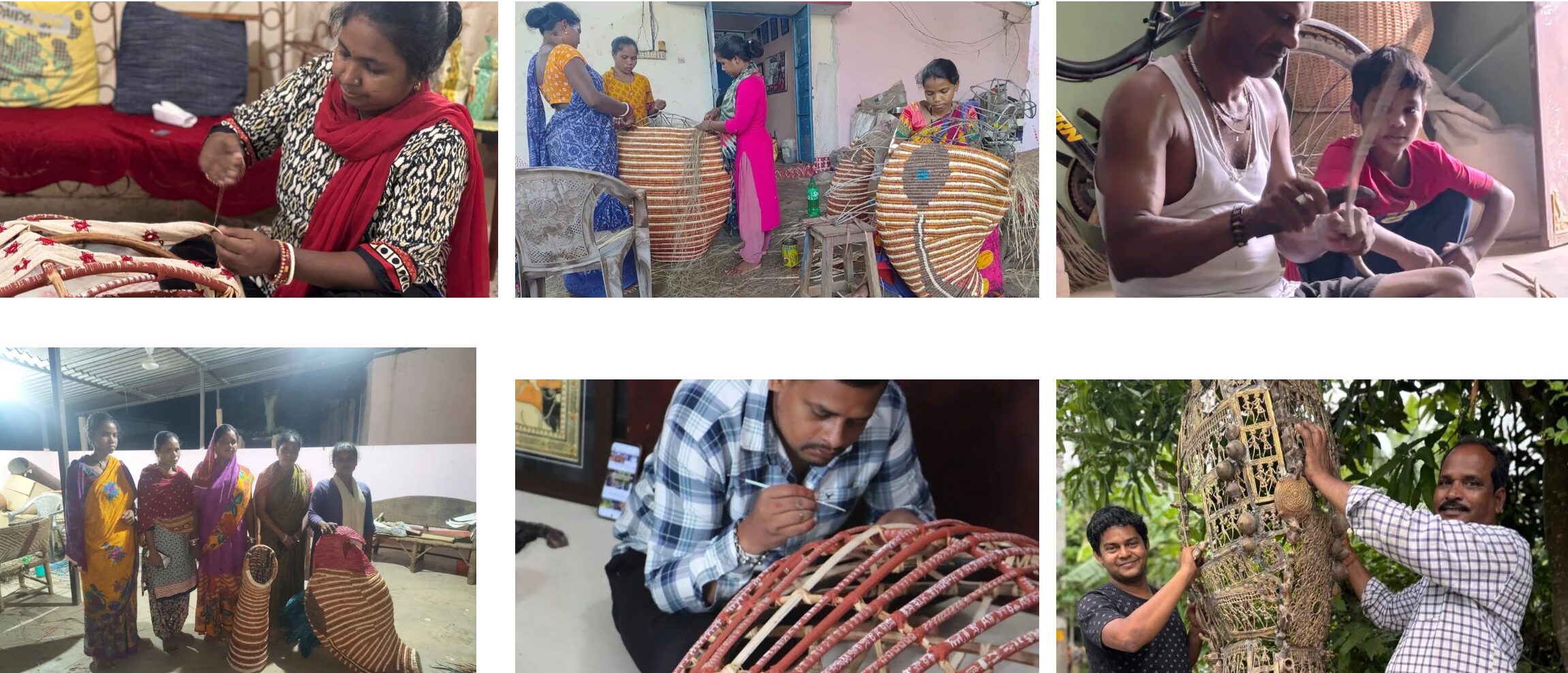}};
\node[above=0.15cm of stage, inner sep=0pt]
  {\includegraphics[height=\thumbsz]{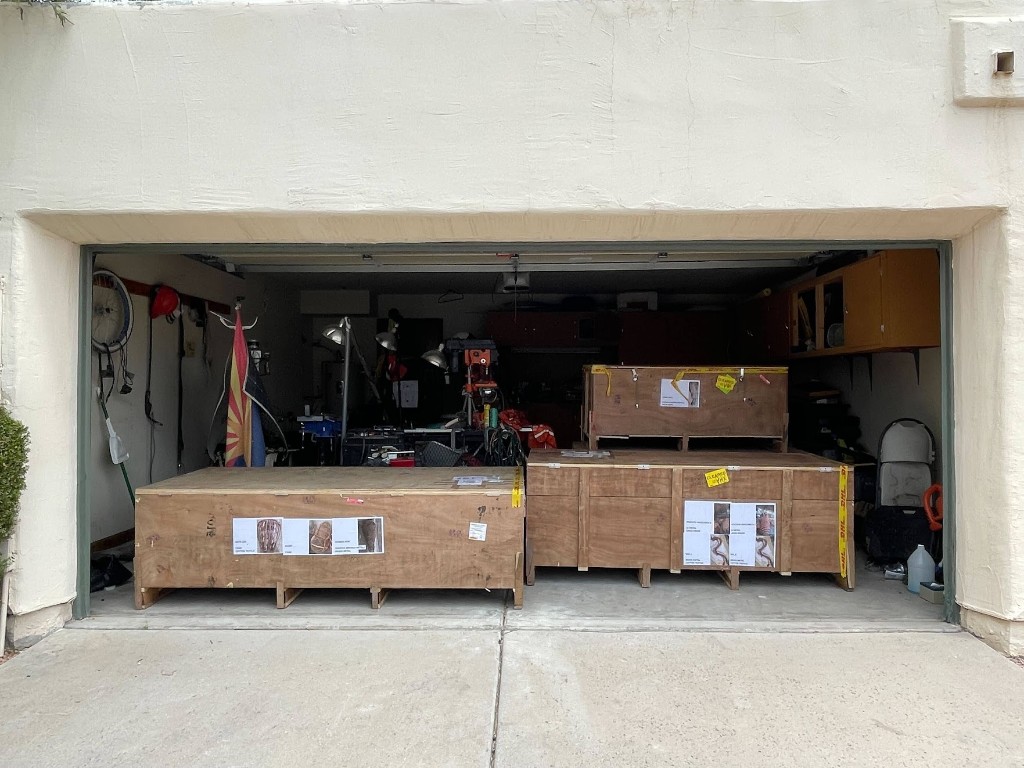}};
\node[above=0.15cm of feafull, inner sep=0pt]
  {\includegraphics[height=\thumbsz]{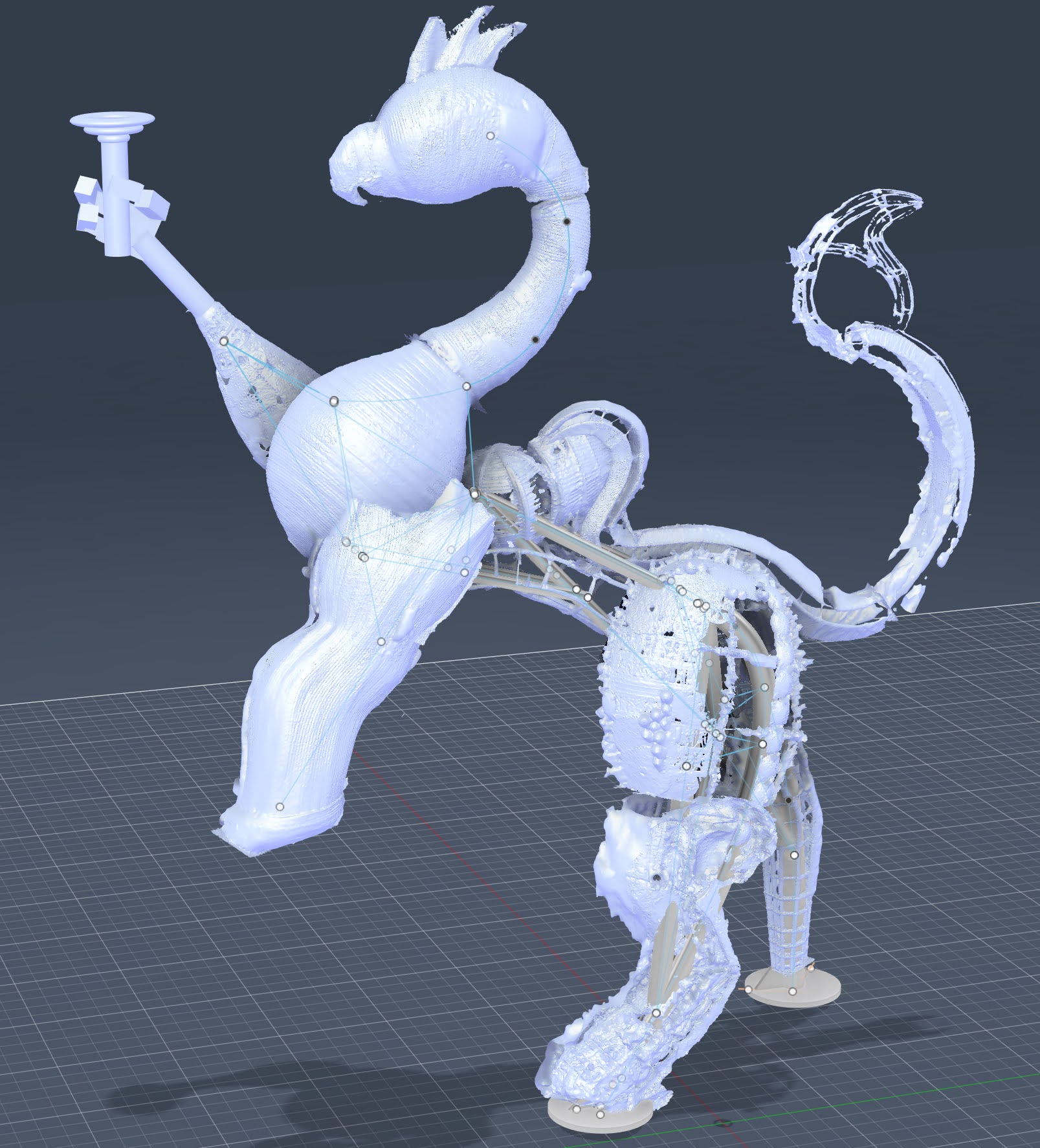}};
\node[above=0.15cm of weld, inner sep=0pt]
  {\includegraphics[height=\thumbsz]{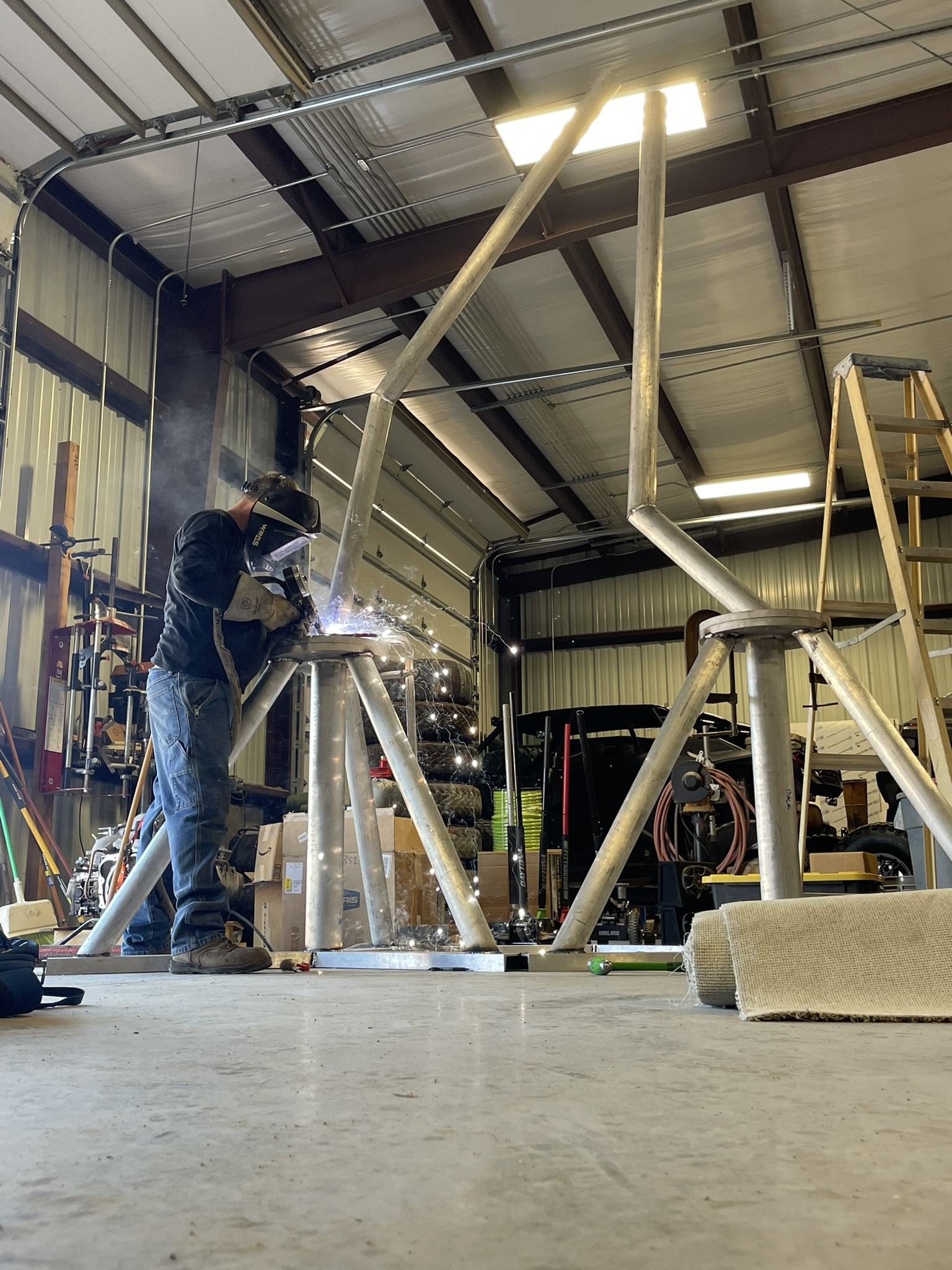}};
\node[above=0.15cm of transport, inner sep=0pt]
  {\includegraphics[height=\thumbsz]{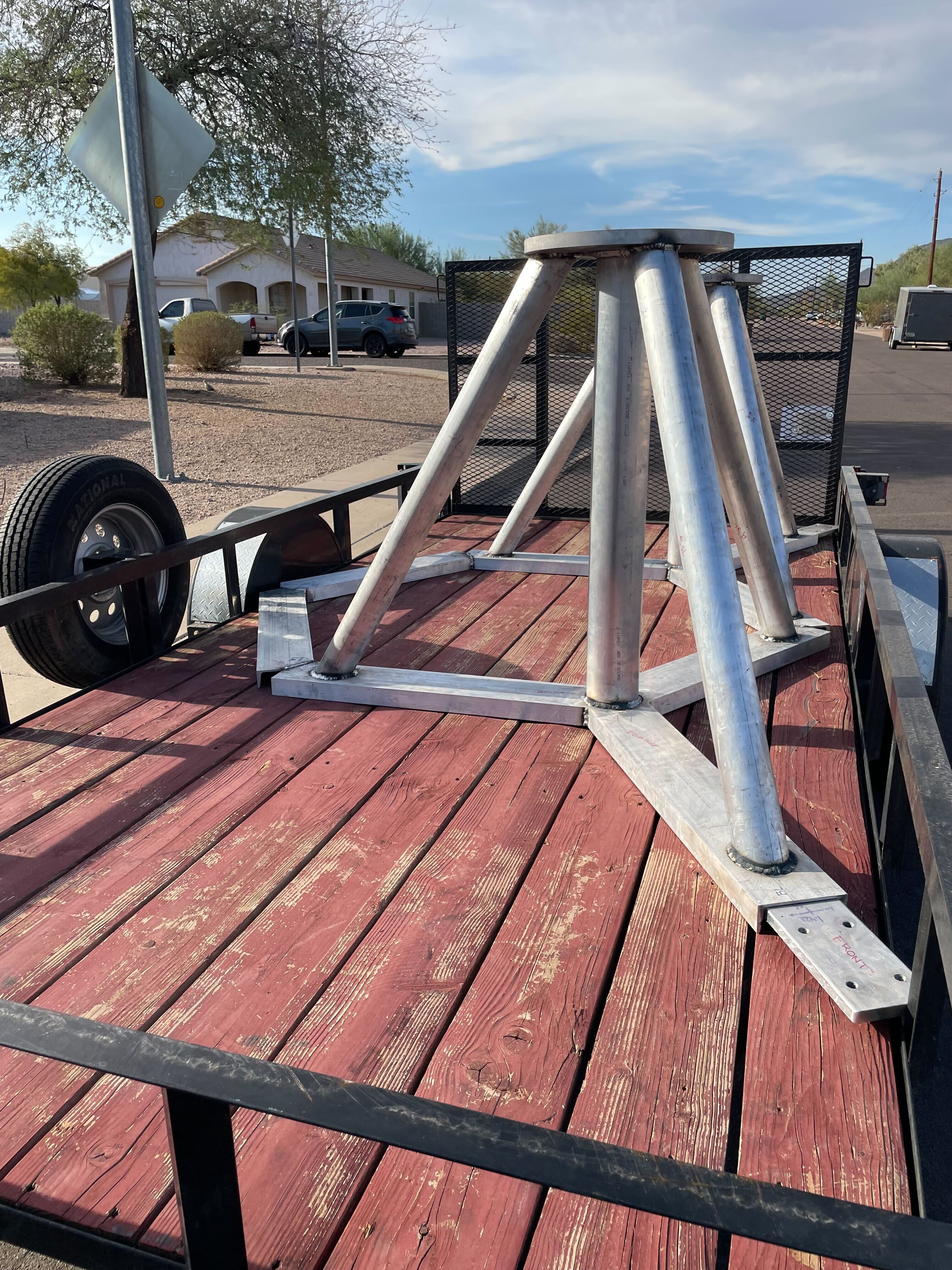}};
\node[below=0.15cm of bmc, inner sep=0pt]
  {\includegraphics[height=\thumbsz]{sculpture-main-photos.jpg}};

\end{tikzpicture}
\caption{\scriptsize{Project timeline for Navagunjara Reborn as two converging branches. \textbf{Inductive Priors} seed both rails. The \textsc{US/Digital} rail carries the approval and design spine. The \textsc{India/Craft} rail carries two independent timelines: the IAF proposal (submitted Oct~1, 2024) seeds the digital twin and executes independently at India Art Fair in Feb 2025; the BM craft fabrication is a separate timeline that begins only after BM Kickoff authorises it (downward dashed). The dotted line between them indicates independent timelines, not a causal dependency. During the craft build, templates flow from the digital twin to artisans and SfM scans return to structural design (teal arrows). Shipping delays compress the structural design window (red). Self-loops mark optimization iterations. Both rails merge at BM Construction---a single-sample draw from the collective design posterior.}}
\label{fig:workflow-schematics}
\end{figure*}

Design strategies shifted in response to external constraints. The original vision was far more radical: the full proposal called for a ceremonial full burn of the entire cane and textile structure at the festival's close—a literal Phoenix rising—leaving only the metal skeleton behind. This was not a vague aspiration but the philosophical centerpiece of the concept, echoing the myth of destruction and rebirth. FAST (Flame Art Safety Team) review and Black Rock City's LNT protocols rendered a full structural burn untenable at the 18-foot scale with embedded crafts. The pivot to controlled poofer flame effects was therefore a significant artistic adaptation—preserving the fire symbolism while abandoning the self-immolation arc. This shift had a structural consequence: since the cane and textile forms would survive the event rather than burn, the engineering emphasis shifted toward durability, modularity, and metal craftsmanship. The adaptation ultimately deepened the cultural integrity of the piece, as the preserved craft elements could be reused in future iterations. In retrospect, this pivot is best read as a constraint-driven philosophical shift rather than a failure of intent: symbolic fire was preserved while cultural material survived. This adaptation strengthened the balance between cultural fidelity and structural durability under festival conditions.

\begin{figure}[htpb]
    \centering
    \includegraphics[width=\linewidth]{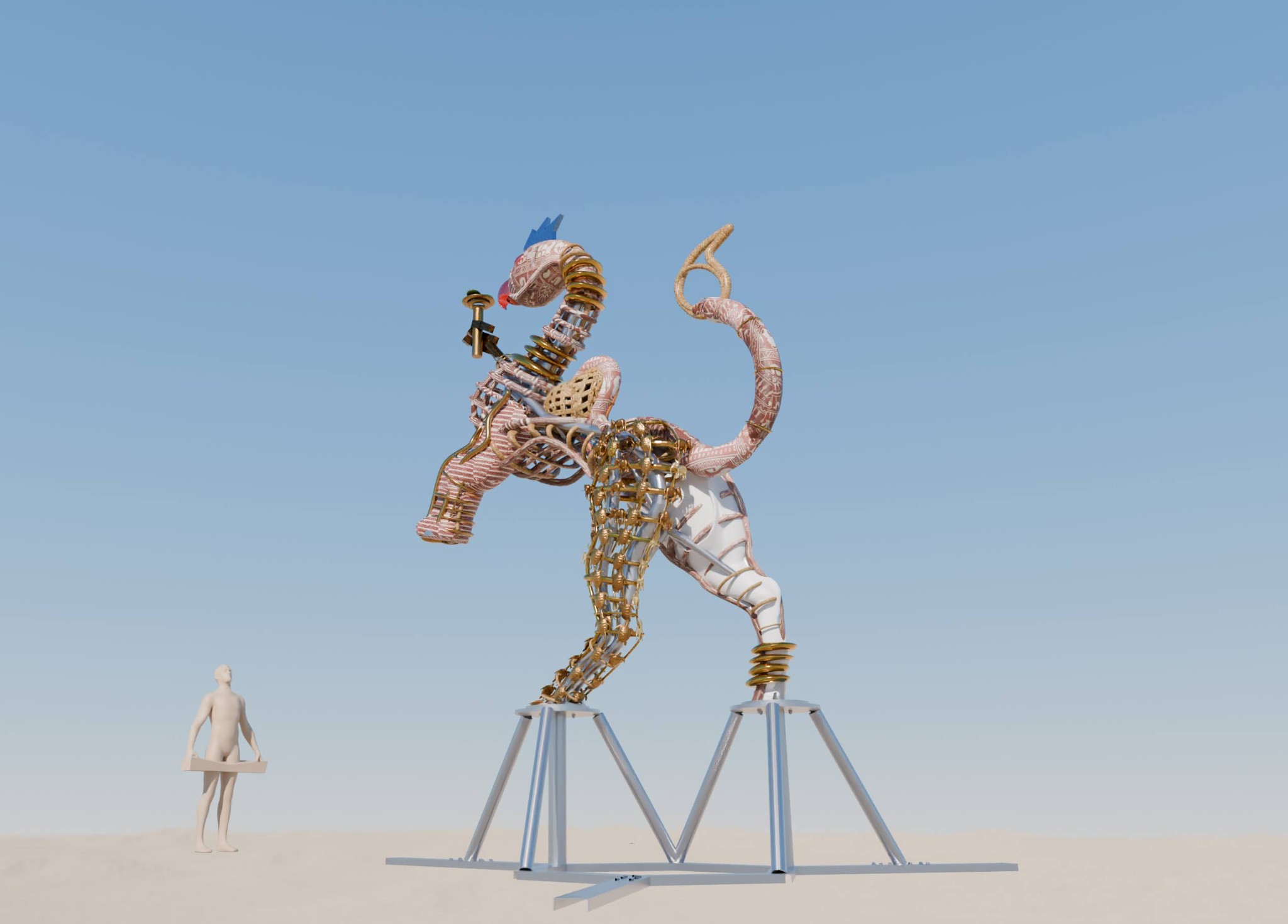}
    \caption{\small{Blender render of the Burning Man 2025 sculpture design with the controlled torch-based poofer flame effects configuration, with a human figure for scale. The hexagonal tripodal aluminum base is visible. This design followed the pivot from the original open-flame burn concept after FAST review and Black Rock City's LNT protocols necessitated a safer approach to fire.}}
    \label{fig:bm2025-flame-effects-render}
\end{figure}

Ultimately, the design framework proved flexible, accommodating dynamic changes in materials and parameters while preserving the integrity of the cultural and mythological foundations. The fire-emitting installation not only references the mythological Phoenix and Navagunjara archetypes, but also transforms ancient symbols into collective, participatory experiences. The act of public spectacle, as witnessed in the dramatic engagement around the sculpture, bridges cultural specificity and universal transformation, embodying the transcultural ambitions discussed herein. Beyond the intrigue and awe at the craftsmanship of the art, interactions at the event highlighted common threads globally on various craft forms, in particular dhokra and lost wax~\cite{higham1996bronze}, and what it might have meant across different cultures. 
\begin{figure*}[h]
    \centering
   
      \begin{subfigure}{0.3\linewidth}
        \centering
        \includegraphics[width=\linewidth]{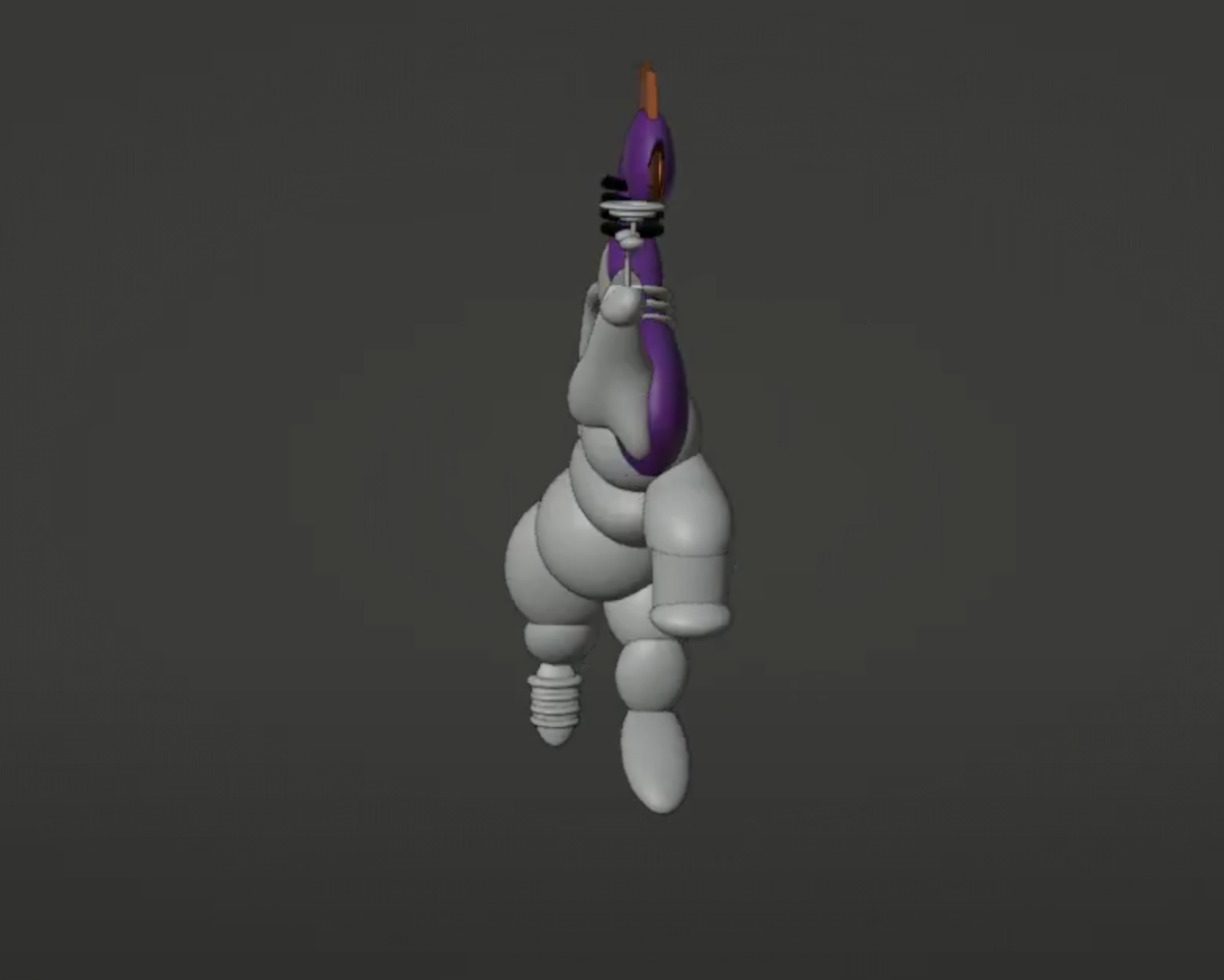}
        \caption{First phase: geometric blending of parametric 3D surfaces such as NURBS sphere, Bezier curves, for the proposed art.}
        \label{fig:spherical-primitives}
    \end{subfigure}
    \hfill
      \begin{subfigure}{0.3\linewidth}
        \centering
        \includegraphics[width=\linewidth]{india-art-fair-8ft-sculpture.jpg}
        \caption{Digital twin of the 8-foot sculpture with minimalist texturing, positioned within the proposed India Art Fair 2025 booth layout---a 30~sqm exhibition stall at Devi Art Foundation, New Delhi---used to validate spatial fit and presentation prior to fabrication.}
        \label{fig:india-art-fair-sculpture}
    \end{subfigure}
    \hfill
     \begin{subfigure}{0.25\linewidth}
        \centering
        \includegraphics[width=\linewidth]{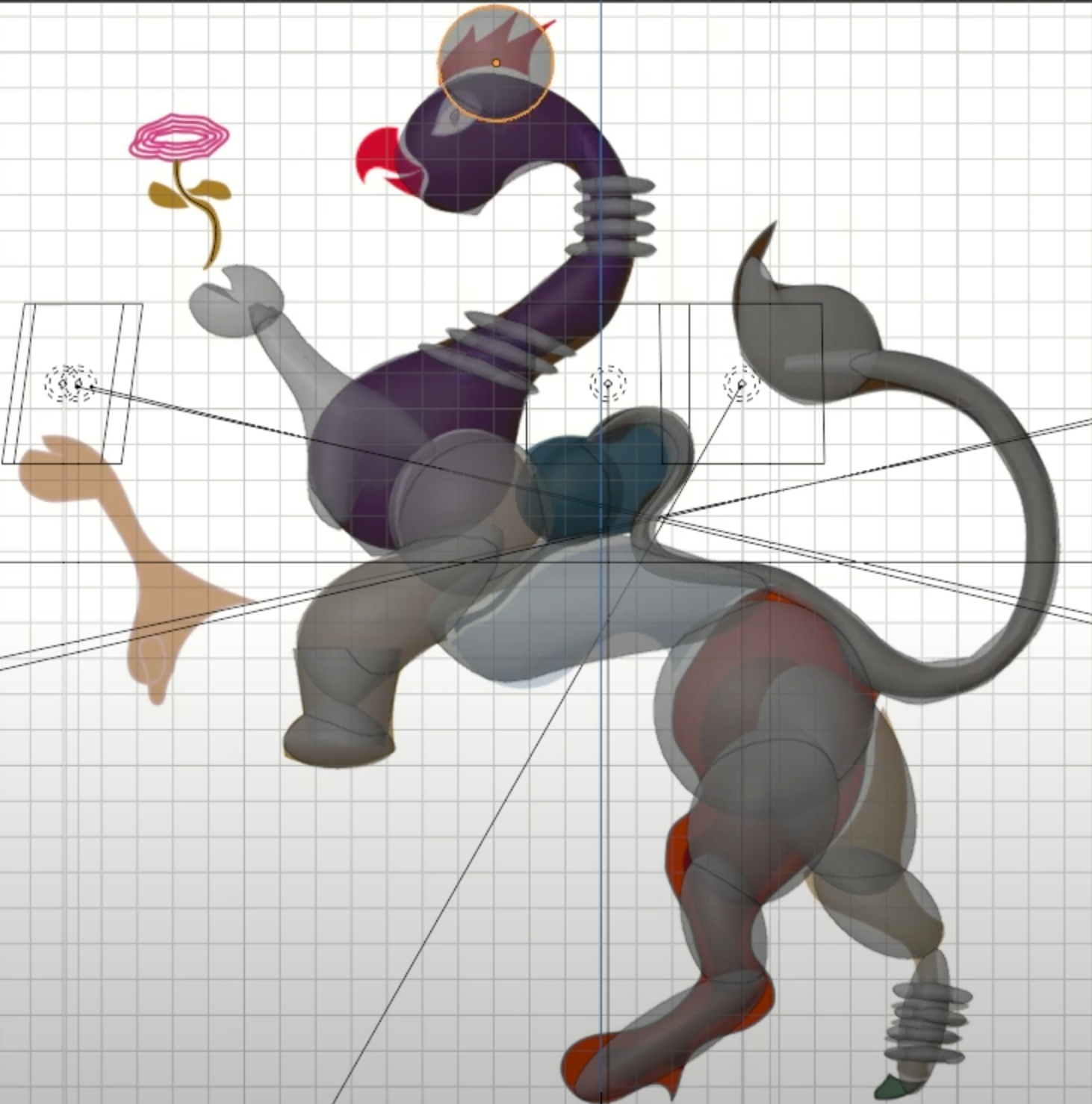}
        \caption{Second phase: Fusion of open-source animal figurine parts into parametric 3D design for India Art Fair 2025,  steps involved iterations in Blender.}
        \label{fig:spheres}
    \end{subfigure}

\hfill 

    \begin{subfigure}{0.3\linewidth}
        \centering
        \includegraphics[width=\linewidth]{transparent-with-glow-digital-twin-nov-2024.png}
        \caption{Animal form shapes were blended into the first-order geometric model for Burning Man 2025 Honoraria Art grant letter of intent (LOI).}
        \label{fig:transparent-glow}
    \end{subfigure}
    \hfill
      \begin{subfigure}{0.3\linewidth}
        \centering
        \includegraphics[width=\linewidth]{bm2025-fullproposal-.png}
        \caption{Burning Man 2025 honoraria art grant artistic 3D renderings for full proposal derived from Blender digital twin of the sculpture.}
        \label{fig:bm2025-proposal}
    \end{subfigure}
\hfill 
    \begin{subfigure}{0.25\linewidth}
        \centering
        \includegraphics[width=\linewidth]{cane-digital-twin-apr-2025.png}
        \caption{Final sculpture design for Burning Man 2025 under the original open-flame burn concept, prior to pivoting to controlled poofer flame effects (see Figure~\ref{fig:bm2025-flame-effects-render}).}
        \label{fig:cane-digital}
    \end{subfigure}

    \caption{Design iterations from Sep 2024 to May 2025: Iterative progression of digital sculpting and structural design for Navagunjara Reborn—from early concepts developed for the India Art Fair, through the Burning Man letter of intent and full proposal, to the finalized design. Each stage reflects adaptive refinement in posture, artistic detail, and engineering, integrating feedback from multidisciplinary teams and evolving constraints across local and international contexts.}
    \label{fig:grouped-all}
\end{figure*}

\begin{figure*}[htpb]
    \centering
    \includegraphics[width=7in]{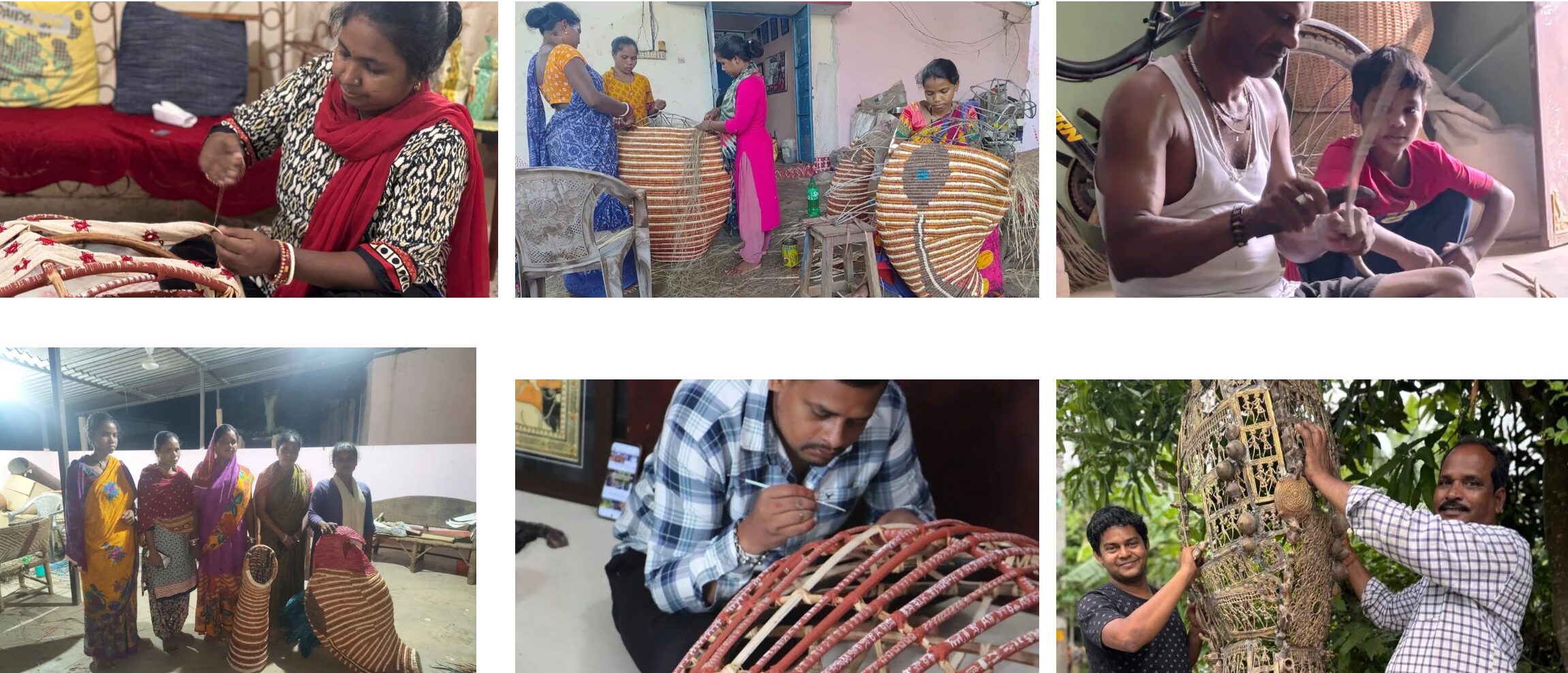}
    \caption{\small{Odia artisans fabricate sculptural components for Navagunjara Reborn using traditional techniques such as wireframing, cane weaving, and metalwork. The integration of digital design references and 3D-printed templates guided the authentic creation of each element, exemplifying a collaborative craft-computation workflow that bridges local knowledge and global innovation.}}
    \label{fig:artisans}
\end{figure*}

\section*{Key Contributions}
The primary contributions of this work are summarized as follows:
\begin{itemize}
    \item Demonstration of one-shot full-scale assembly of the sculpture directly on-site in the Black Rock Desert—never previously assembled, not even partially. In hindsight, this amounts to committing to a single sample drawn from a probabilistic design distribution, without rehearsal. That it stood validates, retrospectively, that the digital-physical pipeline had generated sufficient mutual information to make one draw from the posterior enough.
    \item An adaptive engineering philosophy where the superstructure was designed with consideration to the unique topologies of handcrafted components, rather than forcing traditional crafts to conform to rigid pre-determined topologies of truss structures and such.
    \item A novel global digital-physical feedback loop that used Structure-from-Motion (SfM) photogrammetry to create digital twins of evolving craft pieces, enabling near real-time design updates and situational awareness between distributed teams in India and the U.S.
    \item The dual-use of DeepGIS~\cite{deepgis_xr_github}, a web-based digital twin ecosystem originally for earth and Space sciences, as both an accessible 3D model viewer for artisans and a workflow engine for generating metrically accurate, 1:1 2D paper-printable templates for fabrication.
    \item The development of a practical and reproducible open framework for large-scale, culturally-inspired public art, offering a new model for interdisciplinary collaboration that integrates traditional craft, computational design, and engineering in extreme environments. Source code, Blender tools, FreeCAD truss scripts, and orthographic PDF generation utilities are publicly available~\cite{navagunjara_reborn_github}.
    \item A philosophical framing, developed retrospectively, that interprets the final artifact as a convergence of multiple knowledge systems and constraints rather than a top-down execution of a fixed design.
\end{itemize}

\section{Design and Build Workflow}

The Navagunjara Reborn project demanded a process that could handle logistical uncertainty while integrating digital precision with heritage crafts traditions. The workflow centered on linked digital and physical representations. The 3D model captured artistic intent, informed engineering choices, supported cross-continental communication, and provided a reproducible fabrication reference. Scaled 3D prints enabled tolerance checks and form communication, helping storyboard the sculpture much like clay maquettes in automotive design.

Structural and fire safety requirements for large-scale art installations at Black Rock City—including rigorous wind/safety engineering and flame effects protocols—shaped the iterative development of both digital models and physical assemblies. Structural compliance, from wind load calculations to flame effects related safety, was not a static hurdle but a continuously revisited variable—refined through simulations and prototyping, and small-scale field testing. The digital-physical feedback loop enabled rapid responsiveness to evolving logistical uncertainties and on-site challenges. Storms, material substitutions, and on-playa improvisations were treated as expected operating conditions rather than as exceptions to a fixed plan.

Key elements included:
\begin{itemize}
    
\item 2D-to-3D Translation: Initial sketches were converted into parametric geometry, enabling rapid iteration between artistic concept and engineering feasibility.

\item Digital and Physical Models: 3D printing and scaled prototypes provided tangible feedback for structural testing and fabrication planning.

\item Structure-from-Motion Photogrammetry: Used to digitize evolving sculptural details into meshes, enhancing situational awareness across distributed teams.

\item Optimization for Fabrication: Mesh geometries were optimized for lightweight aluminum tubing and composite structures, balancing aesthetics with load-bearing requirements.

\item Craft–Computation Integration: Orthographic projections generated via DeepGIS linked NURBS and Bézier-based surfaces, SfM-derived polygon meshes of evolving craft pieces, and hand-executed craft panels, uniting Odia artisanship with computational workflows. Meshes reconstructed during fabrication served as both progress records and geometry inputs for structural design.

\end{itemize}

Blender, FreeCAD, and Fusion 360 anchored the modeling pipeline, while DeepGIS extended design intent into metrically accurate paper-based guides. Iterative digital updates synchronized with on-site craft progression in Odisha, ensuring fidelity between vision and execution.

This hybrid workflow exemplifies a systems approach to Big Art, where logistical uncertainty is managed by embedding computational flexibility into craft practice. Flame effects and interactive lighting, choreographed within this framework, symbolized the Phoenix theme of rebirth and transformation.

\begin{figure}[htpb]
        \centering
        \includegraphics[width=0.85\linewidth]{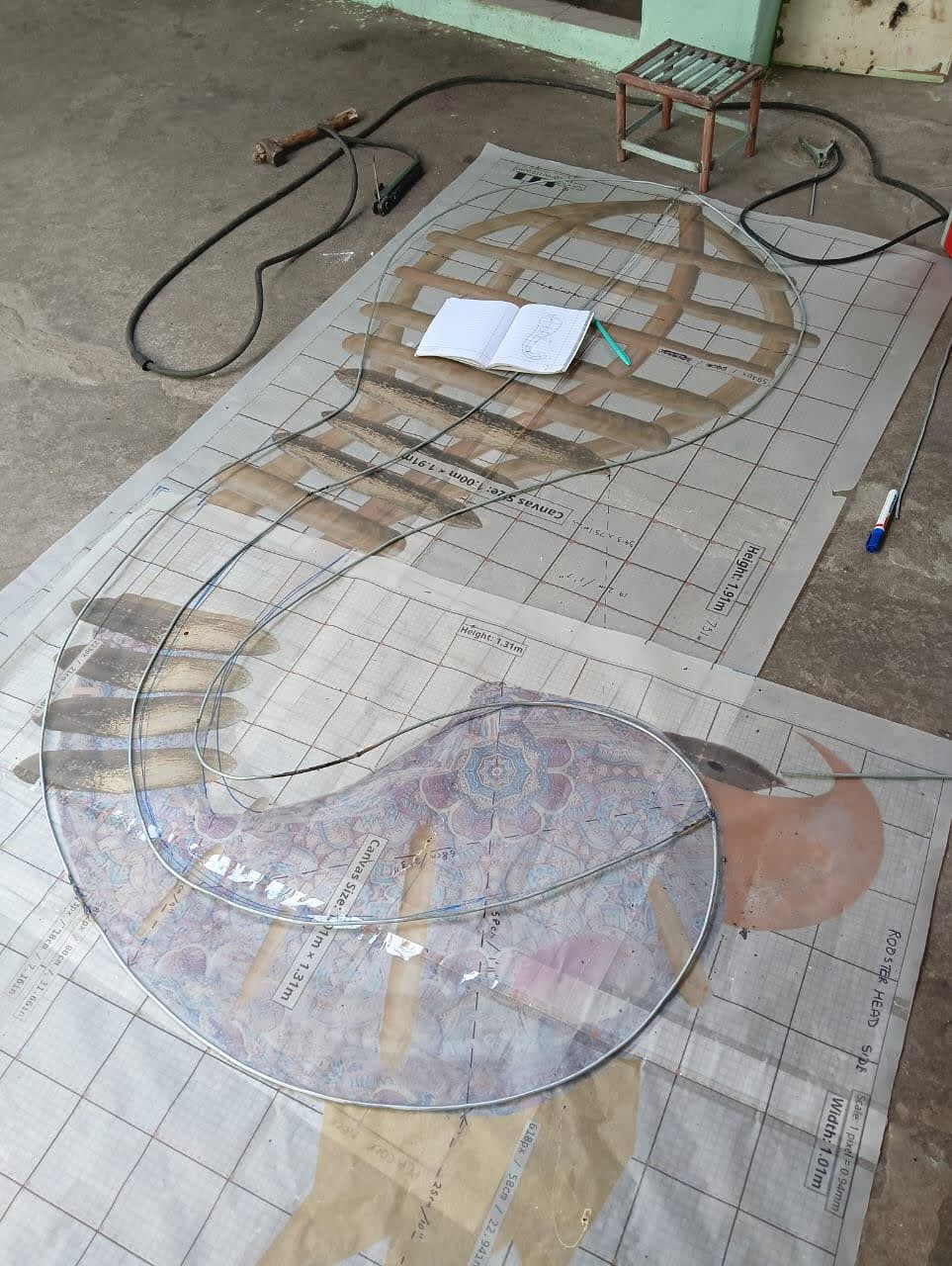}
    \caption{\small{Design translation and craft layout: Digital orthographic projections and printed templates guided Odia artisans in fabricating wireframes and panels. This process demonstrates the integration of computational accuracy with traditional craftsmanship, bridging physical fabrication and digital intent across continents through shared measurement and visualization tools}}
    \label{fig:deepgis-orthomap}
\end{figure}

\subsection{End-to-end 2D-3D-2D design pipeline}
The workflow began with collaboratively generated artistic sketches, translated into Blender 3D models over multiple iterations. At later stages of the build, 3D-scanned crafts meshes were optimized in FreeCAD and Fusion 360, resulting in load-bearing aluminum superstructure designs. 3D-printed scale prototypes allowed for testing mechanical tolerances and identifying potential failure points. 

The 3D model of the sculpture is a heterogeneous composite of three geometry types, all assembled within Blender: (i) \emph{parametric surfaces}—spheres, NURBS, and Bézier curves used to define the overall form and establish proportions; (ii) \emph{public-domain polygon meshes} of animal reference models sourced from Printables.com, remixed and sculpted to match the nine-animal Navagunjara morphology; and (iii) \emph{SfM-derived meshes} reconstructed from photogrammetry scans of actual craft pieces during fabrication in Odisha, integrated back into the model to reflect the artisans' realized forms. The design process in Blender began with parametric modeling, where spheres were positioned, inflated, and sculpted to match 2D projections as sketched collaboratively. The design was accepted for India Art Fair 2025, and the public-domain animal meshes enhanced anatomical realism and proportion. Canewood forms were modeled using Bézier curves, while NURBS surfaces constrained the animal shapes according to aesthetic and structural goals.

The lead artist had previously collaborated on the initial Navagunjara Reborn concept and developed the 3D model for India Art Fair 2025, producing an 8-foot design constrained by the 30~sqm exhibition floor area. The IAF proposal was submitted on October~1, 2024, as a collaborative installation between Boito and Earth Innovation Hub, supported by Devi Art Foundation as the production sponsor. This prior work—the model, orthographic workflow, and artisan relationships established for IAF—became the direct foundation for the Burning Man project. The sculpture scaled progressively across proposal stages: 8 feet for IAF 2025, 12--13 feet for the BM Honoraria Letter of Intent, and 15--18 feet for the BM full proposal, settling at 18 feet for the final build.
Posture was modified to stabilize the stance, with the deer leg (rear right) having a wider stance than the original letter of intent design. 

To ensure sustainability, the Earth Innovation Hub (EIH) initially advocated for the use of digital tools—such as ground outlines, projectors, and on-site 3D prints—for sculpting animal forms in canewood or metal. Despite this, the team opted for the established workflow developed for the India Art Fair 2025, proceeding with full-scale (1:1) orthographic prints. This, coupled with the sheer size of the components (up to 2 meters in length), posed several technical challenges: achieving precise metric accuracy during printing, configuring print widths to match available paper formats, and aligning projections from multiple views with the proper orientation for each individual part.

\begin{figure}[htpb]
    \centering
      \begin{subfigure}[b]{0.48\linewidth}
        \centering
        \includegraphics[width=\linewidth]{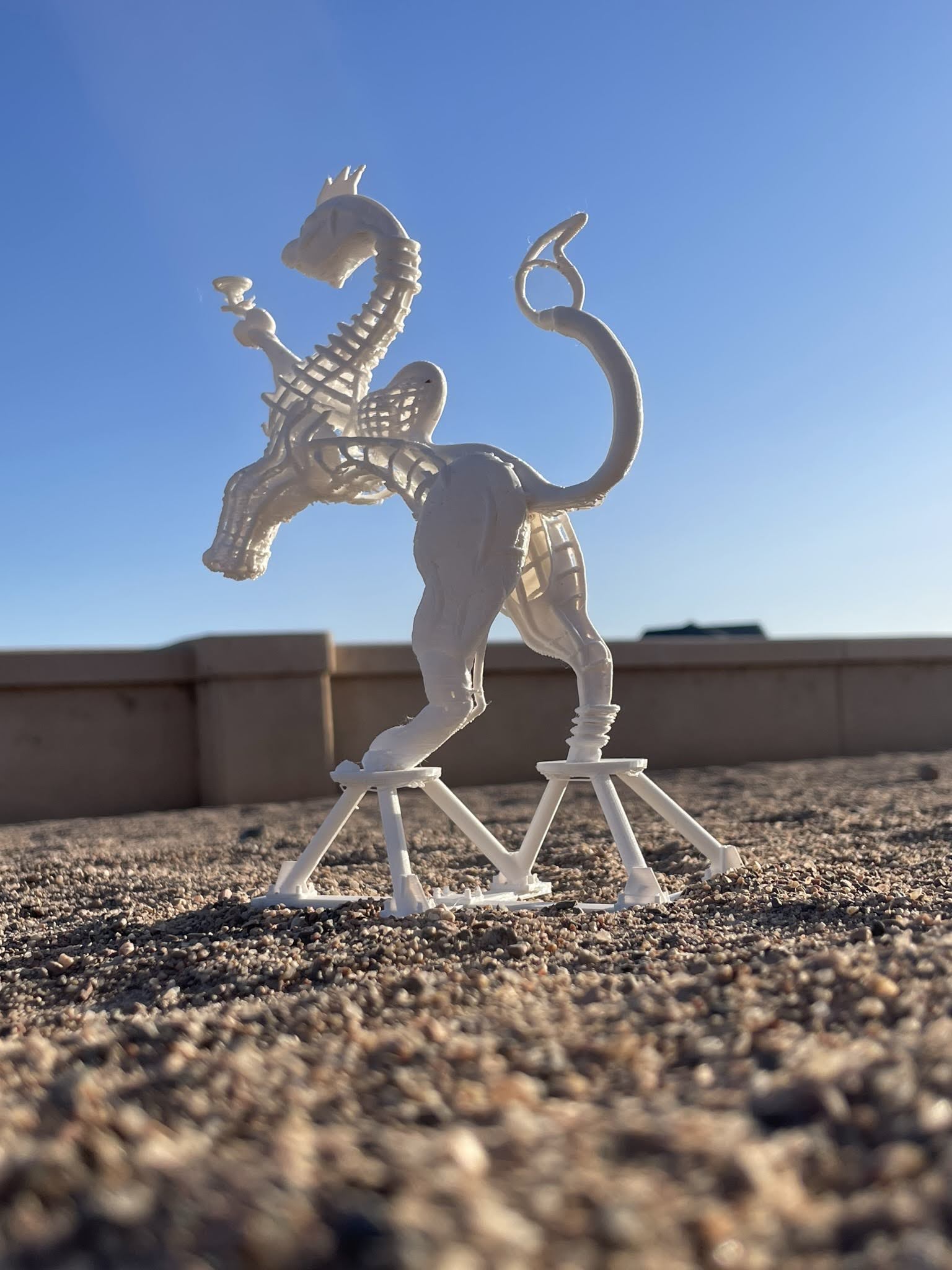}
        \caption{3D-printed scale prototype of Navagunjara Reborn (April 2025), used to validate structural tolerances and communicate design intent to artisan teams prior to full-scale fabrication.}
        \label{fig:deepgis_measurement}
    \end{subfigure}
    \hfill
      \begin{subfigure}[b]{0.48\linewidth}
        \centering
        \includegraphics[width=\linewidth]{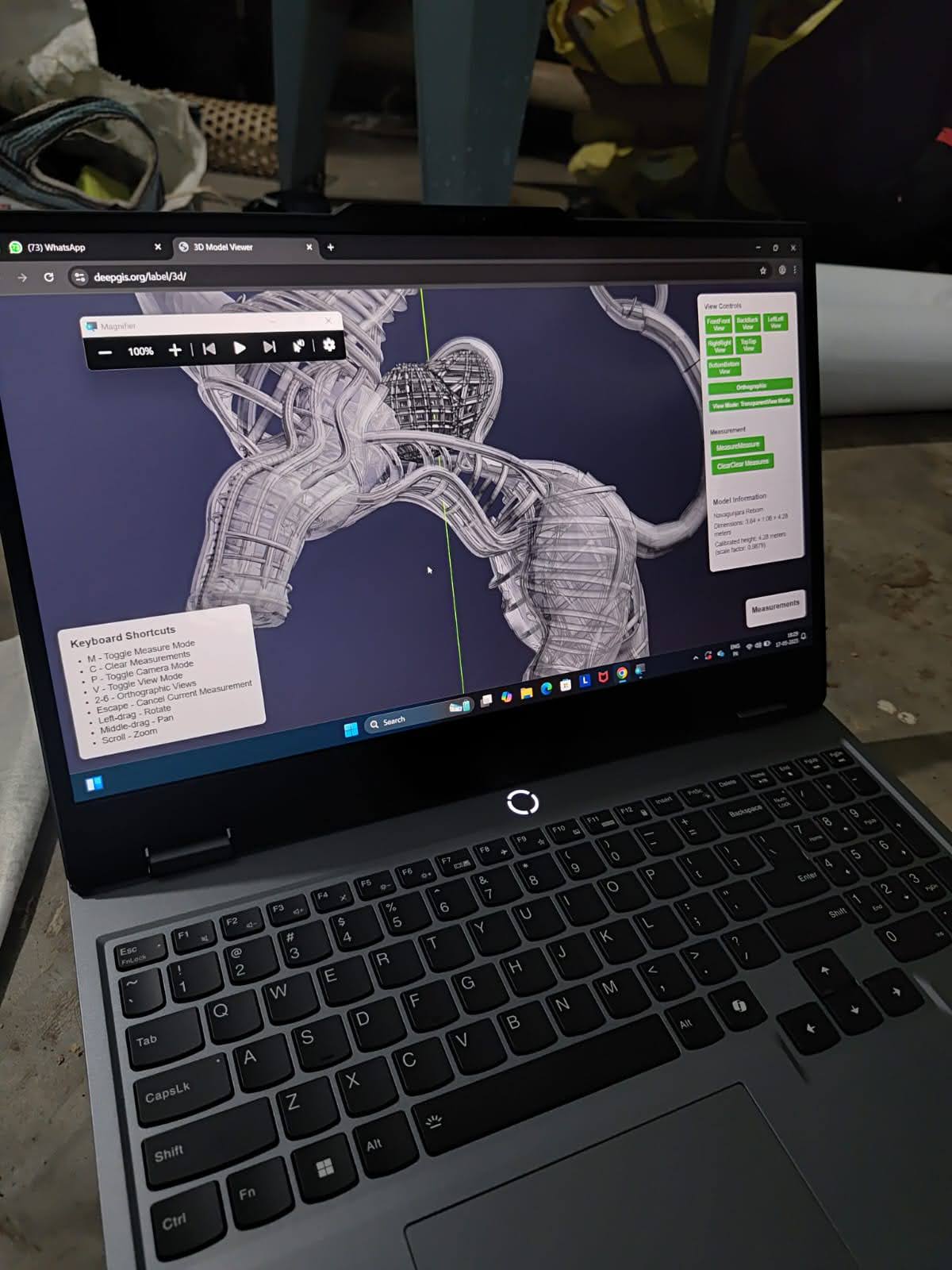}
        \caption{DeepGIS web-based 3D digital twin viewer displaying the Navagunjara Reborn model, enabling artisans in India to access real-time design references and verify component dimensions remotely.}
        \label{fig:deepgis_viewer}
    \end{subfigure}
    \caption{\small{Physical and digital twins in the design pipeline: a 3D-printed scale model of Navagunjara Reborn used for structural validation (left), and the DeepGIS web viewer displaying the digital twin for remote design review and orthographic projection export (right). Together, these tools formed a closed feedback loop between the design team in Arizona and the artisan teams in Odisha.}}
    \label{fig:deepgis_group}
\end{figure}

One critical requirement was that the workflow be dynamic and repeatable---avoiding manual screenshot extraction from Blender and ensuring that all prints accurately represented the latest 3D designs. On-site teams experienced difficulty exporting print-ready files from Blender, prompting the adoption of the DeepGIS web application as a robust and reusable solution. DeepGIS not only facilitated dynamic visualization of 3D models, but also automated the generation of orthographic projections and 1:1 printable templates. Eventually, Blender became available on a laptop computer on-site in India, enabling direct artist access to the digital model as well as verification of dimensions, assembly topologies, and pattern references illustrated in the 3D design files.

DeepGIS~\cite{deepgis_xr_github}, originally developed as a web-based decision support tool for earth and space sciences research and education, demonstrated dual-use as both a high-fidelity 3D model viewer and a workflow engine for printable, metrically accurate animal part templates. The platform allowed for the measurement of geodesic distances along manifold surfaces, which proved essential when estimating the lengths of canewood required for complex wireframes. This integration of DeepGIS streamlined communication between digital designers and craftspeople, ensuring dynamic updates and reducing the risk of \emph{design drift} during iterative fabrication.
\begin{figure}[htpb]
    \centering
    \begin{subfigure}{0.48\linewidth}
        \centering
        \includegraphics[width=\linewidth]{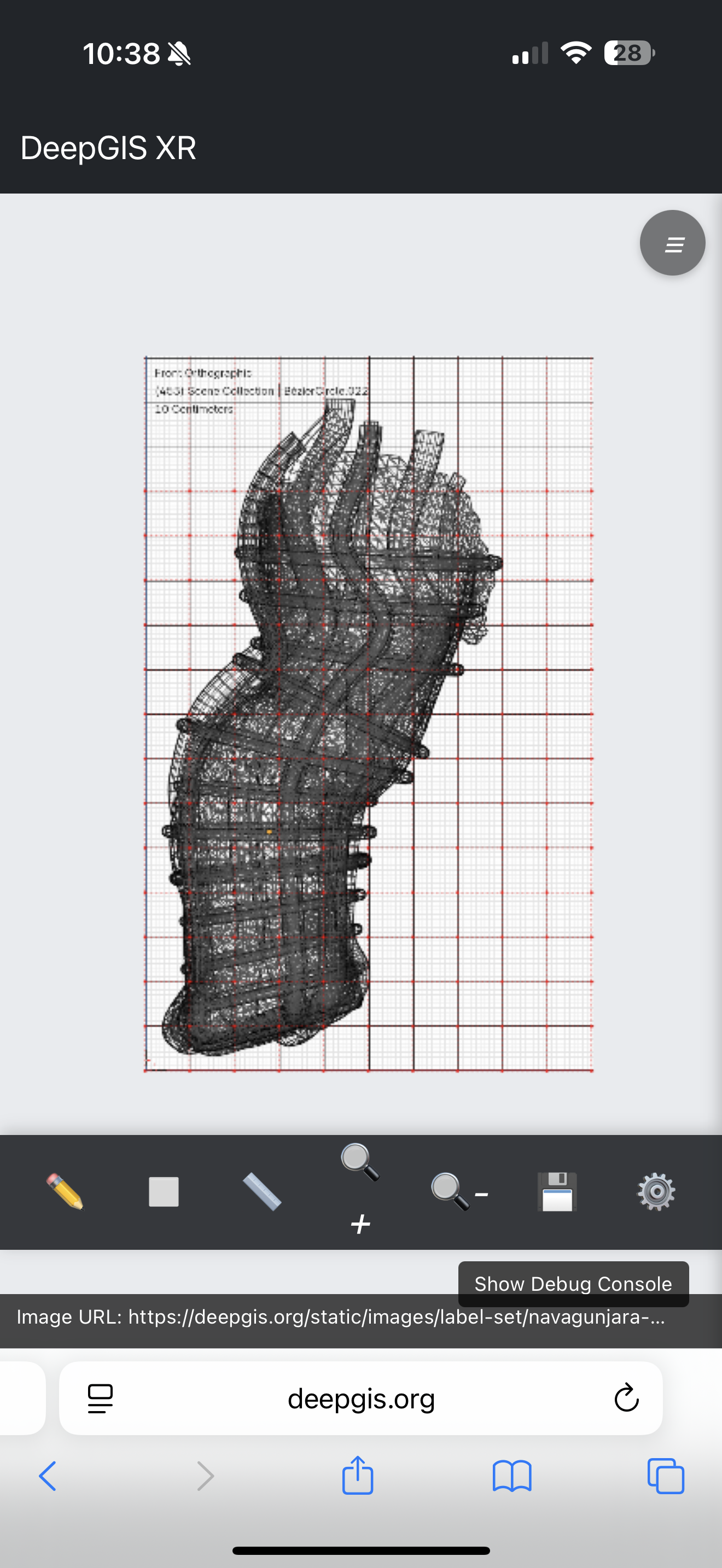}
        \caption{Elephant leg: 2D orthographic projection of the digital twin displayed in the DeepGIS XR app, overlaid on a metric grid for template-based fabrication guidance.}
        \label{fig:sfm-torso-early}
    \end{subfigure}
    \hfill
    \begin{subfigure}{0.48\linewidth}
        \centering
        \includegraphics[width=\linewidth]{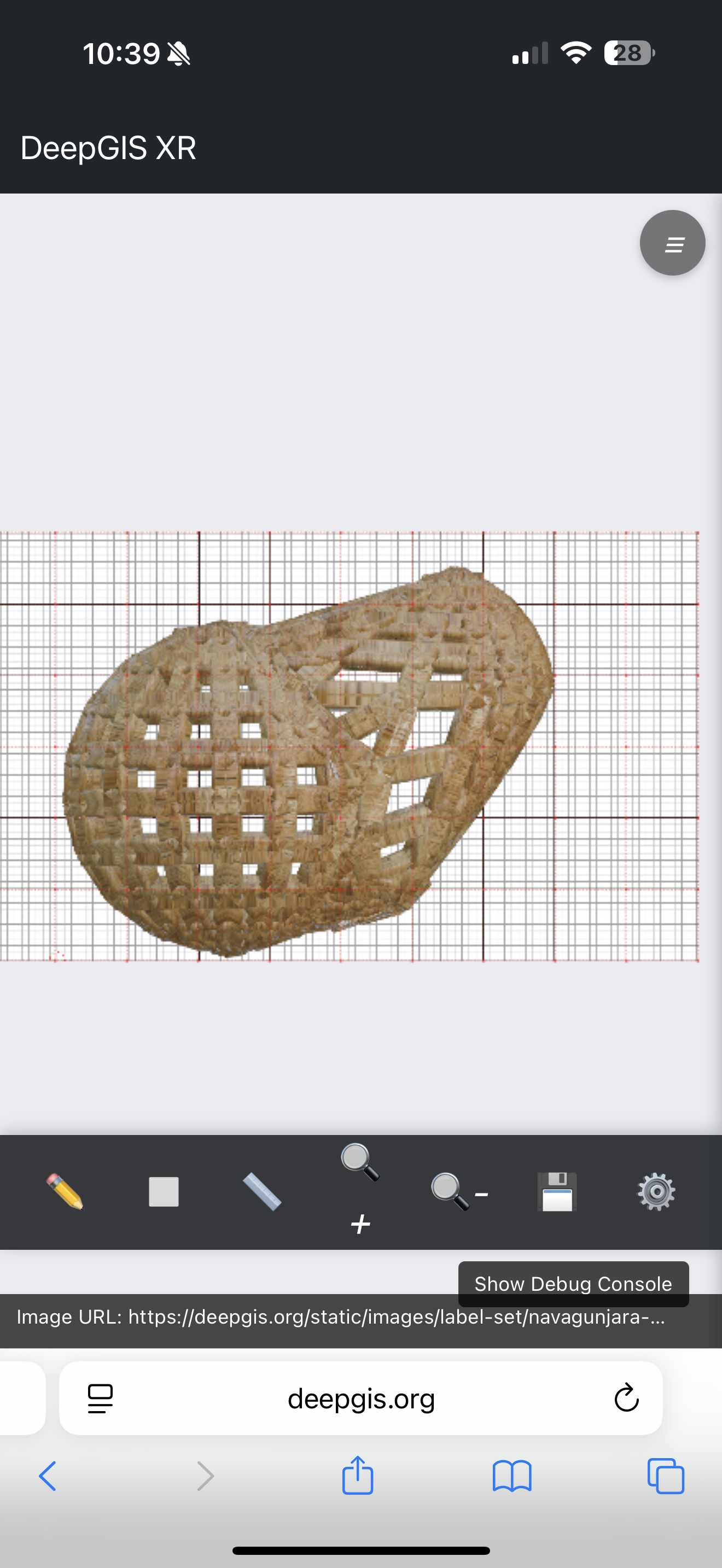}
        \caption{Bull hump: 3D digital twin rendered within the DeepGIS XR app on a metric grid, used to verify dimensions and generate wide-paper orthographic prints for artisans.}
        \label{fig:sfm-leg-early}
    \end{subfigure}
    \caption{\small{DeepGIS XR app displaying the Navagunjara Reborn digital twin with 2D orthographic projection and 3D visualization capabilities. The metric grid overlay enables generation of metrically accurate, wide-paper printable templates directly from the app, guiding artisans in Odisha during the fabrication of cane and metal components.}}
    \label{fig:sfm-cane-group-early}
\end{figure}
DeepGIS was leveraged to automate the generation of metrically accurate PDFs from 3D digital twin models, such that true scales were faithfully reproduced when printed on paper. Experiments were also carried out in automating the generation of 2D orthographic projections for arbitrarily arranged sculpture sections, for instance the deer leg, which has both a roll and pitch angle in the world frame of reference. PCA and related methods were explored to determine the principal component directions of meshes and curves. However, no universal approach was produced during the May 2025 India build, so manual adjustments were made along with Blender's bpy Python coding API.

\begin{figure}[htpb]
    \centering
    \begin{subfigure}[b]{0.3\linewidth}
        \centering
        \includegraphics[width=\linewidth]{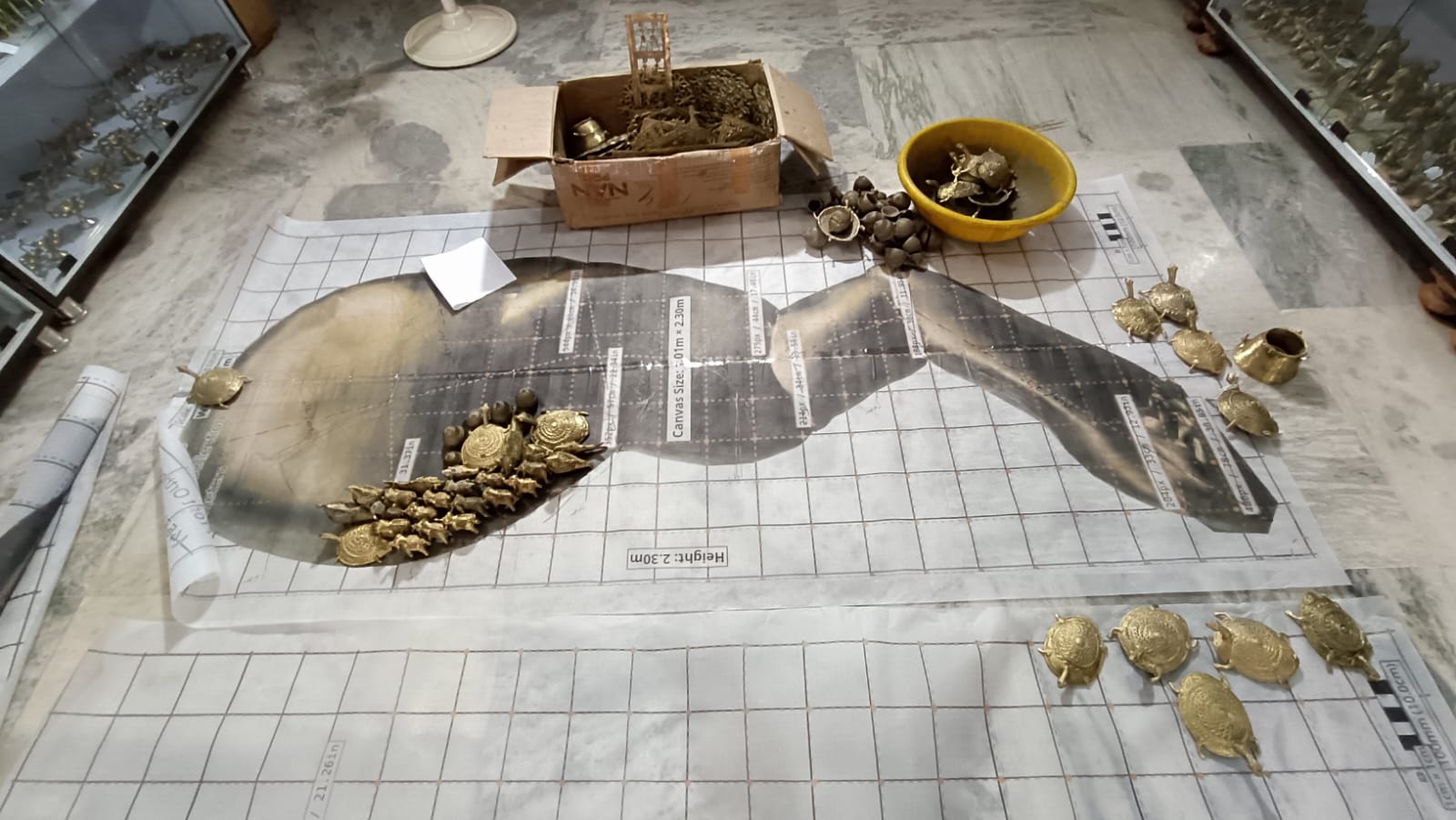}
        \caption{Tigress leg crafts mapping on 2D orthographic prints from sculpture 3D design.}
        \label{fig:tigress-leg-progress}
    \end{subfigure}
    \begin{subfigure}[b]{0.3\linewidth}
        \centering
        \includegraphics[width=\linewidth]{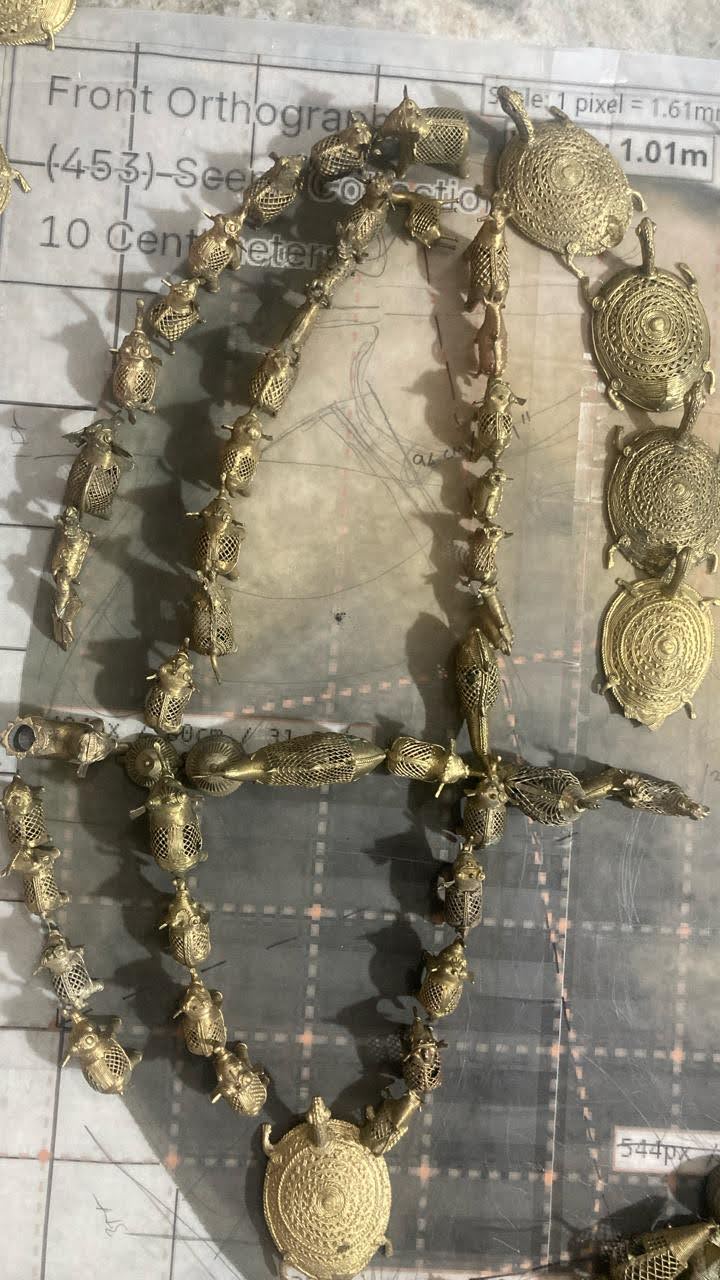}
        \caption{Tigress leg top mapping}
        \label{fig:tigress-leg-top-map}
    \end{subfigure}
    \hfill
    \begin{subfigure}[b]{0.3\linewidth}
        \centering
        \includegraphics[width=\linewidth]{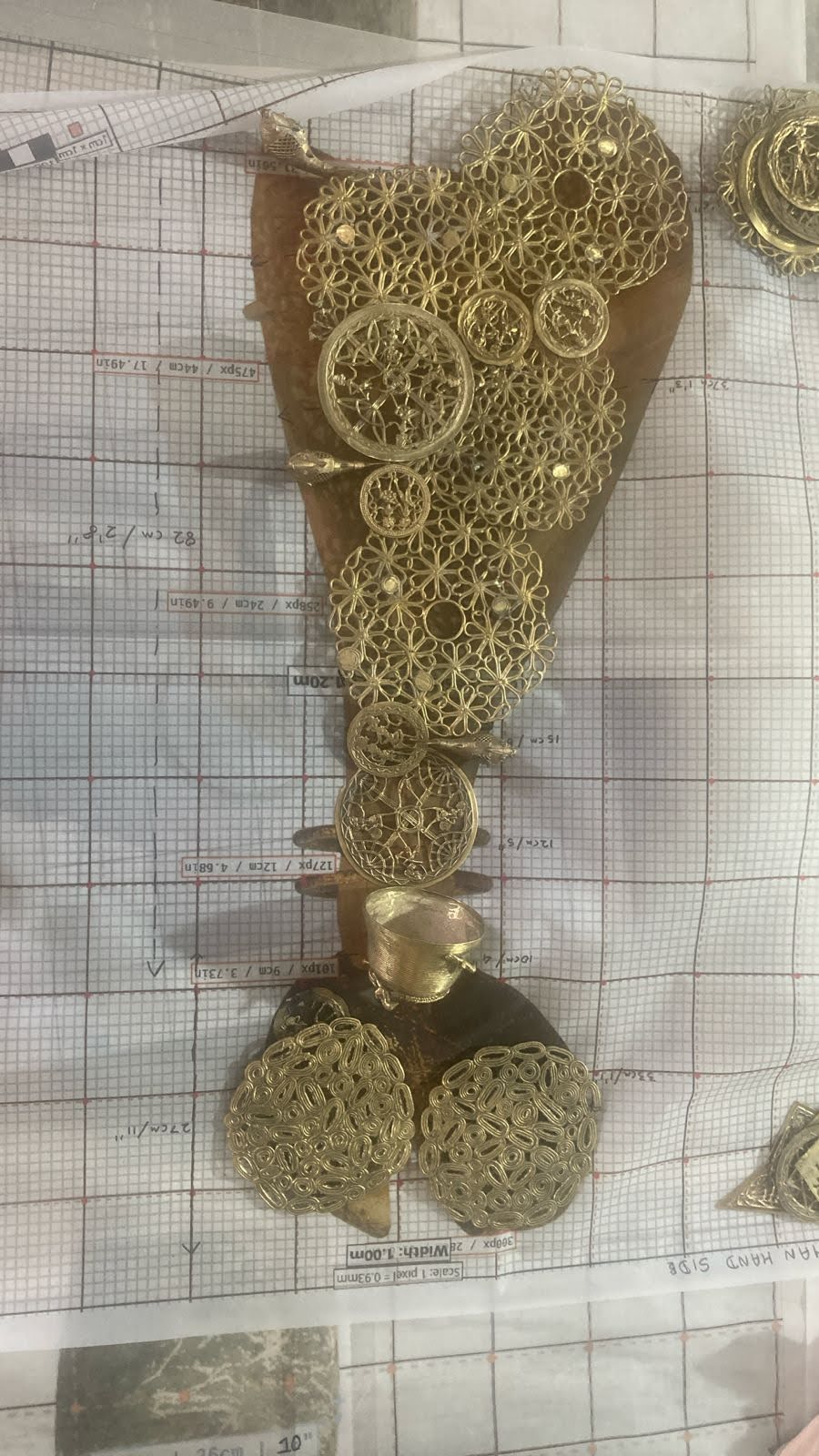}
    \caption{Human arm crafts mapping on orthographic projection}
        \label{fig:arm-crafts-map}
    \end{subfigure}
    
    \begin{subfigure}[b]{0.3\linewidth}
        \centering
        \includegraphics[width=\linewidth]{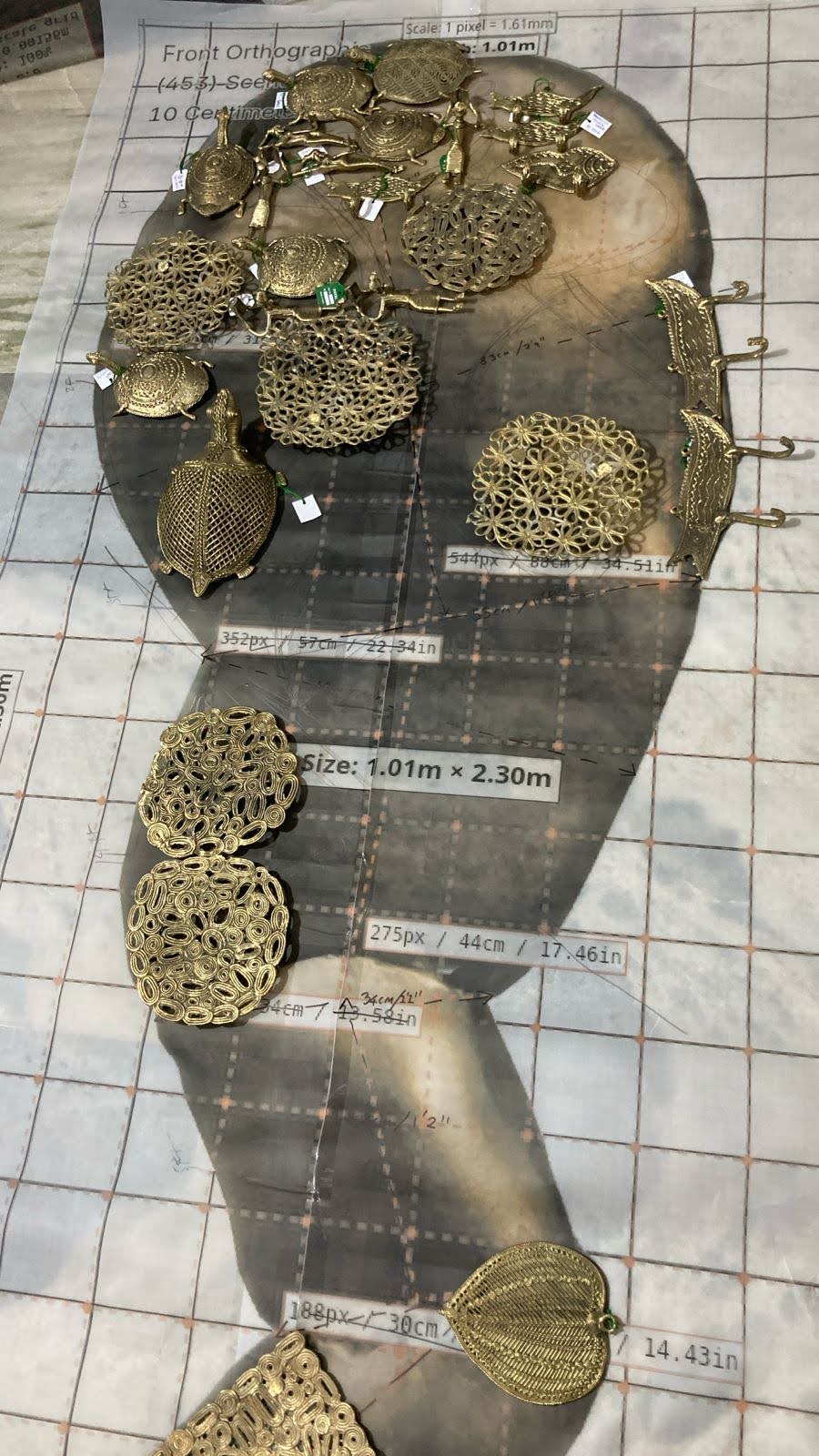}
    \caption{Tigress leg metal crafts mapping}
        \label{fig:tigress-leg-crafts}
    \end{subfigure}
    \begin{subfigure}[b]{0.3\linewidth}
    \centering
    \includegraphics[width=\linewidth]{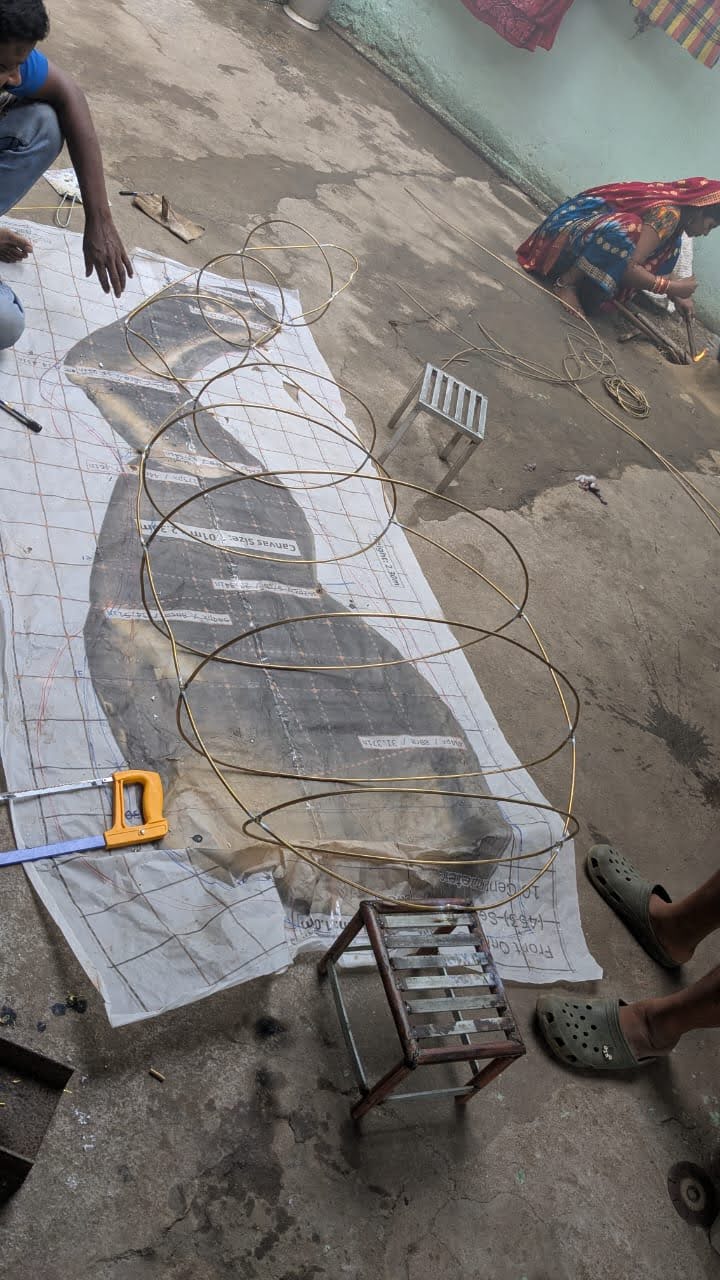}
    \caption{Wireframing of the leg element by Rajesh Moharana and team (Photo courtesy of Bishweshwar Das)}
    \label{fig:placeholder-leg}
\end{subfigure}
        \caption{Dhokra figurine curation and motif atlas for sculptural body parts. Anshu Arora and Badal Satpathy sourced dhokra figurines from local shops---with choices confirmed by the lead artist in real time via video call---and developed a crafts design atlas mapping each figurine to its designated position on the orthographic projections of the 3D digital model. Anshu Arora served as crafts design motif consultant in this stage, bridging traditional iconography and computational form. Rajesh Moharana then drew on his dhokra and metal sculpting expertise to collaborate with the lead artist in translating the atlas into the finished sculptural sections.}
    \label{fig:design-assembly-group}
\end{figure}
\begin{figure*}[htpb]
    \centering
    \begin{subfigure}{0.48\linewidth}
        \centering
        \includegraphics[width=\linewidth]{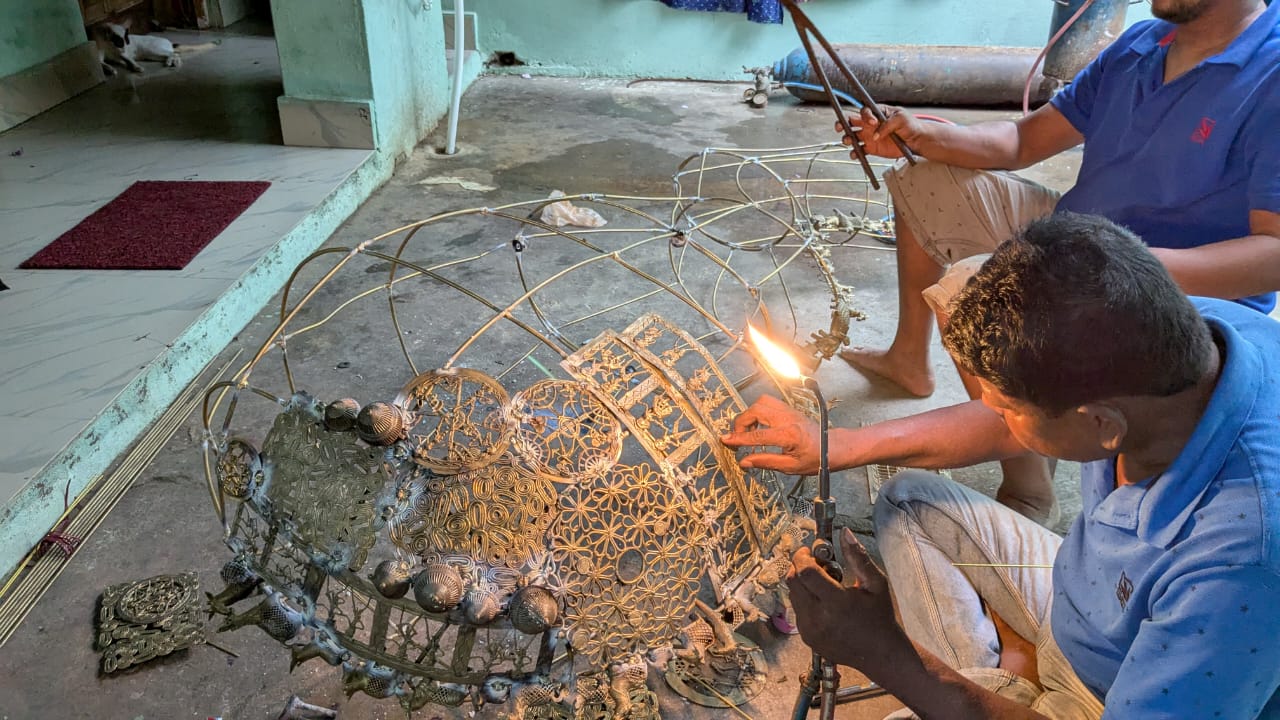}
        \caption{Master craftsman Rajesh Moharana assembles the tigress leg from bronze dhokra metalwork and bronze wireframing, guided by 2D orthographic prints and the 3D digital twin accessed via Blender and DeepGIS, combined with his decades of sculpting expertise.}
        \label{fig:crafts-design-arm}
    \end{subfigure}
    \hfill
    \begin{subfigure}{0.48\linewidth}
        \centering
        \includegraphics[width=\linewidth]{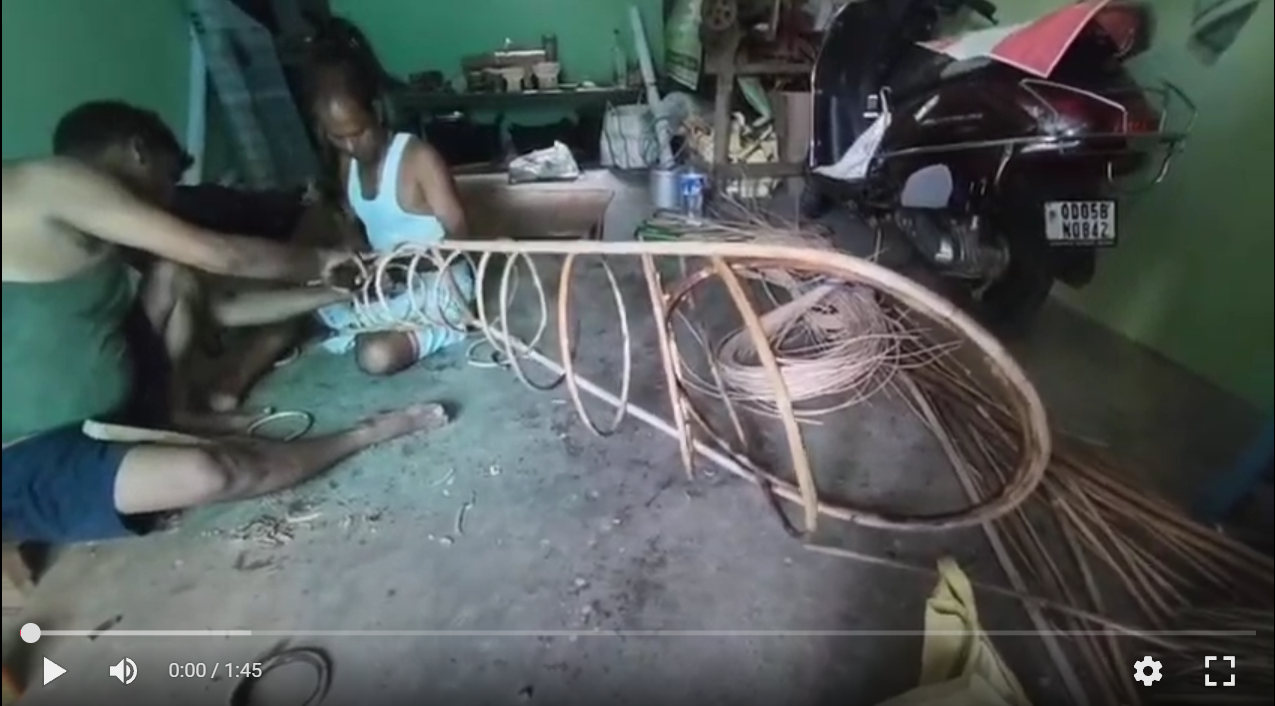}
        \caption{Master craftsman Ekadashi Barik constructs the deer leg from bamboo canewood---carefully bending and assembling longitudinal sections and ring elements into form---guided by orthographic prints and the 3D digital twin.}
        \label{fig:crafts-design-leg}
    \end{subfigure}
    \caption{\small{Master craftsmen Rajesh Moharana and Ekadashi Barik fabricate sculptural components in Odisha, each drawing on decades of craft knowledge alongside digital guides---2D orthographic prints and the 3D digital twin accessed via Blender and DeepGIS. Rajesh constructs the tigress leg from bronze dhokra metalwork and wireframing; Ekadashi forms and assembles the deer leg from bamboo canewood rings and longitudinal sections, carefully shaped and joined. The pairing of computational references with embodied craft expertise is the defining characteristic of the digital-physical workflow at the heart of Navagunjara Reborn.}}
    \label{fig:crafts-design-combined}
\end{figure*}

While artisans had access to direct 3D designs for elliptical wireframe elements, they skillfully combined their own experience in crafting. With 2D orthomaps, 3D-printed models, and digital access via the web app, they constructed cane and metal wireframes, guided by both technical references and their craft intuition. Rajesh Moharana drew on his mastery of bronze dhokra metalwork and wireframing to construct the tigress leg, referencing orthographic prints and the digital twin to guide form while applying his own deep knowledge of lost-wax sculpting traditions. Ekadashi Barik shaped the deer leg entirely from bamboo canewood, carefully bending longitudinal sections and ring elements into form before assembling them---a technique demanding spatial intuition that no CAD tool encodes. The process of mapping the arm is shown above, illustrating an integration of digital tools with traditional expertise. Similarly, cane leg construction demonstrates how these resources and skills were blended to achieve the desired metrically accurate wireframe structures.

\subsection{Rapid-Prototyped 3D Physical Twins}
Panels crafted by Odia artisans were precisely guided using piece-aligned 3D-2D orthographic projections that ensured fidelity to authentic motifs and structural requirements. This digital-physical synergy fostered continuous, creative collaboration between designers and craftspeople. Fused Deposition Modeling (FDM) 3D printing was employed to rapidly prototype sculptural segments, enabling the team to validate mechanical tolerances, explore multiple design iterations, and communicate tangible forms to the Indian artisan teams prior to full-scale fabrication.

\begin{figure}[htpb]
    \centering
    \includegraphics[width=3.5in]{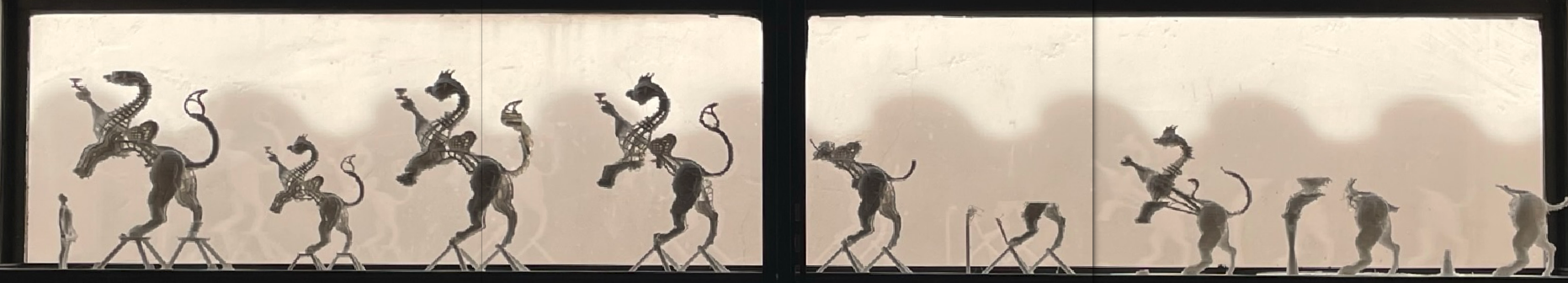}
    \caption{Rapid prototyping through 3D printing: validation of design iterations preceding the Odisha artisan build phase (April 2025).}
    \label{fig:3dprint-evolution}
\end{figure}

\subsection{Situational Awareness and Design Feedback during Craft Building}
\begin{figure}[htpb]
    \centering
    \begin{subfigure}{0.48\linewidth}
        \centering
        \includegraphics[width=\linewidth]{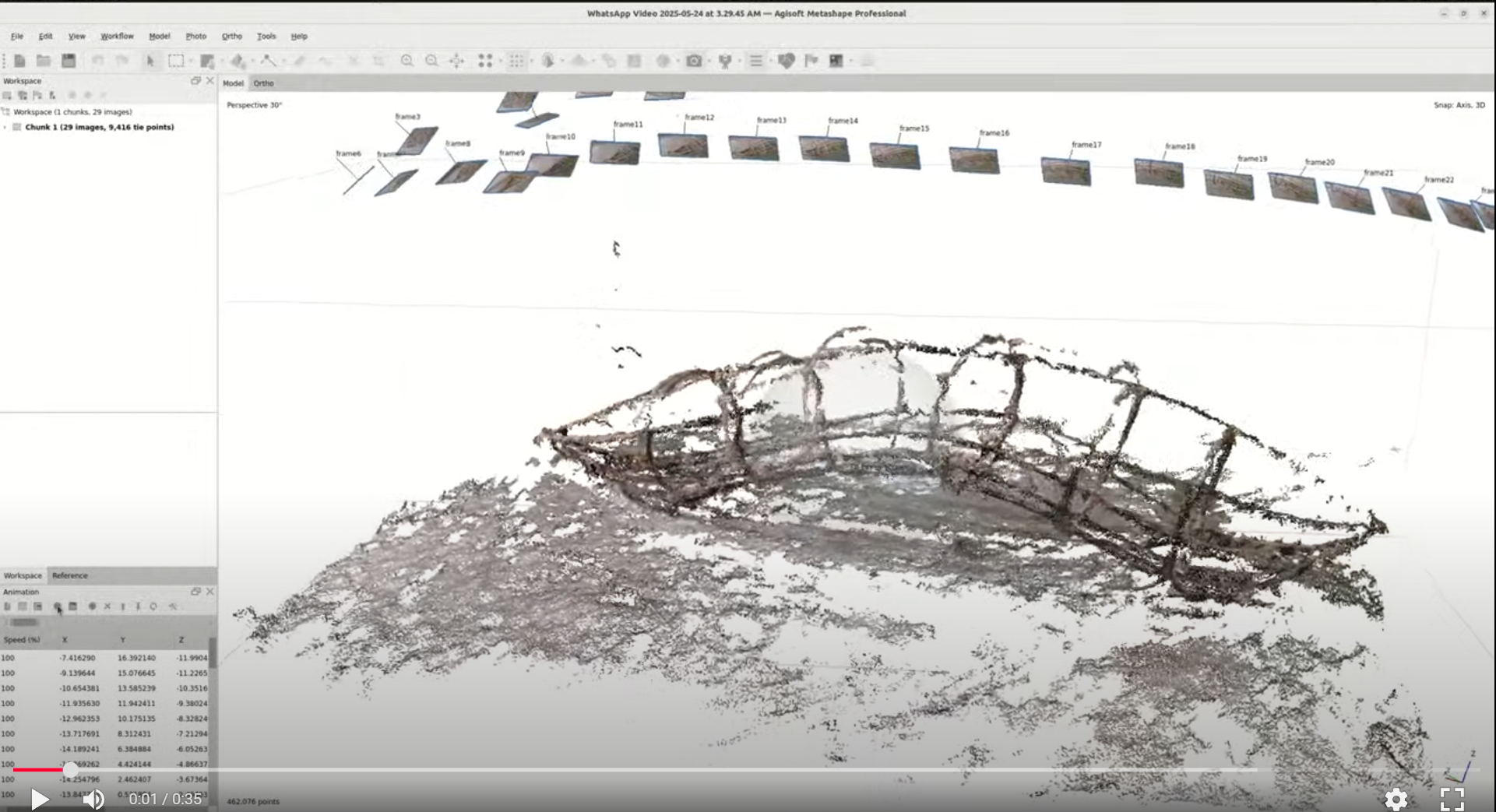}
        \caption{Torso SfM Reconstruction}
        \label{fig:sfm-torso-cane}
    \end{subfigure}
    \hfill
    \begin{subfigure}{0.48\linewidth}
        \centering
        \includegraphics[width=\linewidth]{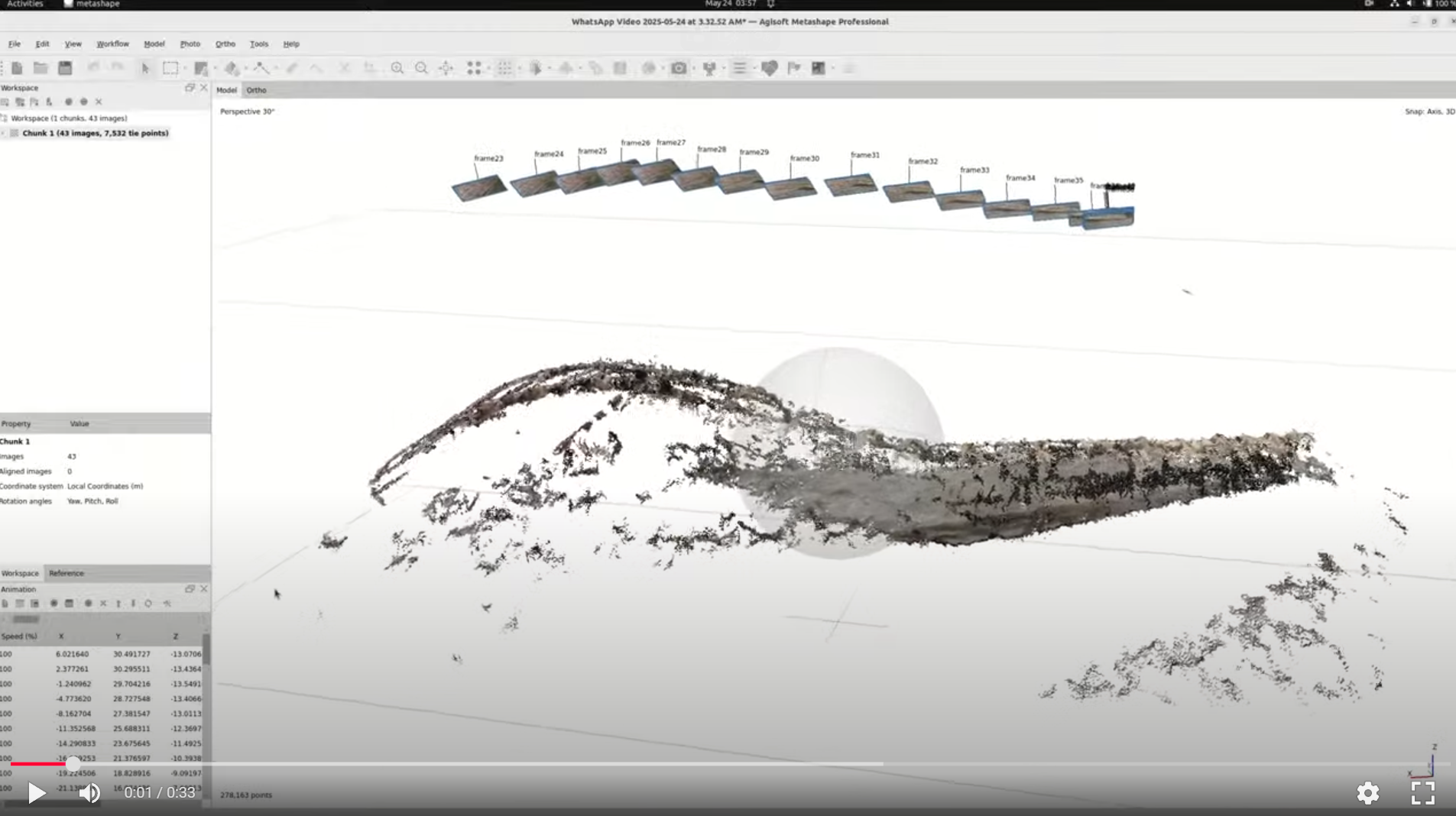}
        \caption{Leg SfM Reconstruction}
        \label{fig:sfm-leg-cane}
    \end{subfigure}
    \caption{\small{SfM reconstructions of cane artifact: torso and leg. Images were captured by Bishweshwar Das}}
    \label{fig:sfm-cane-group}
\end{figure}

Photogrammetry was central to design feedback as the craft build unfolded in India. Using phone cameras and structured guidance on lighting and motion, a team member captured systematic image sequences to generate accurate 3D records of evolving sculptural elements. Agisoft Metashape software~\cite{agisoft_metashape_2023} was used for structure-from-motion (SfM) reconstruction. Each scan cycle added geometric evidence to the model, reduced mismatch between design intent and physical reality, and enabled timely structural decisions. These reconstructions also served as a living archive and reference for quality control. Due to pace and logistical constraints across multiple sites, data for 3D reconstruction could not be acquired for the final pieces for head, neck, and chest. Precise weights were also not available for individual components; however, weights of combinations of pieces from shipping records allowed rough estimates prior to arrival of the sculpted pieces in the US.

As fabrication advanced in Odisha, the 3D model was updated to reflect the workmanship and geometry of each panel. Public-domain animal meshes sourced from Printables.com—including reference deer, tiger, and elephant anatomies—were remixed and incorporated as geometry scaffolds, such as in the development of the tigress leg wireframe, where the mesh provided joint topology reference for Rajesh Moharana's dhokra metalwork. While initial ambitions to involve naturalists and wildlife conservationists could not be realized, anatomical fidelity was achieved through the artisans' deep familiarity with regional wildlife and their established sculpture traditions—exemplified by Moharana's probing questions about joint articulation, which shaped critical design refinements in ways no reference mesh alone could have determined. Similarly, Ekadashi Barik's canewood construction of the deer leg—bamboo rings and longitudinal members carefully formed and joined—embodied a structural logic of curvature and tension that emerged from craft practice rather than specification, producing a geometry the model could describe but not prescribe.

\section{Structural Design}
The structural design of Navagunjara Reborn evolved through iterative feedback between artistic sculpting, digital modeling, and engineering verification. The goal was to preserve the integrity of hand-crafted panels from Odisha while embedding them within a robust, climbable-ready superstructure capable of meeting all Black Rock City safety regulations and environmental demands. This goal operationalizes the intersection of two distinct kernels: the \emph{craft kernel}, carried in the hands and intuitions of Odia artisans whose forms were already fixed by the time they arrived in Arizona, and the \emph{structural kernel}, encoded in FEA simulations and Black Rock City's 75~mph wind-load requirements. The engineering challenge was not to impose one kernel on the other, but to find the structural topology that satisfied both simultaneously—enveloping each handcrafted component without forcing it to conform to a geometry it was never made to match. This is the product-kernel principle applied to engineering: the superstructure is the configuration that scores high under both the craft and structural kernels at once.

\subsection*{Black Rock City Structural Requirements}
For the 18-foot structure with flame effects, Black Rock City mandated the following:
\begin{itemize}
    \item Full engineering documentation, wind/load analysis for 75~mph gusts, and certified anchoring, spaced at least 3 feet apart. For this project, we chose 22-inch aluminum alloy helical ground anchors and forged rigging hardware.
    \item Full detail of layout, and high-visibility marking/illumination.
    \item Flame systems plan.
  \end{itemize}

\begin{table}[h]
\centering
\begin{tabular}{|l|l|}
\hline
\textbf{Requirement}         & \textbf{Protocol} \\
\hline
Height                      & 18 feet \\
Wind Gust Design            & 75 mph \\
Anchor Spec                 & 3,000+ lbs tension, 22$''$ helical \\
Anchor Spacing              & $\geq$3’ apart \\
Factor of Safety            & $\geq$ 3 (see FEA) \\
Climb Load                  & 250 lbs/pt, dynamic 400 lbs \\
FAST Approval               & Dossier/inspection required \\
Fire Ext./Burn Kit          & Per FAST, staged on site \\
\hline
\end{tabular}
\caption{Summary of structural and flame effect design parameters for the 18-foot-tall installation.}
\label{tab:blackrockcity_requirements}
\end{table}

\subsection*{Anchor and Rigging Documentation}
Anchors were laid out in a hexagonal grid with $>$3~ft separation (never ganged), and eye-bolted to base plates using 1$''$ ratchet straps and forged shackles. All anchor points exceeded the maximum calculated loads from simulated wind+uplift. Night visibility was ensured by LED-lights and perimeter marking cones.

\subsection*{Installation \& Disassembly Sequence}
\textbf{Installation steps} included: 
\begin{enumerate}
    \item Placement and anchoring of the base structure.
    \item Bolting of the legs to the base structure.
    \item Sequential panel and feature attachment, each checked using assembly drawings.
    \item Flame system installation by flame artist; full safety perimeter and kit setup inspection.
\end{enumerate}
Teardown followed this sequence in strict reverse; all hardware returned to inventory, site swept/magnet-raked (Leave No Trace).

\begin{figure}[htpb]
    \centering
    \begin{subfigure}[b]{0.48\linewidth}
        \centering
        \includegraphics[width=\linewidth]{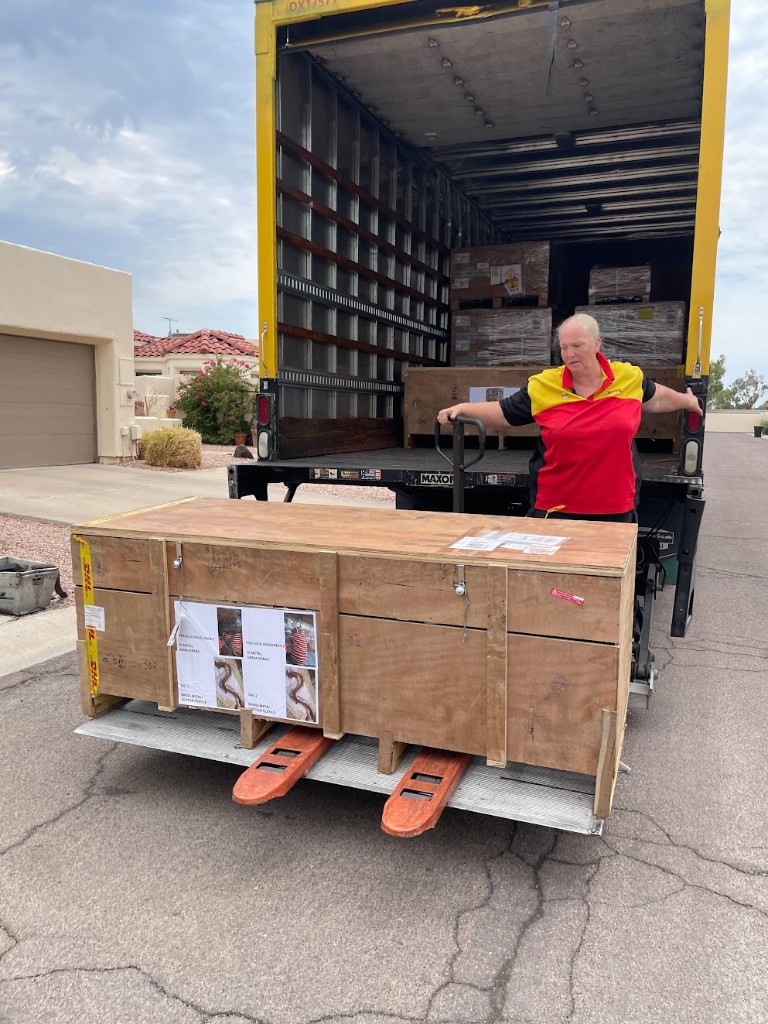}
        \caption{Labeled crates with orthographic prints of contents visible on the exterior, facilitating identification without opening.}
        \label{fig:dhl-unload}
    \end{subfigure}
    \hfill
    \begin{subfigure}[b]{0.48\linewidth}
        \centering
        \includegraphics[width=\linewidth]{crates-garage-phoenix.png}
        \caption{Crates staged in the Phoenix garage workshop for structural retrofit.}
        \label{fig:crates-garage}
    \end{subfigure}
    \caption{\small{Air freight arrival and staging in Phoenix, AZ (mid-July 2025). Craft components shipped via air freight from Odisha arrived in labeled wooden crates, each marked with photographic references of its contents. The garage workshop served as the staging and structural retrofit facility during the compressed six-week window before Burning Man.}}
    \label{fig:shipping-arrival}
\end{figure}

\begin{figure*}[t]
    \centering
    \includegraphics[width=7in]{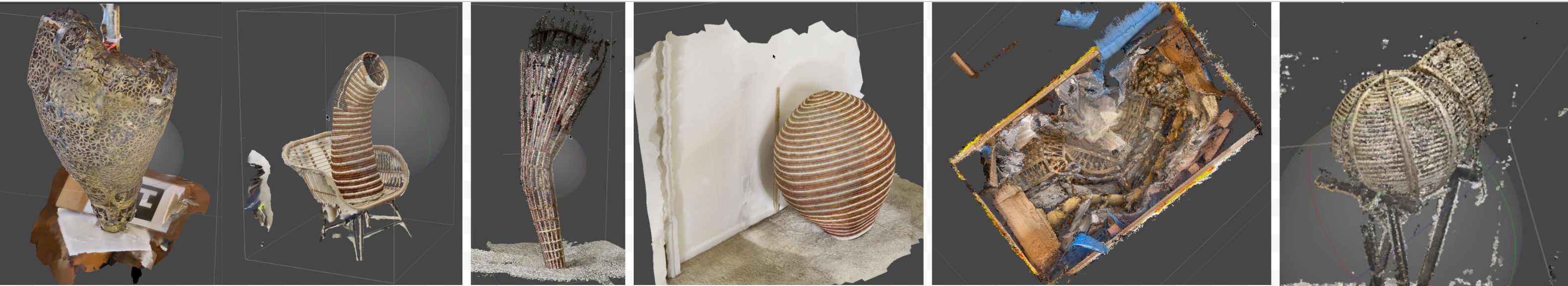}
    \caption{\small{Structure-from-Motion photogrammetry and smartphone imaging workflows enabled rapid, high-fidelity 3D scans of artisan-crafted panels. GoPro and iPhone image sets, processed in Agisoft Metashape, generated digital twins of sculpted components—including elephant legs, arms, torsos, and more. When combined with CAD tools, these 3D scans facilitated non-intrusive truss and assembly design, allowing the engineering team to optimize structural support around each unique handmade element with minimal cutting and maximal preservation of artistic form. This integrated process ensured both the safety and the distinct material character of the installation, from shipping to final assembly on playa.}}
    \label{fig:received-crafts-sfm}
\end{figure*}

\subsection*{Fusion of Crafts and Structure}

The integration of artisan-crafted components with engineered aluminum superstructures posed manageable challenges due to variability in handcrafted panels relative to original 3D dimensions. Instead of forcing crafts to conform to rigid frames, our structural design strategy enveloped and worked around the topology and tolerances of each piece. This approach minimized material waste and preserved the authenticity of Odia motifs while ensuring robust structural support. SfM scans acquired before installation enabled remote engineering sign-off, while physical tolerances were managed in real time during structural build. Joints and fixings were all rated for a minimum factor of safety $>$3 under combined loads. Craft sections and panels were only minimally adapted to ensure fit.

\begin{figure}[t]
    \centering
    \begin{subfigure}[b]{0.48\linewidth}
        \centering
        \includegraphics[width=\linewidth]{sfm-cad-dt.jpg}
        \caption{SfM scan of a craft component for archival and engineering alignment.}
        \label{fig:sfm-cad}
    \end{subfigure}
     \begin{subfigure}[b]{0.48\linewidth}
        \centering
        \includegraphics[width=\linewidth]{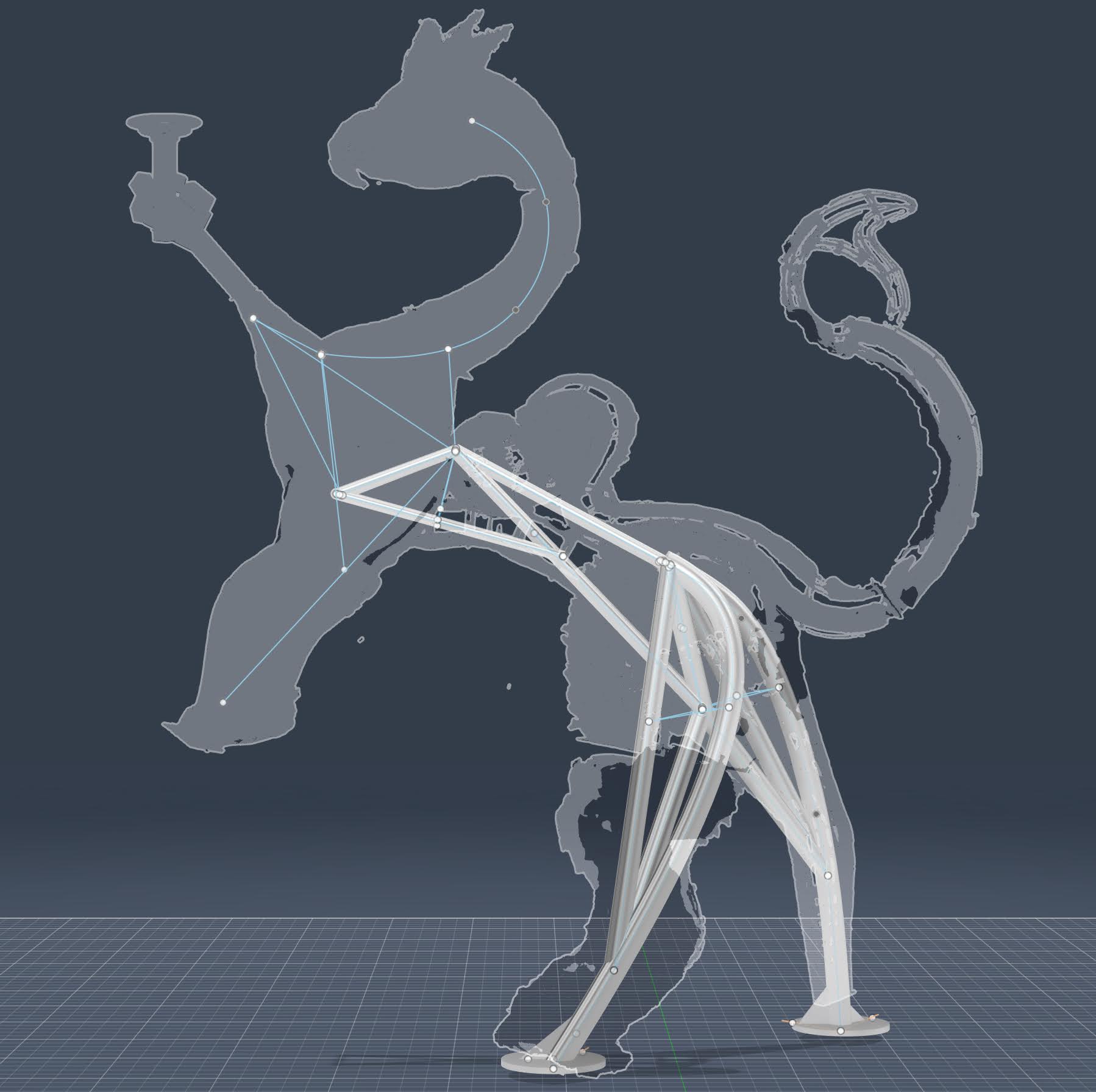}
        \caption{\small{CAD alignment of scanned panels within the superstructure for minimal intervention fit.}}
        \label{fig:fem-design}
    \end{subfigure}

    \caption{\small{Integration of artisan-crafted components with structural engineering minimizing the need for modifications or cuts to crafted panels, maximizing both structural integrity and visual impact in the final assembly.}}
    \label{fig:sfm-truss-integration}
\end{figure}

A continuous loop using SfM photogrammetry and 3D modeling enabled remote verification of component dimensions and refinement of assembly design. For example, on-site teams selectively reduced steel framing within the elephant leg to reduce weight and substituted some welds with bolts and cable ties to accommodate last-minute fit variations, enabling safe and rapid assembly under harsh desert conditions.

Modularity proved vital. Telescoping joints and stage clamps facilitated on-playa adjustments and recovery from weather-related disruptions, balancing precision assembly with manual fabrication methods necessitated by limited power access. Blender~\cite{soni2023review} and DeepGIS bridged cultural and disciplinary gaps, allowing Odisha artisans to work from metrically accurate orthographic projections and 3D-printed templates, and helping keep craft practice aligned with computational design.

\begin{figure*}[htpb]
    \centering
    \begin{subfigure}[b]{0.48\linewidth}
        \centering
        \includegraphics[width=\linewidth]{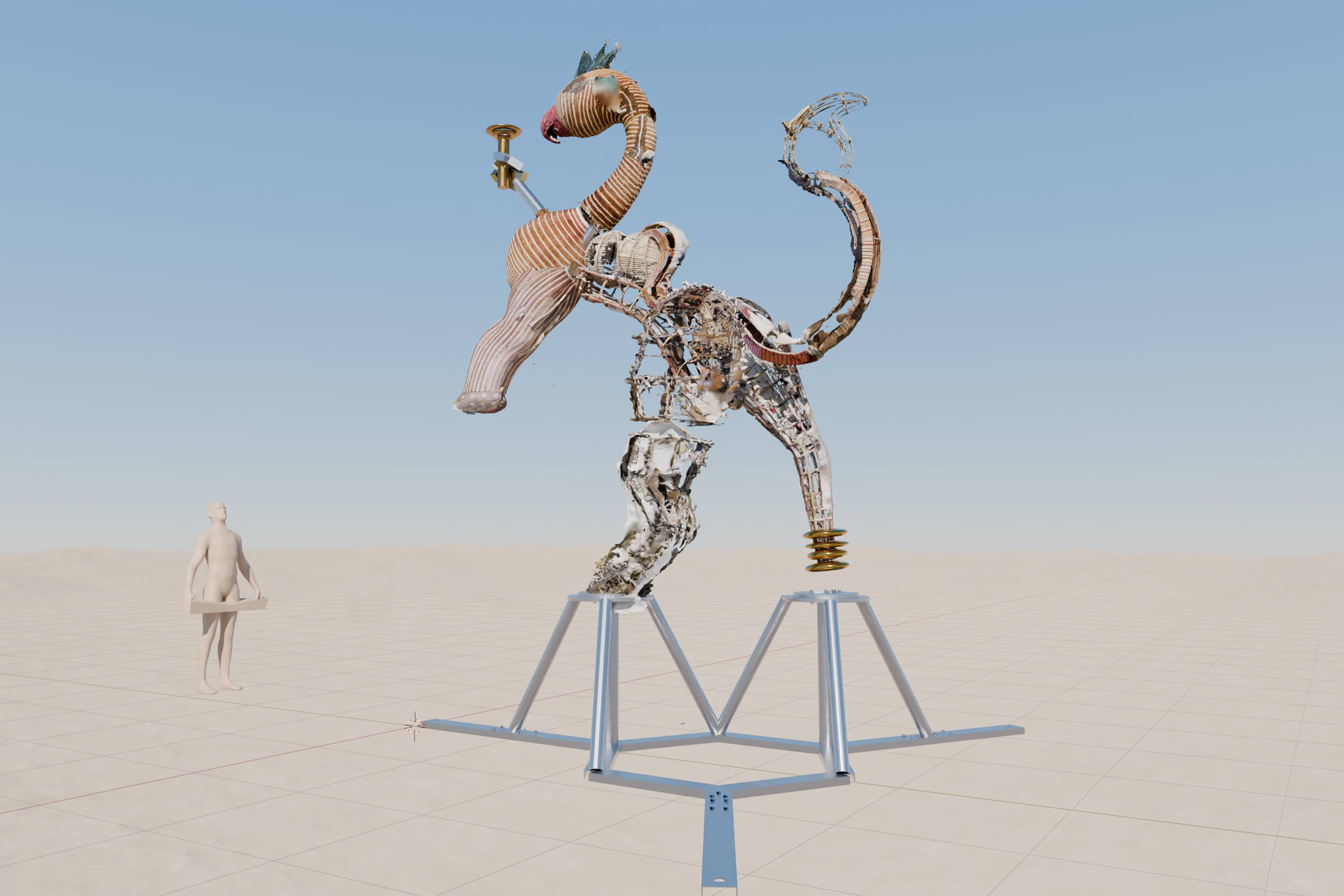}
        \caption{Structure-from-Motion photogrammetry, used for 3D reconstruction of the physical crafts panels and sections, and assembly of a digital twin, prior to build. Combined use of Blender and Autodesk Fusion, allowed iterative optimization of final structural design prior to actual cutting and fabrication of metal elements.}
        \label{fig:sfm_dt}
    \end{subfigure}
    \hfill
    \begin{subfigure}[b]{0.48\linewidth}
        \centering
        \includegraphics[width=\linewidth]{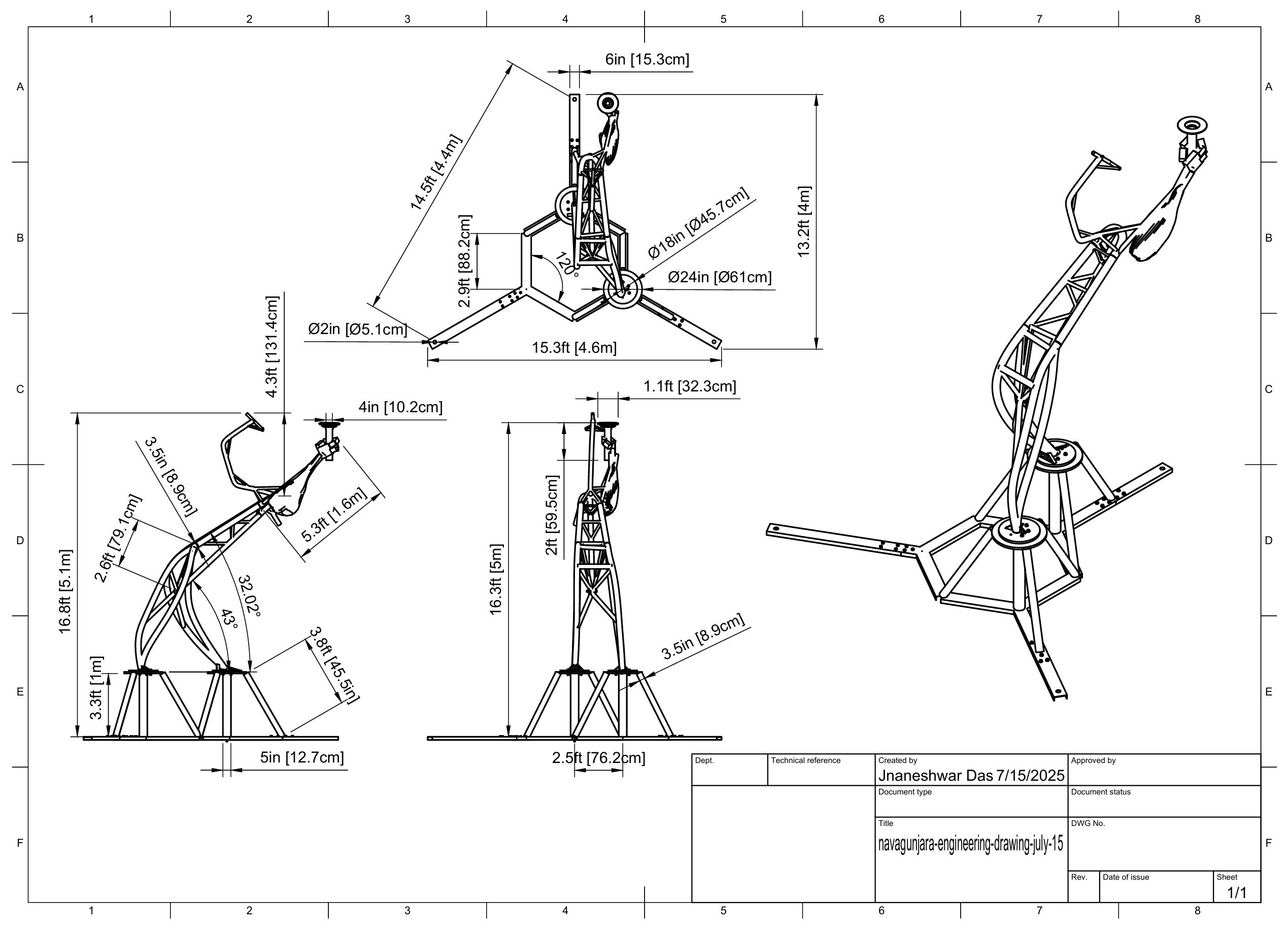}
        \caption{CAD-based design drawings}
        \label{fig:cad_drawings}
    \end{subfigure}
    \caption{Comparison of SfM-derived models with CAD design drawings, used for validation of sculptural components.}
    \label{fig:sfm_cad_group}
\end{figure*}

\subsection*{Gravity and Wind Load Assessment}
Finite element modeling (FEA) validated the truss system under both static (wind) and dynamic (potential climb/human interaction) loads:
Potential climb load of 250~lbs at each primary foothold and up to 400~lbs dynamic loads at panel joints were validated; all critical connections meet $>$3 FOS. Although for this version of the sculpture, we did not allow climbing, the base superstructure with dual-tripod layout has been designed to withstand loads from interaction through climbing. Future iterations of the sculpture will explore climbing options up to the neck of the sculpture.

\subsection{Detailed 3D Archiving of Received Crafts}

After unboxing and unwrapping, the sculpted pieces were digitally archived using a GoPro action camera and a smartphone. Graduate student involvement during this process highlighted the potential for future educational projects or coursework, particularly those focused on verification and validation of 3D designs in the context of art installation building.

Based on the material, shape, and dimensions, each sculptural component exhibited unique structural and aesthetic properties:
\begin{table*}[h!]
\centering
\begin{tabular}{|p{3.5cm}|p{10cm}|}
\hline
\textbf{Component} & \textbf{Materials and Techniques} \\
\hline
Elephant leg & Two vertical sections; sabai grass woven around 6 mm steel wire frame; central support lines later removed to reduce weight before mounting. \\
\hline
Human arm & Entirely metal fabrication; no lost-wax or wound investment casting; produced as a single unit with port openings for interior access. \\
\hline
Tigress rear right leg & Two parts, metallic; Dhokra~\cite{sethi2016dhokra} metalwork; animal motifs along wireframe; upper section has truss openings functioning as cross-members. \\
\hline
Deer rear left leg & Structural cane-wood mesh; adorned with pattachitra paintings; textile strips interwoven between cane elements. \\
\hline
Lioness torso & Cane-wood mesh for 3D volume; decorated with pattachitra paintings; appliqué textile strips between cane sections near base. \\
\hline
Rooster head & Steel wire framing for shape, with dyed sabai grass for space filling woven patterning.   \\
\hline
Peacock neck and chest & Steel framing with sabai grass woven and dyed for coloration; creates volumetric form and texture. \\
\hline
\end{tabular}
\caption{Summary of sculpted crafts and materials and methods used.}
\label{tab:3DArchive}
\end{table*}

Photogrammetry and DeepGIS projections enabled accurate digital replicas of artisan-crafted panels. These scans informed CAD-based alignment and optimization, ensuring geometric fidelity while minimizing intrusive adjustments. Remote inspection workflows supported structural updates without compromising the authenticity of craft topology.

\subsection*{Superstructure Optimization}
Two months prior to the exhibition at Burning Man 2025, the design iterations shifted from the use of Blender to Autodesk Fusion 360~\cite{Fusion3602025}. The engineering strategy was intentionally adaptive: rather than forcing the crafts to conform to predetermined frames, the aluminum tubing and truss systems were designed around the forms provided by the artisans. This approach minimized material waste, reduced cutting and welding requirements, and highlighted a symbiotic dialogue between computational engineering and traditional craftsmanship. Several base configurations were evaluated, with a hexagonal tripodal design ultimately selected for both its aesthetic qualities and structural stability. The three-legged hexagonal base, which supports two elevated footholds, can be repurposed for various two-legged animal poses by adjusting height and stride length. The hexagonal geometry additionally offers \(\binom{6}{2}\) foothold combinations, enhancing design flexibility. Viewed retrospectively, the hexagonal base can be understood as a maximum-entropy structural choice: it preserves the most future design options subject to stability constraints, deferring commitment to specific animal poses until the latest possible moment—a pattern the team recognized only after the build revealed how much that flexibility mattered.

Aluminum tubing and schedule-40 pipes were selected for their balance of strength, weight, and cost-effectiveness. Telescoping joints and welds provided flexibility during assembly, while design inspiration from the lead author’s prior Burning Man experience (A Journey Aquatic, 2023) informed practical approaches to modularity and field assembly.

\begin{figure}[htpb]
    \centering
    \subfloat[Base plate scoping]{
        \includegraphics[width=0.35\linewidth]{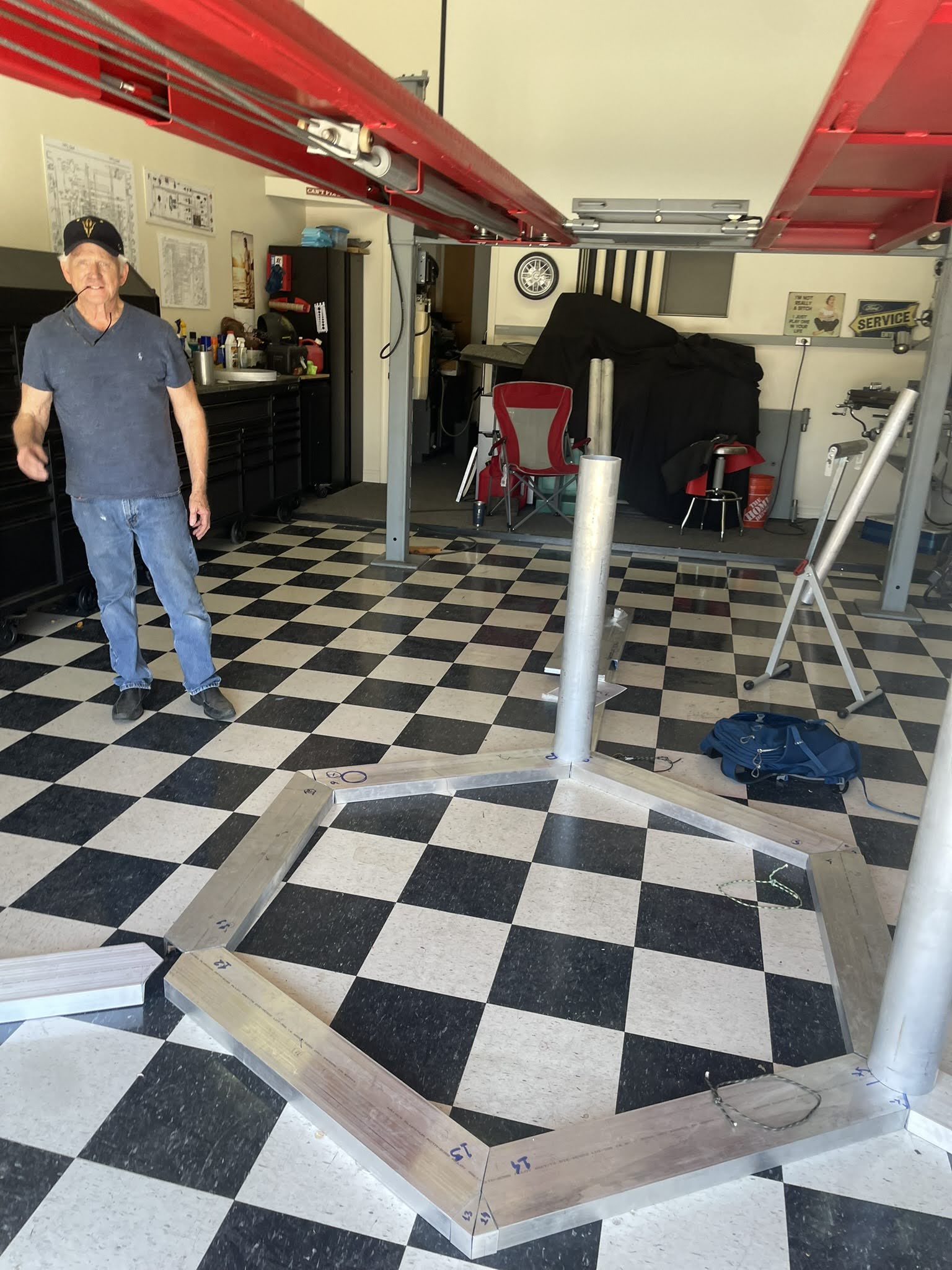}
        \label{fig:skiba-base-scoping}
    }
    \hspace{0.05\linewidth}
    \subfloat[Anchor hole drilling]{
        \includegraphics[width=0.35\linewidth]{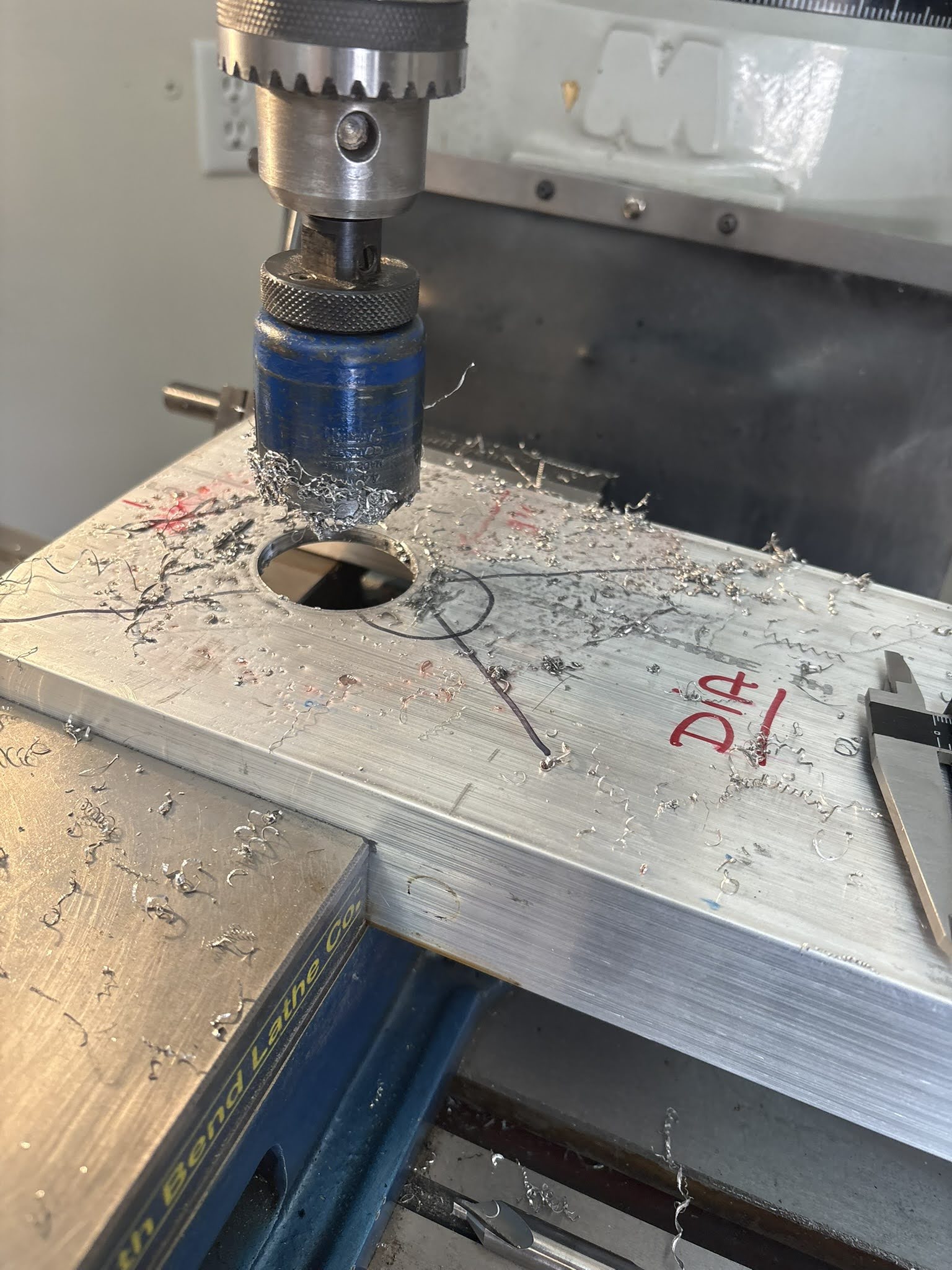}
        \label{fig:skiba-anchor-hole}
    }
    \hspace{0.05\linewidth}
    \subfloat[Disc hole machining]{
        \includegraphics[width=0.35\linewidth]{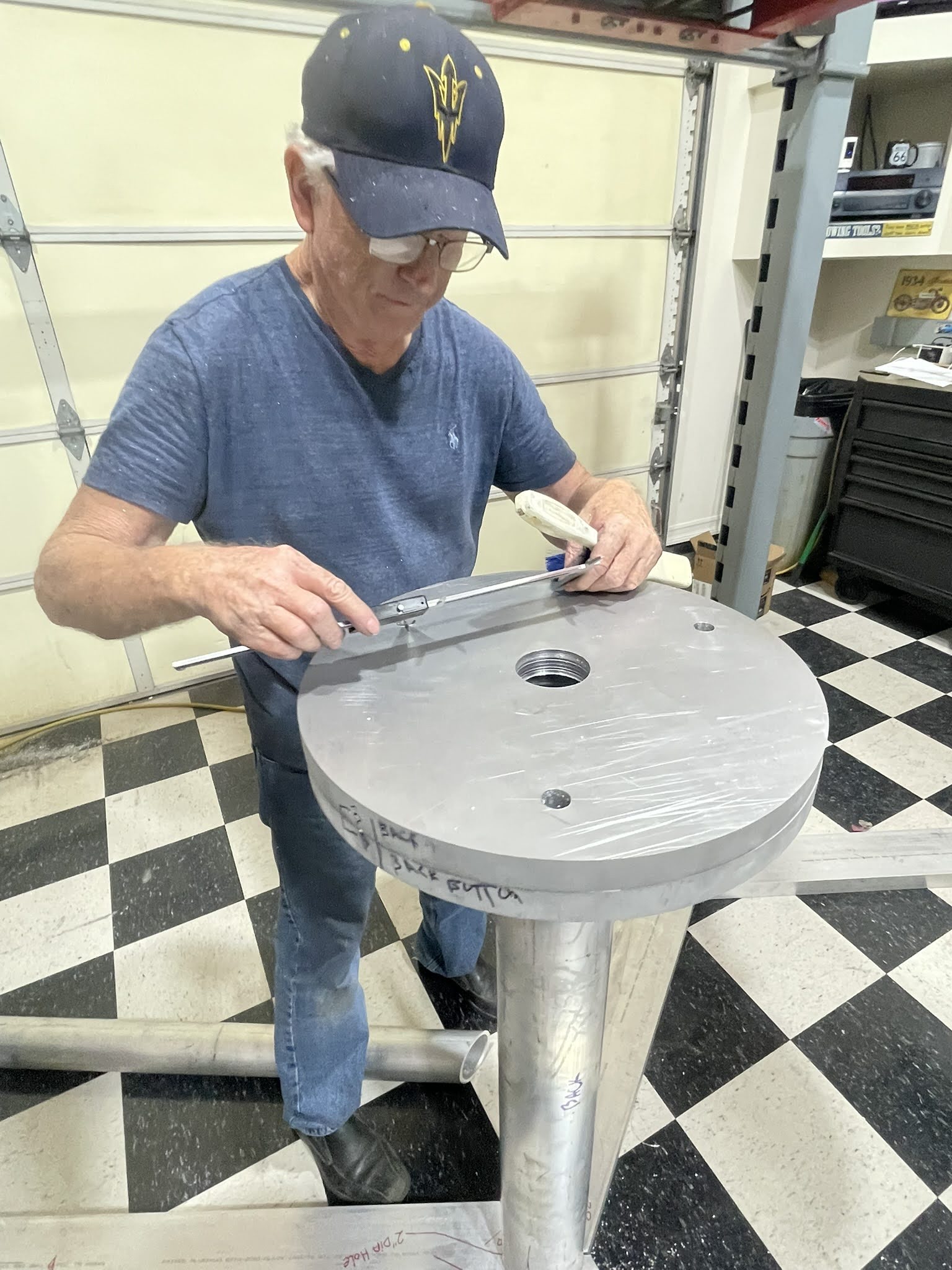}
        \label{fig:skiba-disc-holes}
    }
    \subfloat[Workshop base assembly]{
        \includegraphics[width=0.35\linewidth]{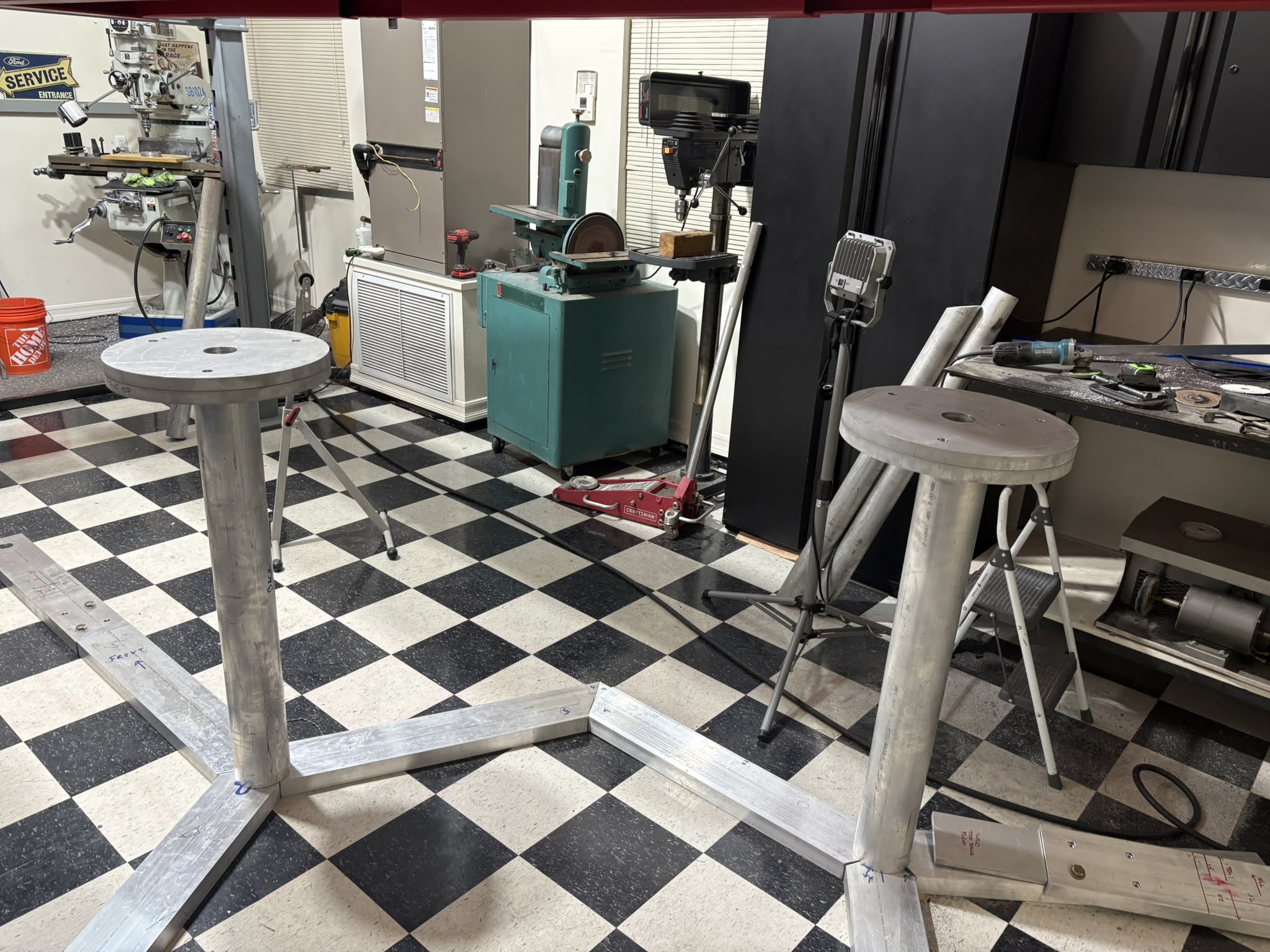}
        \label{fig:skiba-workshop}
    }
    \caption{ASU facilities management alumni Christopher Skiba fabricates pieces of the superstructure at his Tempe, AZ workshop. }
    \label{fig:skiba-fabrication}
\end{figure}

\begin{figure}[htpb]
    \centering
    \subfloat[MIG welding the deer leg truss column to the foot disc. Cut angle computed in Fusion 360; precise alignment held before welding to lock in the intended joint geometry.]{
        \includegraphics[width=0.4\linewidth]{welding-deerleg.JPEG}
        \label{fig:welding-deer}
    }
    \hspace{0.05\linewidth}
    \subfloat[MIG welding the tigress leg truss column to the foot disc. The paired leg angles together with the hexagonal base determine the sculpture's final heading.]{
        \includegraphics[width=0.4\linewidth]{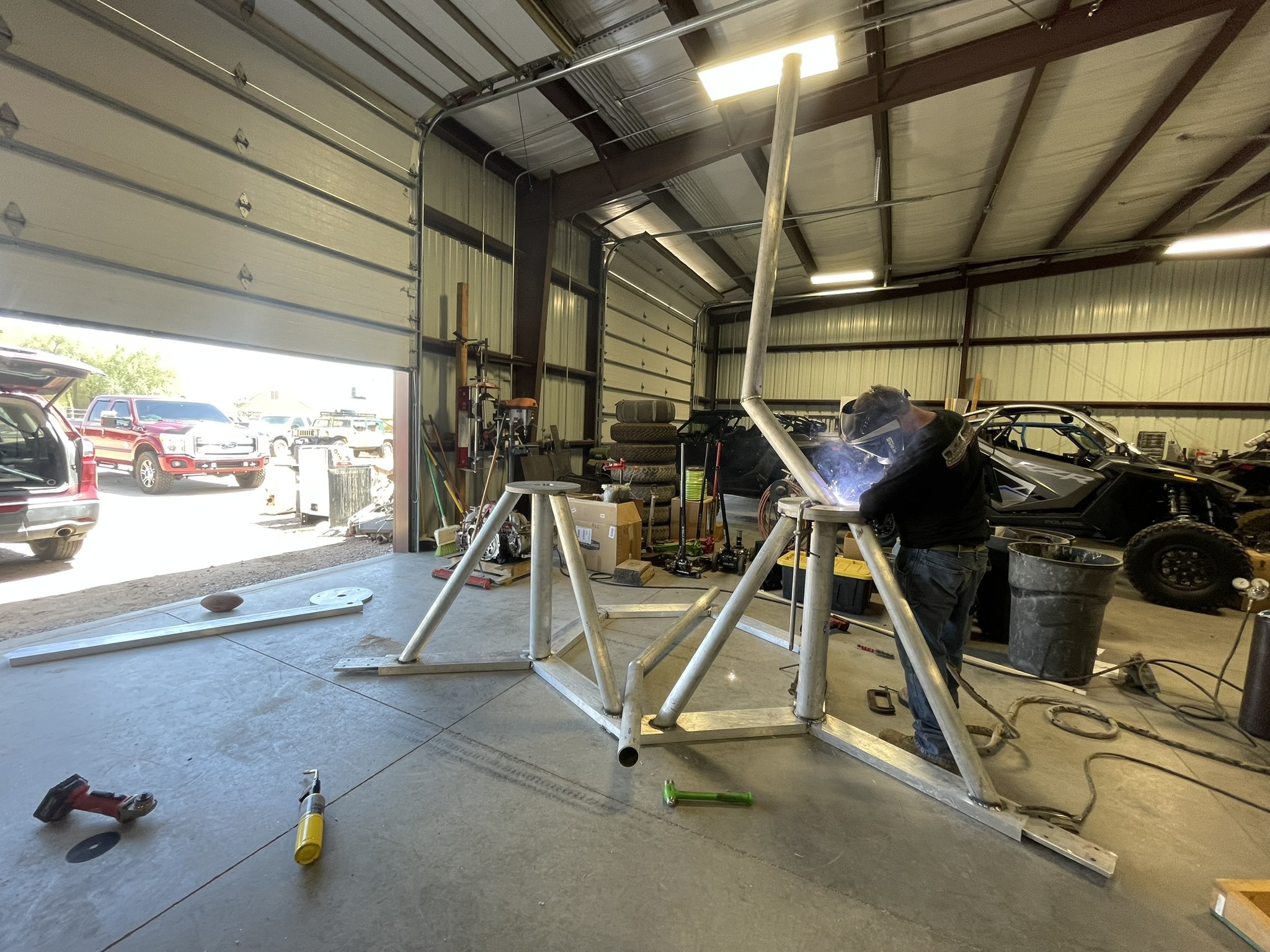}
        \label{fig:welding-tiger}
    }
    \caption{Structural welder Shane Till performs precision MIG welding of the deer leg and tigress leg aluminum truss columns to their foot discs at his Tempe, AZ workshop. Each leg required cut planes optimized in Autodesk Fusion 360, followed by careful angular alignment before welding to lock in the correct joint angles. These angles---together with the hexagonal base---determined the heading of the completed sculpture, ensuring the raised human arm, the rooster head, and the flame effects would all point precisely toward The Man. The final installation achieved exactly this orientation on playa, with residual compliance in the torso-to-leg attachment providing a degree of fine adjustment during on-site assembly.}
    \label{fig:welding-legs}
\end{figure}

\subsection*{Finite Element Analysis (FEA)}

Finite element modeling tested the superstructure under vertical and lateral loads, including worst-case wind scenarios. Drag equations and displacement analyses ensured stability with a factor of safety greater than 5 for the base, and greater than 3 for the sculpture skeleton, intentionally exceeding minimum requirements to prepare for dynamic loading should the piece be adapted as climbable art in the future.
\begin{figure}[htpb]
    \centering
    \begin{subfigure}[b]{0.48\linewidth}
        \centering
        \includegraphics[width=\linewidth]{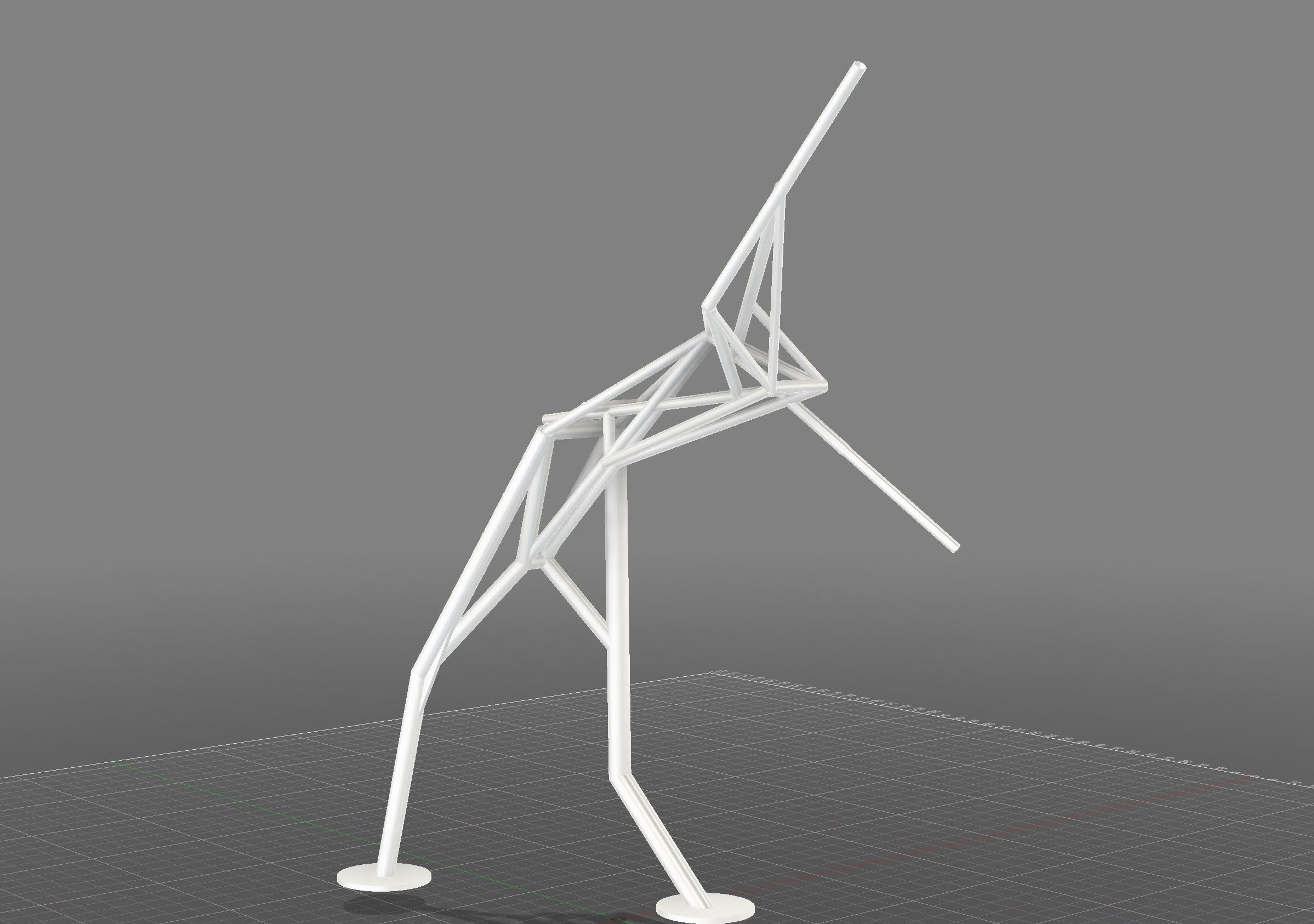}
        \caption{Digital model of the core superstructure.}
        \label{fig:truss_design_front}
    \end{subfigure}
    \hfill
    \begin{subfigure}[b]{0.48\linewidth}
        \centering
        \includegraphics[width=\linewidth]{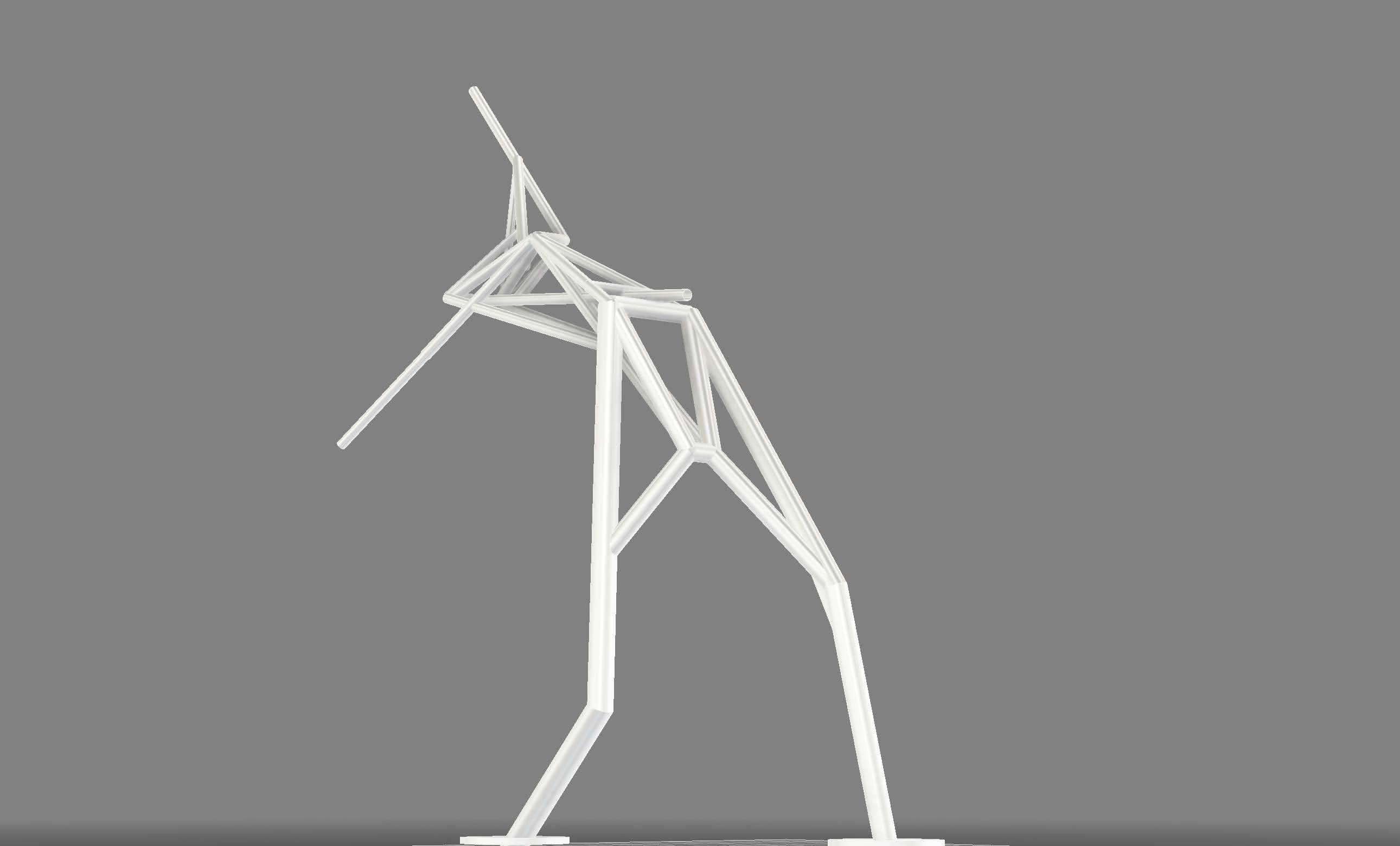}
        \caption{Alternate view of the truss framework.}
        \label{fig:truss_design_back}
    \end{subfigure}
    
    \vspace{0.5cm} 
    
    \begin{subfigure}[b]{0.8\linewidth}
        \centering
        \includegraphics[width=\linewidth]{fem-final-straight-lines-design.png}
        \caption{Finite Element Analysis (FEA) model.}
        \label{fig:fem_analysis}
    \end{subfigure}

    \caption{\small{Structural design and validation of the Navagunjara Reborn installation: Digital models (a, b) illustrate the core aluminum tubing and truss system designed to support the artisan-crafted components. Finite element modeling (c) was used to test the superstructure under vertical and lateral loads, including worst-case wind scenarios. The analysis confirmed factors of safety greater than 5 for the base and above 3 for the sculpture skeleton, intentionally exceeding minimum requirements to prepare for dynamic loading and future adaptations for climbability.}}
    \label{fig:structural_analysis_combined}
\end{figure}
\begin{figure}[htpb]
    \centering
    \begin{subfigure}[b]{0.29\linewidth}
        \centering
        \includegraphics[width=\linewidth]{trailer-base-fit.jpg}
        \caption{Base sized for direct placement on an 8' trailer.}
        \label{fig:trailer-base}
    \end{subfigure}
    \begin{subfigure}[b]{0.29\linewidth}
        \centering
        \includegraphics[width=\linewidth]{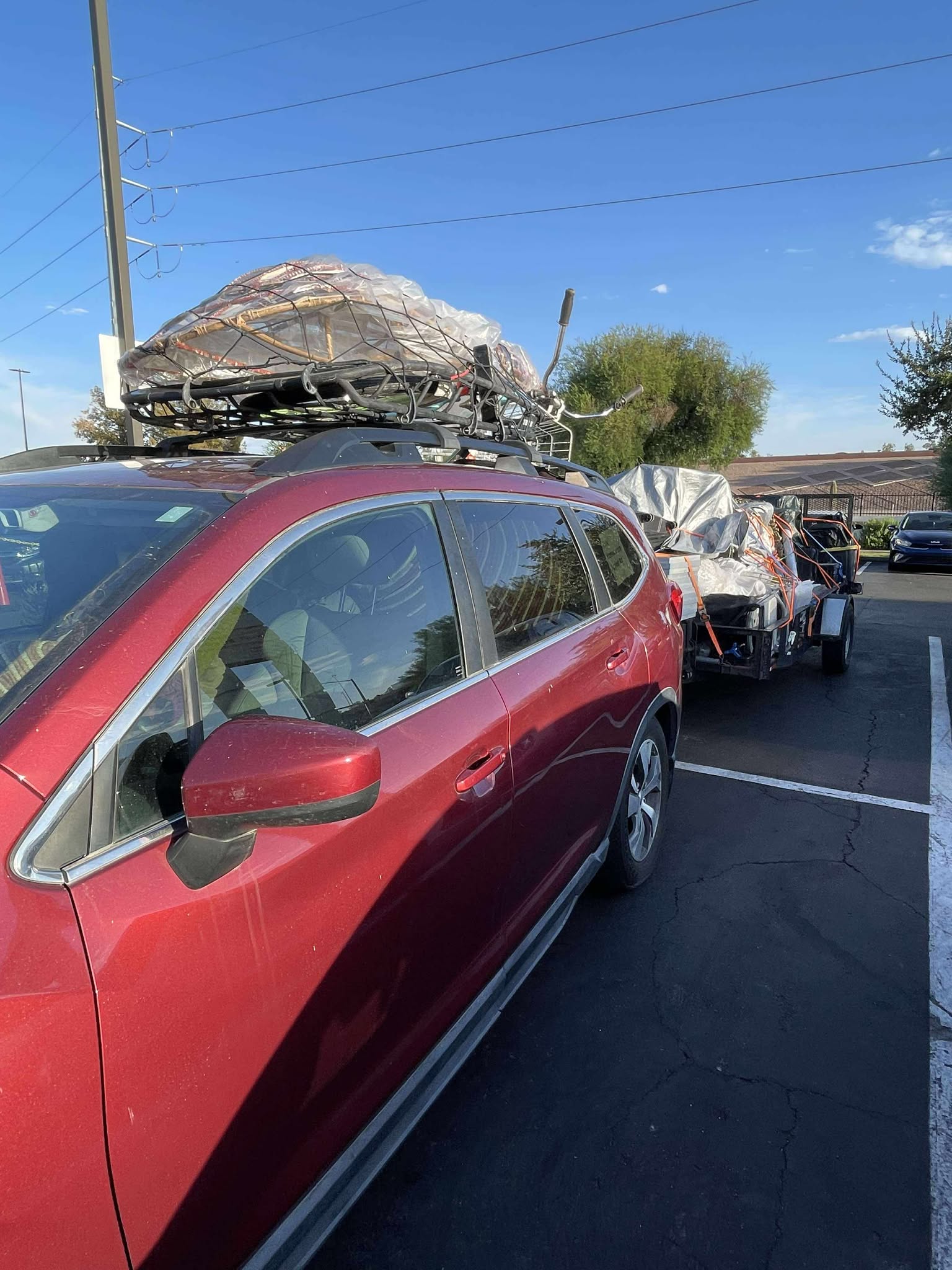}
        \caption{Loaded trailer and SUV carrying craft panels, superstructure, and tools.}
        \label{fig:trans1}
    \end{subfigure}
    \begin{subfigure}[b]{0.29\linewidth}
        \centering
        \includegraphics[width=\linewidth]{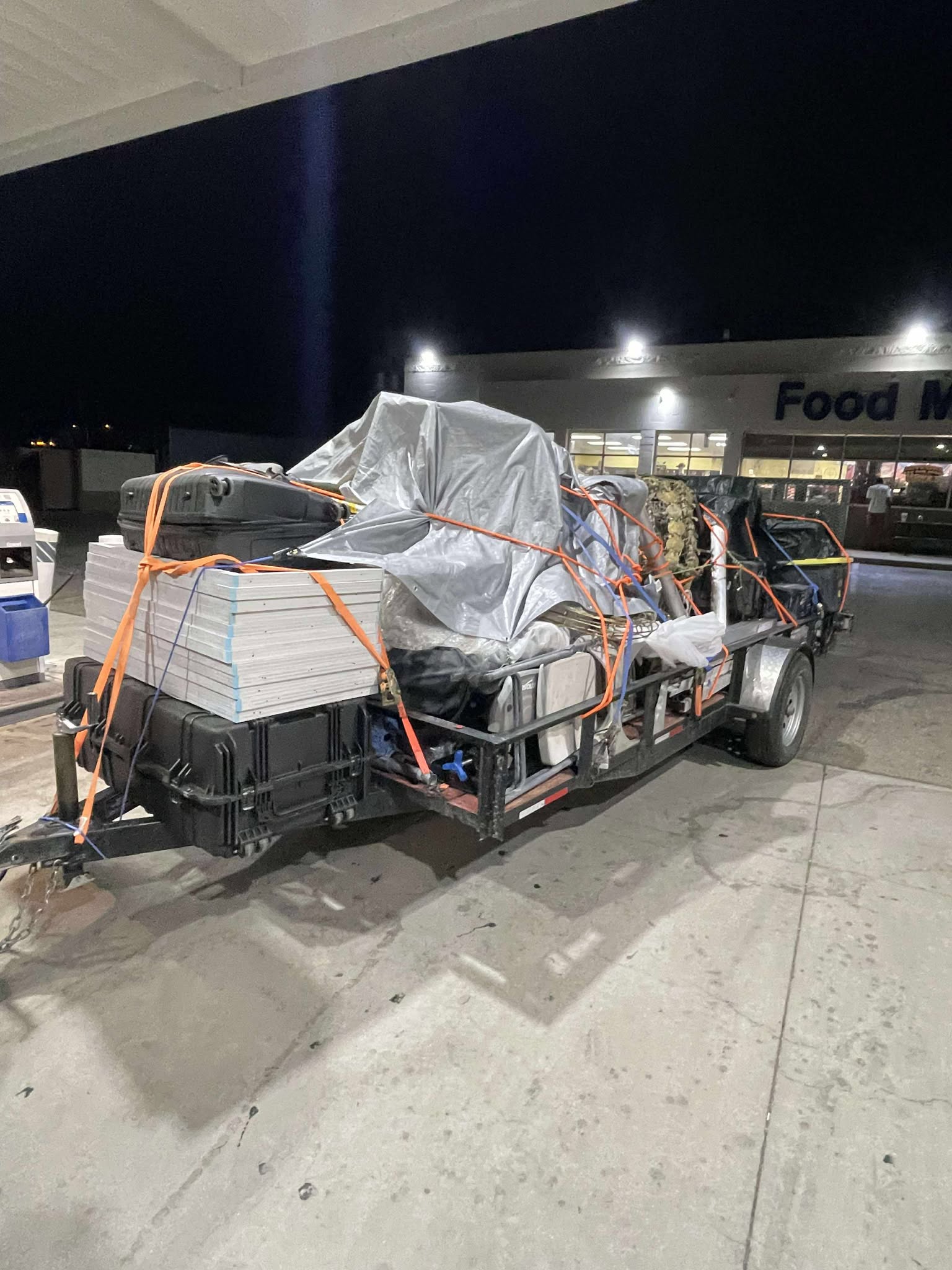}
        \caption{Closeup of the packed trailer during transit to playa. Stacked solar panels are visible in the front of the trailer.}
        \label{fig:trans2}
    \end{subfigure}
    
    \caption{Careful packing and logistical coordination enabled the safe transport of the intricate sculptural components and superstructure elements from Odisha to the Nevada desert. This figure illustrates the organized stacking of lightweight yet sturdy panels, metal frameworks, and craft materials into protective crates and trailers. By optimizing container sizes and leveraging modularity, the team ensured efficient shipping, minimized handling risks, and facilitated rapid on-site assembly under challenging environmental and schedule constraints.}
    \label{fig:grouped_transport}
\end{figure}
In sum, the structural design reflects a dual commitment: protecting the cultural authenticity of Odia craft while engineering for resilience, safety, and scalability under demanding environmental conditions and Black Rock City's regulatory protocols. Our digital-physical framework enabled production of visuals, engineering drawings, and structural details at different stages of the build.

\subsection*{Logistics and Supply Chain: A Constraint Cascade}

The shipping and logistics timeline illustrates how a single adaptive decision can cascade through budget, measurement strategy, and schedule in ways that compress the entire downstream design space.

The India fabrication build ran from mid-April through the end of May 2025. The original plan called for sea freight at approximately \$5,000---roughly a third of the total Honoraria grant. The ${\sim}$45-day transit window was not idle time: SfM-derived meshes of completed craft pieces would be processed during shipping, structural designs optimized against those scans, and truss fabrication procedures planned so that the bulk of the assembly-ready work would be complete before the pieces arrived in late July to early August. The plan kept the maximum number of downstream options open for as long as possible---lower cost, concurrent engineering during transit, and high-quality 3D models of the actual crafts in hand on arrival. The team was working intuitively and agilely, not from a formal framework; in retrospect, this mirrors the logic of the Sagawa-Ueda principle~\cite{sagawa2010}---acquire measurement information, perform engineering work during the information-transit window, and act upon arrival.

The decision to switch to air freight, driven by a desire for schedule compression, did not deliver the expected gain: pieces still arrived mid-July---only marginally ahead of the sea freight estimate---but at \$13,000---against an original shipping budget of \$5,000---consuming nearly the entirety of the \$14,870 Honoraria grant in a single line item. More consequentially, it eliminated the concurrent-engineering window the sea freight plan had built in. Every piece also had to be evaluated against strict weight and volume limits, introducing a combinatorial packing problem with trade-offs between cultural completeness and shipping feasibility. The remaining budget gap was carried collectively by the lead and co-artists and the crowdfunding community---a fitting reflection of the decommodified spirit of the work. Monumental art of this kind does not proceed on budget; it proceeds on commitment.

The switch also changed the scanning calculus. With sea freight and 45 days in transit, thorough photogrammetric coverage of all completed pieces would have been both urgent and feasible. With air freight assumed to be days away, careful scan coverage felt less critical---and was deprioritized toward the end of the India build. In practice, reasonable SfM scans were obtained for the torso, deer leg, metal human hand, and tigress leg, as well as early-stage wireframe scans for the elephant leg and peacock neck before sabai grass weaving was complete. The pieces completed latest---the rooster head, peacock neck in final form, and chest---received the least coverage, as the compressed air-freight timeline crowded out the systematic video acquisition needed for photogrammetry. This created a measurement gap later compensated for through bounds estimations derived from shipping weights, known craft dimensions, and artisan reports, rendering the structural optimization stack probabilistic rather than deterministic for those components.

The result was that the sea freight and air freight plans represented topologically distinct solution paths through the same design space---different in budget headroom, information quality, and schedule---with different downstream assembly sequences contingent on each. The air freight path, having consumed the budget buffer and the concurrent-engineering window simultaneously, left the team navigating a narrower corridor with Burning Man less than six weeks away. That the sculpture was completed validates the resilience of the modular digital-physical framework through improvisation, modularity, and collective commitment.

\section{Deployment at Black Rock City}

The Black Rock City event timeline spans three phases: \emph{build week} (the week before the festival opens, during which Honoraria artists access the playa to assemble their installations), \emph{festival week} (the main event, culminating in the ceremonial burn of The Man on Saturday night), and \emph{stroke days} (the period from Sunday through Wednesday of the following week, during which artists are required to complete teardown and Leave No Trace restoration of their site). For Navagunjara Reborn, team members started arriving on Tuesday of build week, while the trailer carrying the artwork from Phoenix with the lead artist arrived on Thursday. Festival week then saw the sculpture operational with active flame effects through the final night. Teardown and site restoration were completed within the stroke days window in accordance with Burning Man's LNT protocols.

\begin{figure}[htpb]
    \centering
    \begin{subfigure}[b]{0.4\textwidth}
        \centering
        \includegraphics[width=\linewidth]{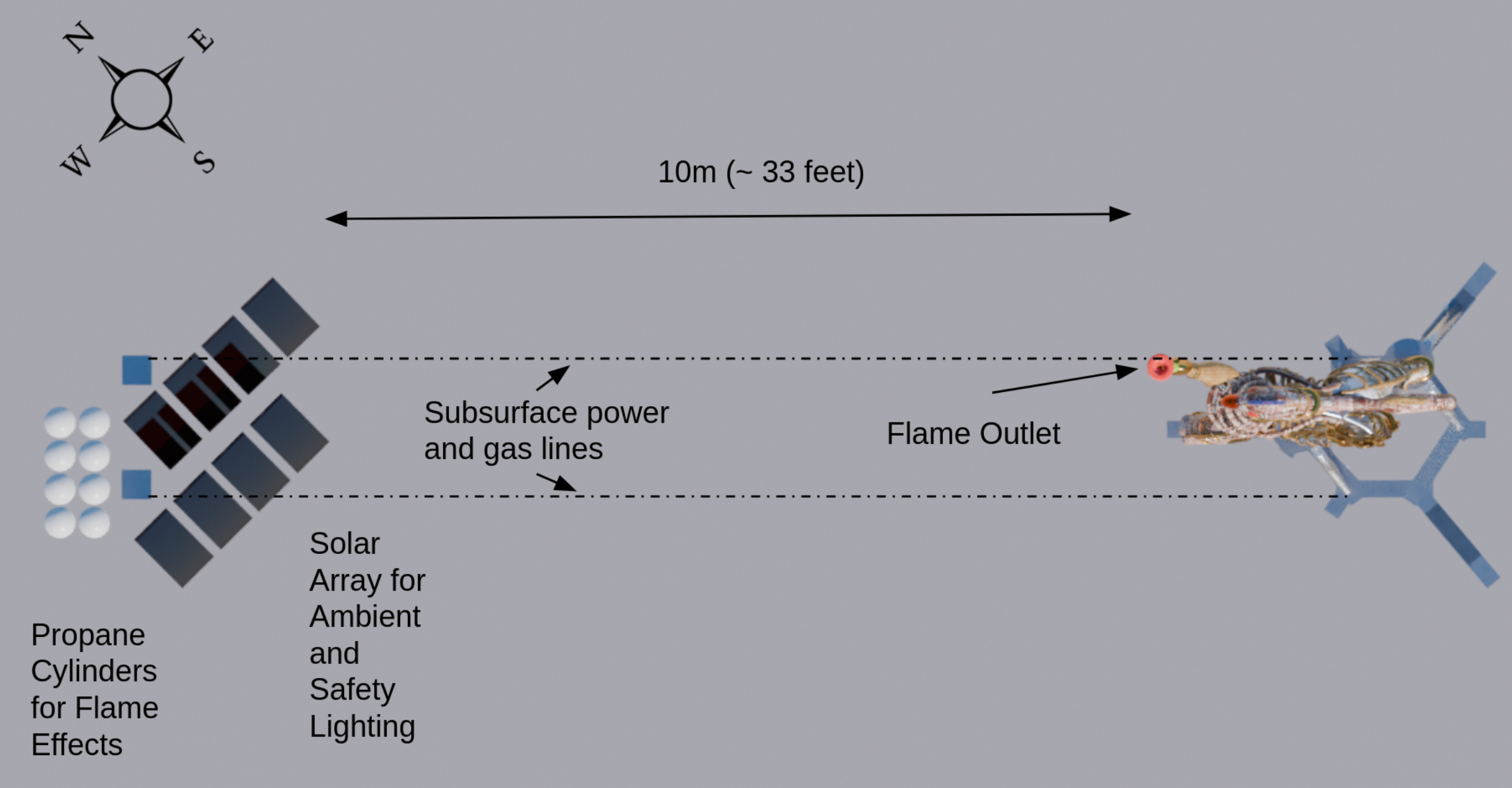}
        \caption{Navagunjara Reborn installation layout, showing sculpture placement, equipment staging zones, power and propane routing, and anchor positions.}
        \label{fig:pwr-gas-schema}
    \end{subfigure}
    \hfill
    \begin{subfigure}[b]{0.4\textwidth}
        \centering
        \includegraphics[width=\linewidth]{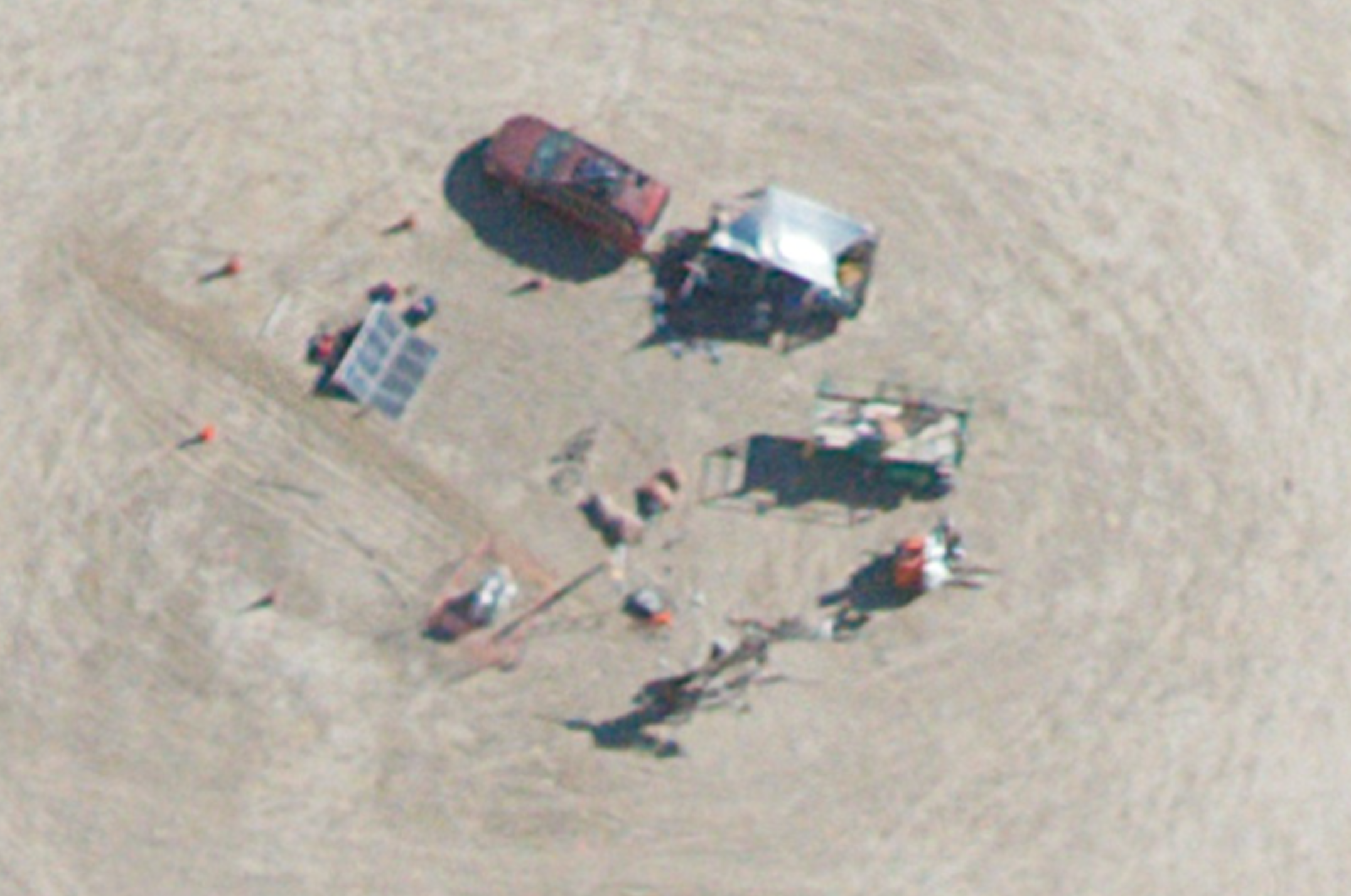}
        \caption{High-resolution aerial panorama of the mid-build Navagunjara Reborn site at Black Rock City, showing the sculpture's superstructure and surrounding work zones during on-playa assembly.}
        \label{fig:gigapan-build}
    \end{subfigure}
    \caption{On-playa construction at Black Rock City: The first image shows a planned spatial layout for assembly of the sculpture components, equipment staging, and work zones designed to accommodate storm conditions and facilitate secure anchoring. The next image offers a mid-build aerial perspective, capturing the logistical complexity, community collaboration, and modular workflow that enabled rapid resilience and adaptation in the demanding environment of the Nevada desert. Solar panel array and propane cylinders are visible in this image.}
    \label{fig:navagunjara-combined}
\end{figure}The installation was placed 700 feet from The Man at the 9 o'clock position within the Black Rock City grid, providing a central location. Despite storms, the sculpture was completed and exhibited. An 8x16' trailer was acquired for \$2200 and used for transportation of the disassembled installation to the playa art site. Its width allowed precise placement of the base superstructure without leg extensions.

\begin{figure}[htpb]
    \centering
    \begin{subfigure}[b]{0.48\linewidth}
        \centering
        \includegraphics[width=\linewidth]{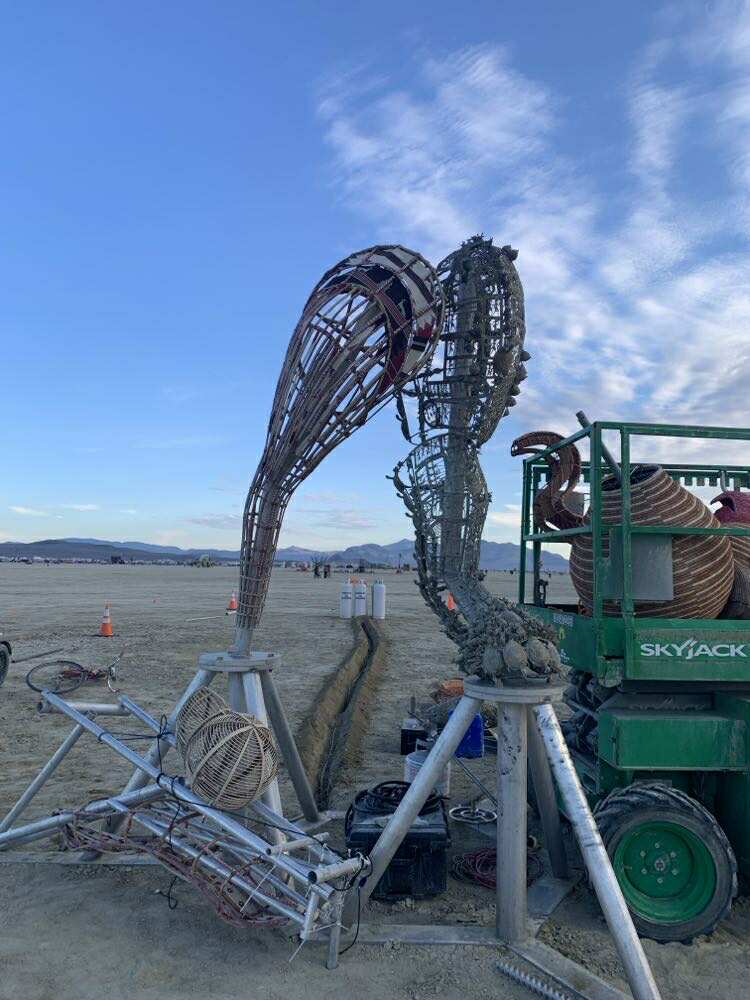}
        \caption{FAST crew trenching for propane and power lines at the sculpture base.}
        \label{fig:propane-trenches}
    \end{subfigure}
    \hfill
    \begin{subfigure}[b]{0.48\linewidth}
        \centering
        \includegraphics[width=\linewidth]{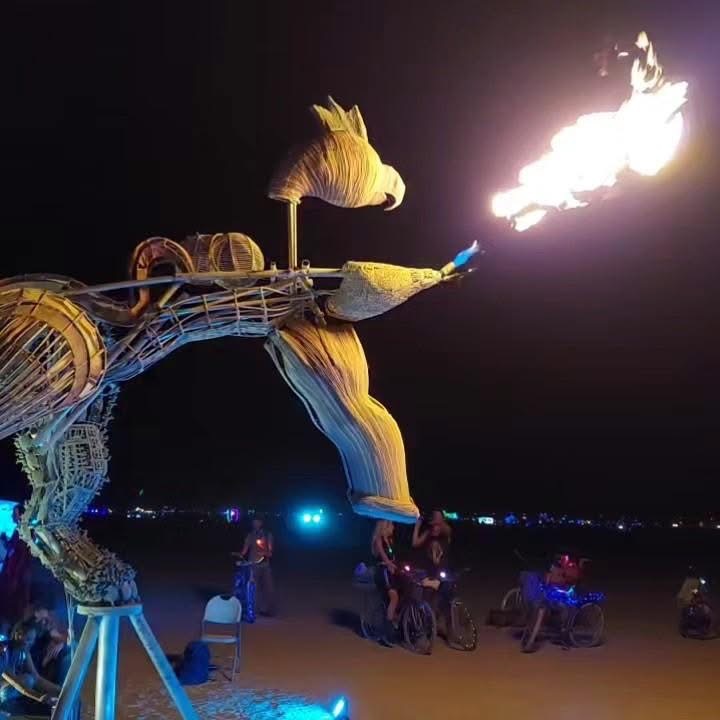}
        \caption{Night view with active poofer flame effects by Swig Miller.}
        \label{fig:phoenix-flame-effects}
    \end{subfigure}
    \caption{Preparing infrastructure for interactive art: BMORG's fire support (FAST) crew excavates trenches to safely route propane and power lines during on-playa assembly of Navagunjara Reborn. Concurrently, the completed torso—protected and poised for lifting—awaits hoisting and attachment to the assembled leg substructure. This scene highlights the coordination of fuel, flame effects, electrical supply, and heavy-lift operations required to safely integrate large-scale sculpture in a temporary desert city.}
    \label{fig:playa-infra}
\end{figure}

Eight 100W solar panels were arranged with four in series and two sets in parallel, managed by a 40A MPPT charge controller, a 400Ah LiFePO4 battery, and a 1000W inverter. A 2kW inverter-generator was kept as backup. One 100~lb propane cylinder provided fuel for the poofer effects through an accumulation chamber. A 25~lb propane cylinder served the pilot flame, constructed from an off-the-shelf weed-removing flame thrower.

The installation drew two key validation outcomes:
\begin{itemize}
    \item \textbf{Community Engagement:} Diverse audiences connected deeply to the fusion of mythology and craft, with conversations at the site spanning dhokra traditions, Phoenix symbolism, and the collaborative process.
    \item \textbf{Technical Validation:} Modularity enabled rapid post-storm build adaptations and precise fits, affirming the value of flexible, contingency-aware design.
\end{itemize}

\begin{figure}[htpb]
    \centering
    \includegraphics[width=3.2in]{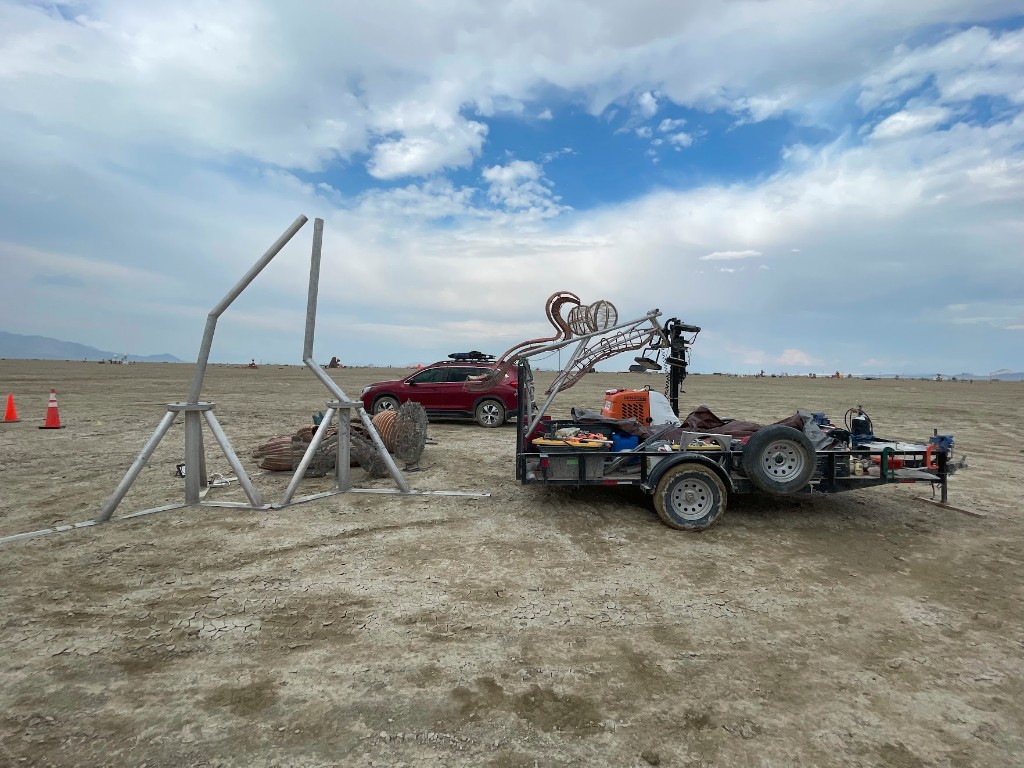}
    \caption{\small{On-playa pre-staging: the aluminum base superstructure stands erected with legs attached, while the cane-and-metal torso (foreground, right) awaits lifting and threading onto the truss. The trailer serves as both transport and mobile workshop. This intermediate state captures the moment between structural assembly and craft integration---prior to the addition of flame effects apparatus. Photo: Black Rock Desert, August 2025.}}
    \label{fig:playa-prestaging}
\end{figure}

\begin{figure}[htpb]
    \centering
    \includegraphics[width=0.55\linewidth]{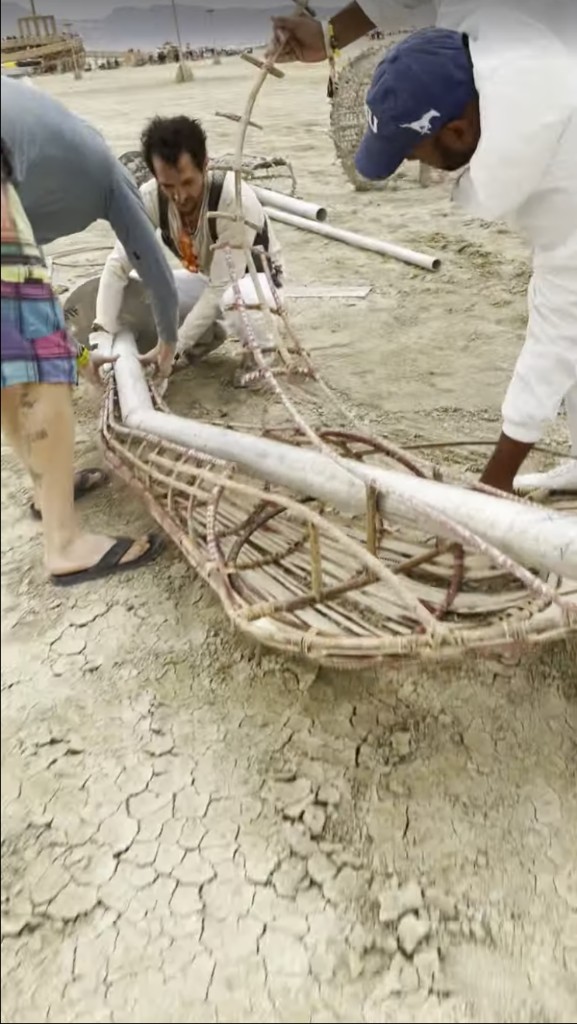}
    \caption{\small{Precisely formed aluminum leg truss being inserted into the cane deer leg on playa, with minimal cuts to the craft element.}}
    \label{fig:deer-truss-insert}
\end{figure}

\subsection{On-Playa Improvisation}  

Installation was interrupted by three storms, yet the modular design enabled rapid recovery and adaptation. The aluminum spine successfully supported a variety of materials, demonstrating both structural robustness and design flexibility. Interactive flames and integrated artwork further fostered participant engagement and storytelling throughout the installation.  

Adverse weather introduced significant challenges. Dust and rain often made power tools impractical, prompting a shift to manual methods such as hacksaws, always with attention to playa safety. MIG welding, protected by fireproof wraps, proved effective in select torso areas. However, the majority of upper-body connections—including those forming the truss tubes of the torso, head, and front legs—were secured using bolts, stage clamps, and steel cable ties.  

Weight reduction opportunities were also leveraged during assembly. For example, the steel framing inside the elephant leg was selectively cut using a bolt cutter, reducing overall weight by several kilograms and easing the final mounting process.

\subsection{Sound and Acoustic Experience}
The original proposal envisioned a curated soundscape of tribal folk music commissioned from Odisha musicians. While this was not realized in the 2025 installation, on-playa observations revealed an emergent acoustic phenomenon: wind passing through the hollow aluminum tubes of the superstructure produced tonal resonances—an unplanned but evocative sound the lead artist documented as a design possibility for future iterations. On days when the sculpture was active, passing art cars and ambient playa music provided an organic acoustic context. The intersection of wind-driven structural resonance and community-sourced sound points toward a future design direction in which the sculpture functions not only as a visual artifact but as a passive acoustic instrument.

\subsection{Teardown Process}
The teardown of Navagunjara Reborn was guided by principles of modularity, ecological responsibility, and operational safety, ensuring that all components were removed efficiently and in accordance with Burning Man’s Leave No Trace (LNT) ethic. The process leveraged the sculpture's modular engineering, enabling the systematic disassembly of sculptural panels, structural hardware, and support infrastructure without damaging handcrafted elements or the desert environment.

    The process began with the removal of major sculptural elements using lifting in stages, which minimized physical strain and safeguarded the integrity of large, delicate craft panels. Mechanical fasteners such as bolts, clamps, and steel cable ties—selected during assembly for their adaptability—enabled quick reversal of connections without the need for power tools, especially critical under the playa’s adverse weather conditions. Steel cable ties were retained for future reuse and recycling.

Smaller hardware, fixtures, and flame system components were methodically collected in labeled containers, and each section of the superstructure was demounted in sequence to avoid material pileup or risk to artwork. Team members maintained a strict inventory and separation of debris, reclaiming all metal, composite, and craft materials for post-event reuse or return.

The final stage involved thorough inspection and sweeping of the build site to ensure that no minor debris, hardware, or material fragments remained, reinforcing the team’s commitment to festival-wide ecological stewardship and accountability. Throughout, careful photographic documentation enabled verification of the LNT protocol and provided a reproducible template for future teardown operations in large-scale art installations.

\begin{figure*}  
    \centering
    \includegraphics[width=7in]{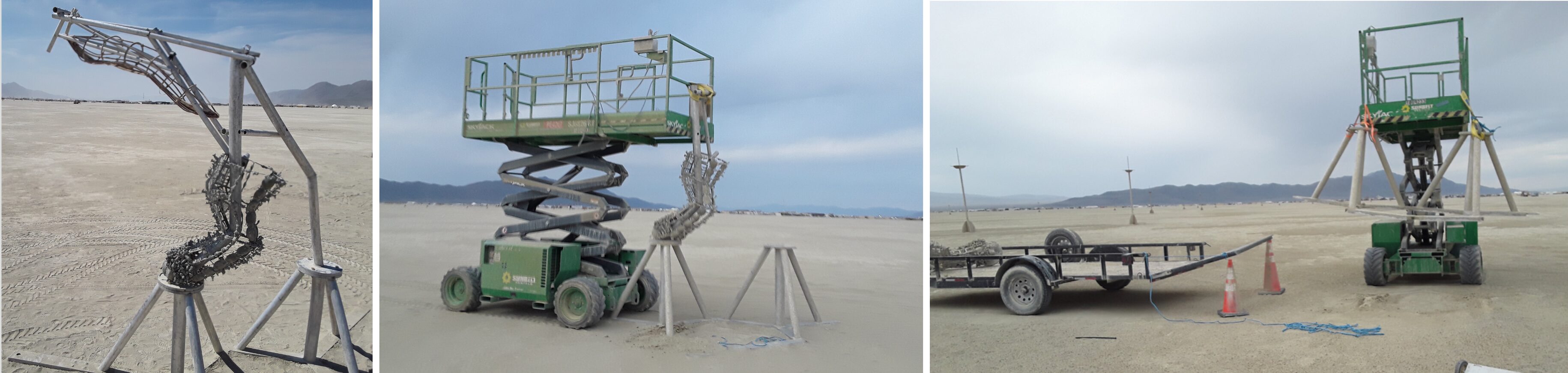}
    \caption{\small{Dismantling and Leave No Trace (LNT) on playa: Team systematically disassemble the Navagunjara Reborn structure, carefully recovering panels, hardware, and infrastructure to ensure minimal environmental impact. The stages of teardown—from lifting major elements to collecting small debris—reflect the project’s commitment to efficient, sustainable art practices and festival-wide principles of ecological stewardship and radical accountability.}}
    \label{fig:teardown-sequence}
\end{figure*}

Figure~\ref{fig:teardown-sequence} visualizes a few steps of this multi-stage process, documenting lifting of sculptural elements, organizing reclaimed materials, and collectively ensuring a clean restoration of the site. This systematic, respectful teardown marks the closing loop of the project's digital-physical framework and underscores the importance of sustainability in public art practice.

\begin{figure*}[t]
    \centering
    \begin{subfigure}[b]{0.48\textwidth}
        \centering
        \includegraphics[width=\linewidth]{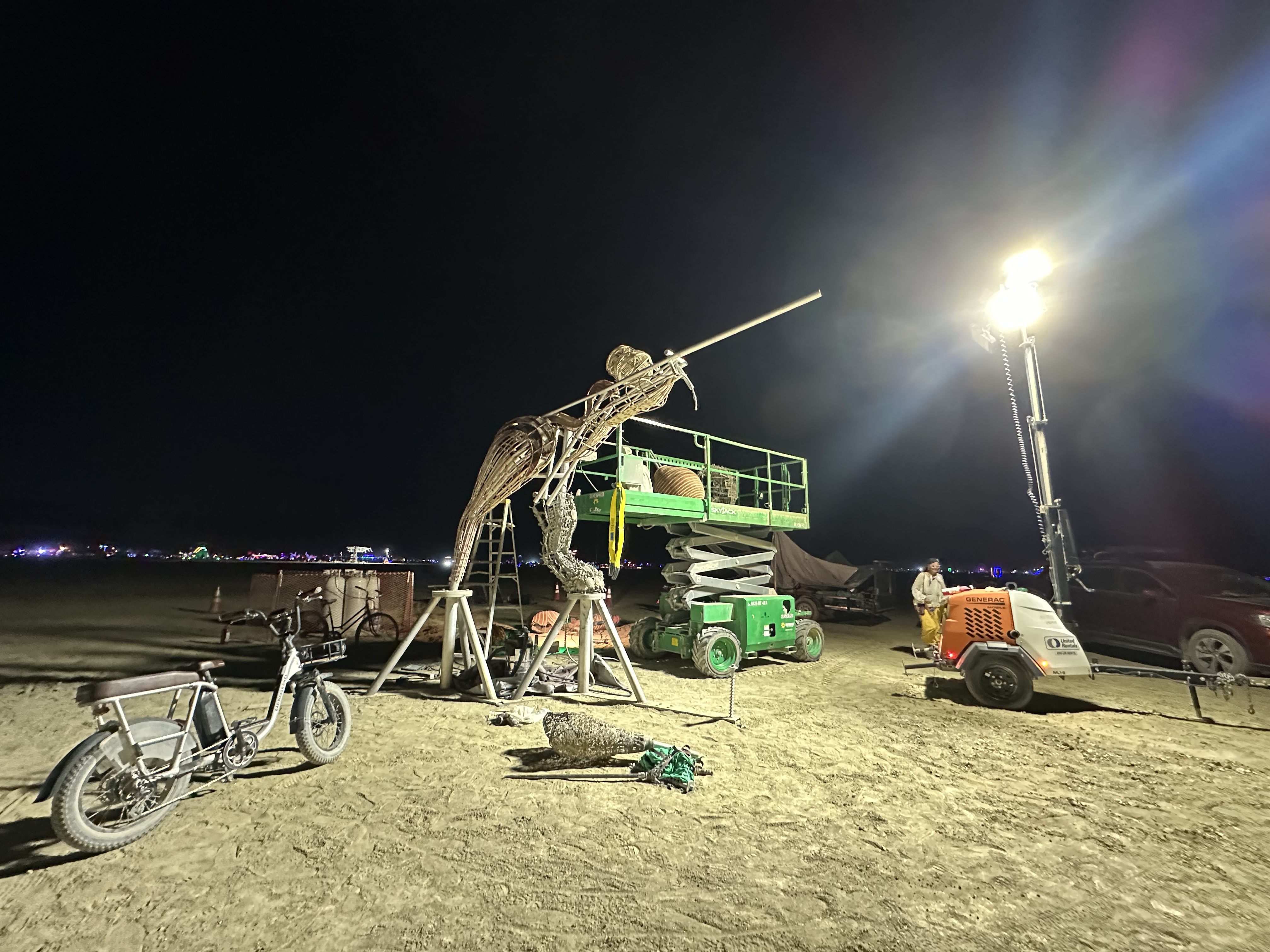}
        \caption{Mid-build assembly on playa. Photo: Kyle Breen.}
        \label{fig:mid-build}
    \end{subfigure}
    \hfill
    \begin{subfigure}[b]{0.48\textwidth}
        \centering
        \includegraphics[width=\linewidth]{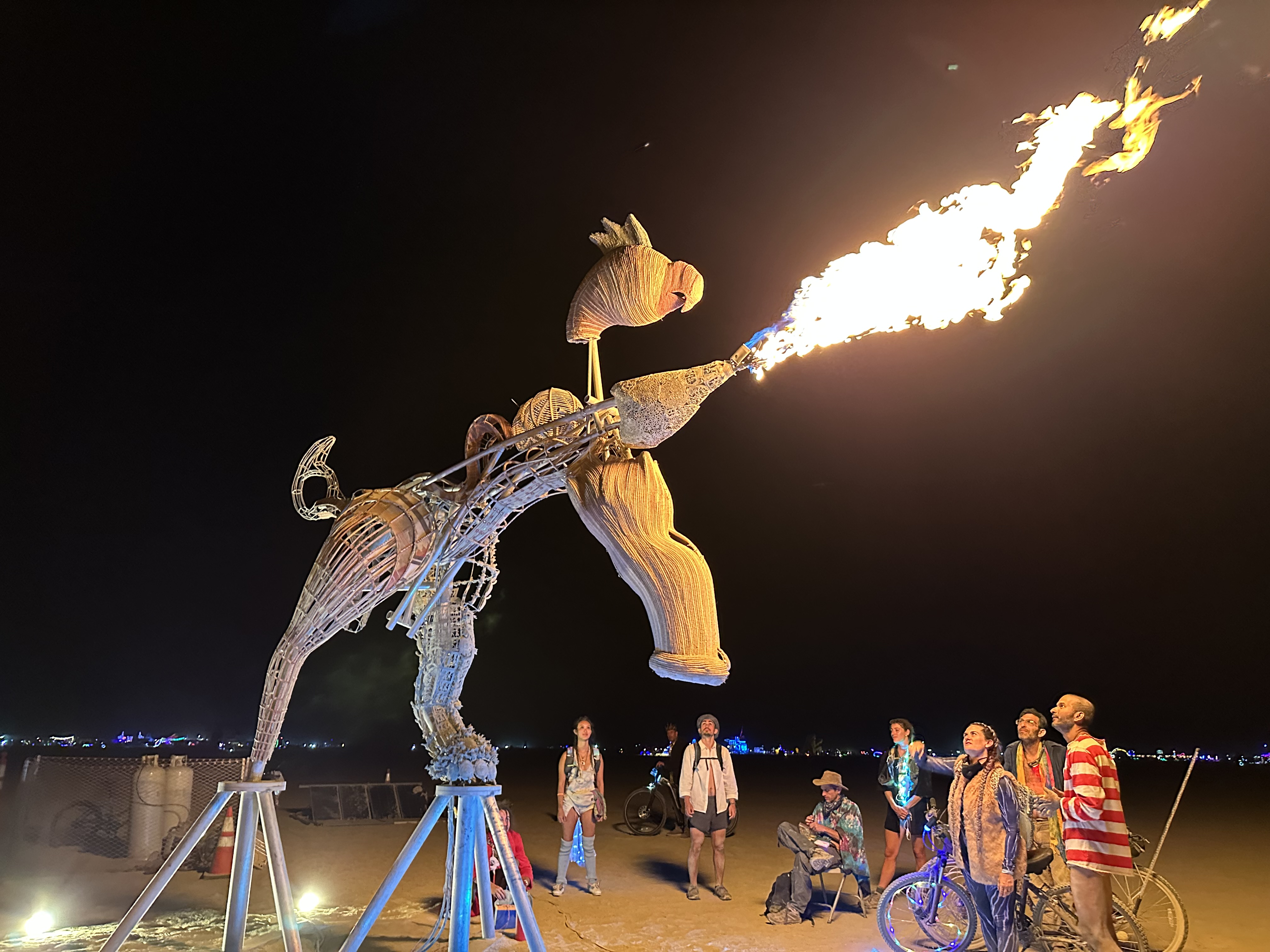}
        \caption{Completed sculpture illuminated at night during interactive flame effects. Photo: Kyle Breen.}
        \label{fig:full-build}
    \end{subfigure}
    \caption{\small{Navagunjara Reborn on playa at Black Rock City, August 2025: mid-build assembly (left) and the completed sculpture with active poofer flame effects (right).}}
    \label{fig:playa-build-complete}
\end{figure*}

\subsection{Deviations in final installation from original design plans}
Several aspects of the final installation diverged significantly from the original design intent, each driven by the compounding effects of shipping delays, storms, and compressed iteration time on playa.

\paragraph{Assembly strategy: from self-erecting crane to slip-on threading.}
The original engineering plan called for a swing-arm self-assembly sequence in which the sculpture would erect itself in stages using motorized pulleys attached to its own growing structure—effectively functioning as its own crane. The torso would be swung up onto the base first, followed by the elephant leg, then the head, arm, and remaining components in a deliberate staging order. This approach assumed that the structural trusses connecting the legs to the torso could be pre-welded and tested in Arizona before playa, and that sufficient iteration time would be available for CAD refinement, welding, and fit verification while the photogrammetry-scanned craft models were still in transit through June and July. In practice, precise SfM scans were not available for all components, and precise weights were also unknown---the pieces turned out to be significantly heavier than estimated, in part because GI (galvanized iron) steel had been used for the neck, head, and chest wireframes due to material constraints during the India build. These compounding uncertainties forced structural design commitments to be made probabilistically, against bounds rather than exact measurements. In retrospect, tighter approximations could have been made---but the optimization space was difficult: each variable (weight, geometry, material stiffness, artisan-side substitutions) was coupled to others through constraints that were themselves evolving, and the information needed to resolve them was distributed across continents and arriving asynchronously.

The central challenge was the rapid transition from fabrication to shipping in India, followed by shipping itself arriving late---compressing two sequential phases that the original plan had assumed would overlap productively. Design optimizations that depended on SfM data from the craft pieces were crunched into a narrowing window, with each decision point requiring optimal stopping: commit to a structural cut now with incomplete data, or wait for better information that might never arrive in time. The lead artist ultimately applied optimal stopping to pre-playa activities in Phoenix---halting structural iteration and arriving on playa on Thursday of build week rather than the planned Tuesday, two days behind schedule into an already storm-disrupted build site. Craft pieces had arrived mid-July with Burning Man six weeks out, and the structural trusses crucial to the leg-to-torso connections could not be fully pre-welded or tested in the time available. Without these pre-welded connections, the swing-arm self-assembly was not feasible, and no crane had been budgeted.

The team adapted by threading the aluminum truss tubes \emph{through} the canewood torso on playa—inserting structural elements through the cane mesh without cutting the craft at any point—and securing them with bolts, stage clamps, and cable ties. This was done through storms, by hand, with manual tools. For the deer leg, the canewood mesh was cut at a minimal number of optimally chosen points to allow insertion of the pre-welded leg truss as originally planned. The same approach was used for the metal-structured legs. The result was a structurally sound assembly that preserved the integrity of every craft element, achieved without the planned self-erecting crane sequence.

\paragraph{Unattached components.}
The neck and chest sculptural sections, although fabricated, were not attached during the final installation. Time constraints during the compressed playa build schedule, compounded by storm interruptions, precluded their integration. These components were safely transported back for potential inclusion in future iterations of the work.

\paragraph{Flame effects evolution.}
The original flame design called for a single propane torch. In practice, flame artist Swig Miller evolved this into a two-poofer-and-pilot configuration: two finger-style poofers for dramatic upward bursts and a thumb-style element serving as the continuous pilot flame. This arrangement offered greater visual dynamism and improved operational control, and was approved through the Burning Man FAST (Flame Art Safety Team) inspection process.

\section{Lessons Learned}

The realization of \textit{Navagunjara Reborn} illuminated several principles for fusing traditional craft and computational engineering within a globally distributed, monumental art project:
\begin{itemize}
    \item \textbf{Adaptability Enables Authentic Collaboration.} Craft-centric modular workflows and real-time digital-physical feedback preserved cultural authenticity, but demanded technical flexibility and dynamic problem-solving across continents.\vspace{2pt}
    \item \textbf{Digital Equity is a Core Challenge.} Reliable and user-friendly digital platforms (e.g., Blender~\cite{Blender2025}, DeepGIS) were essential for cross-team integration; however, constraints in hardware access and connectivity underscored the need for streamlined interfaces, targeted training, and resilient offline protocols for distributed teams.\vspace{2pt}
    \item \textbf{Resilience Depends on Built-in Contingency.} Environmental disruptions and logistical complexity highlighted the importance of modular design, versatile assembly strategies, and robust backup plans, especially when incomplete records or delays occurred.  Detailed 3D designs allowed the bulk shipping failsafe option of shipping of soft-form materials and metal crafts in loose form, and use of scanned crafts through SfM, to adapt a basic representative design around the planned superstructure design. \vspace{2pt}
    \item \textbf{Hybrid Documentation Mitigates Data Gaps.} Limitations in photogrammetric capture and in-transit documentation exposed weaknesses in digital archiving. Combining manual record-keeping with post-shipment full scans provided partial coverage, but future practice requires routine, standardized validation at each stage.\vspace{2pt}
    \item \textbf{Impact Requires Deliberate Evaluation.} Extensive engagement resulted from the project, but systematic measurement of community impact and artifact legacy remains an open challenge due to limited resources and time.\vspace{2pt}
    \item \textbf{Retrospective Philosophical Framing.} The final sculpture can be read as a convergence of mythology, craft tradition, structural constraint, and desert environment, distributed across contributors rather than centered in any single authorial domain.
\end{itemize}

\section{Conclusions}

The \textit{Navagunjara Reborn} project establishes a reproducible framework for creating large-scale, culturally grounded public art through international interdisciplinary collaboration. By integrating traditional craft with digital and engineering workflows, this effort not only preserved artisanal heritage but also pioneered new modes of participatory art. At an epistemological level, the project demonstrates that monumental culturally-inspired art can be understood as a configuration satisfying a \emph{product kernel}—the simultaneous intersection of every contributor's inductive priors: the mythological knowledge of the lead artist, the embodied craft intuitions of dhokra metalworkers, sabai weavers, cane craftspeople, and pattachitra painters, the structural logic of engineers and builders, and the environmental demands of the desert itself. No single contributor holds the whole kernel; the sculpture is the one configuration that scores high under all of them at once. Its evolution toward birth followed a Maximum Caliber trajectory~\cite{DixitMaxCal2018}—the most open, constraint-consistent path through collective design space—with each storm and improvisation absorbed as a fluctuation, not a failure.

A final reflection is warranted. The success of this project is a \emph{presence-only sample}. We observe the single realization in which the sculpture stood on playa, illuminated by fire, witnessed by thousands. We do not observe the counterfactual trajectories—the paths where shipping was one week later, where one more storm hit during the torso lift, where the budget gap was not covered, where the cane cracked under the truss threading. Those trajectories are unobserved absences in the design space. In maximum-entropy modeling of species distributions~\cite{phillips2006maxent}, presence-only data is all an ecologist has: one sees where the organism exists, never where it tried and failed. This project is the same. The Navagunjara stands as a single observed occurrence in a vast space of possible outcomes, and the framework documented here—the product kernel, the MaxCal trajectory, the digital-physical feedback loop—is best understood not as a guarantee of success but as the set of conditions that made this particular draw from the posterior survivable. That it survived is the datum. That it might not have is the context. Future practitioners should read this paper with both in mind.

In retrospect, if Maximum Caliber and Maximum Entropy fit the project's arc so naturally, it is worth asking: what, exactly, was being optimized? The answer is clarifying. Nothing was being optimized in the conventional sense—there was no single utility function, no loss to minimize, no target form to converge on. What was being maximized, at every decision point, was the \emph{number of remaining trajectories that still lead to a coherent artifact by the hard deadline}. The hexagonal base kept pose options open. The modular joints deferred commitment. The probabilistic structural bounds allowed design to proceed without exact measurements. The threading of trusses through uncut cane preserved craft integrity without foreclosing structural soundness. Each of these choices widened the funnel of paths that converge on ``sculpture stands on playa.'' The Navagunjara that emerged is, in this sense, the maximum-entropy artifact: not the most ornate version, not the most faithful to the original proposal, not the structurally optimal form, but the \emph{most probable} form given all the kernels acting simultaneously—the one that required the least additional information to specify beyond what the constraints already determined. MaxEnt finds the least-presumptive distribution. MaxCal finds the most probable path. The sculpture is both.

\begin{figure}[t]
    \centering
    \includegraphics[width=\columnwidth]{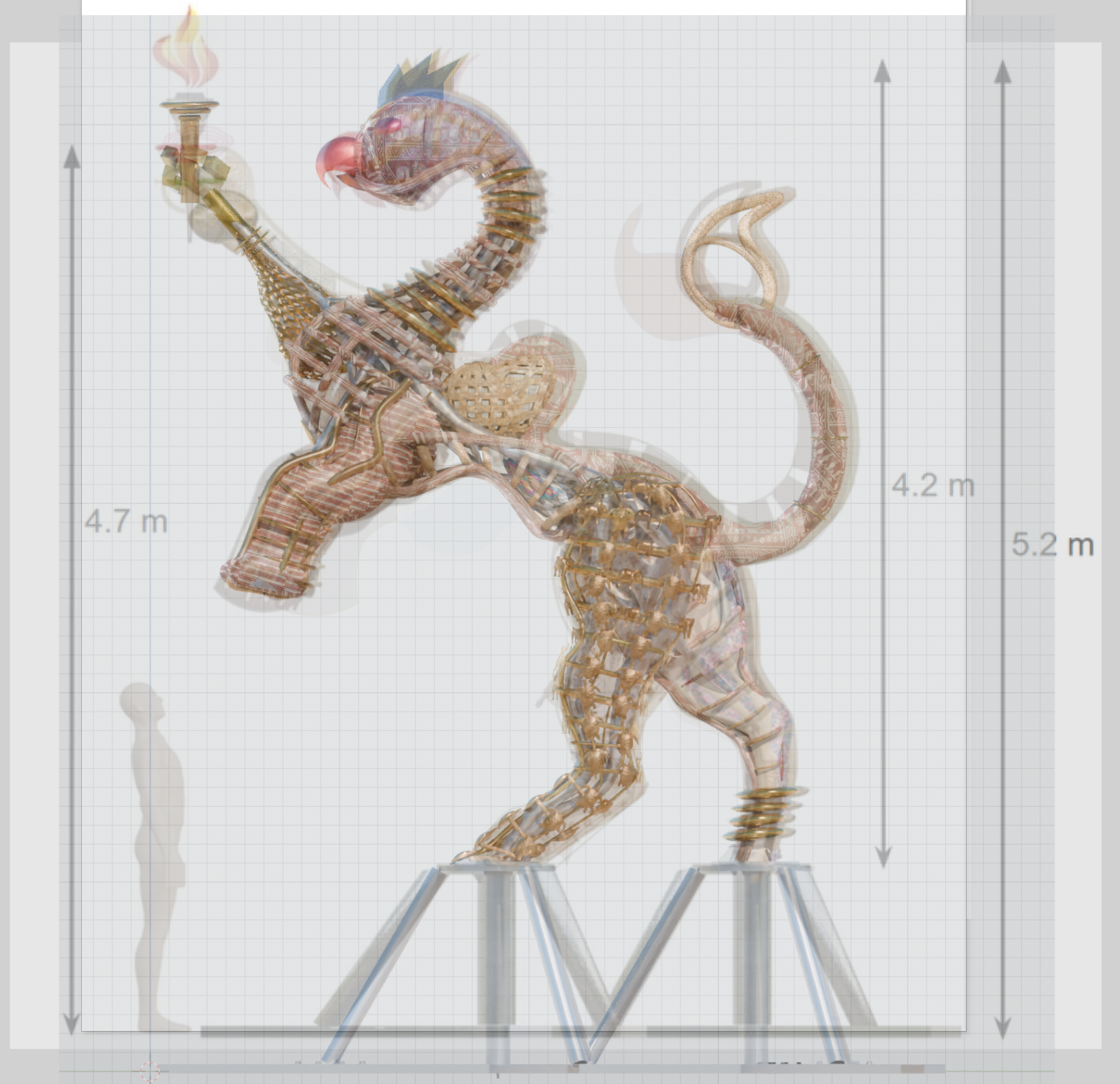}
    \caption{\small{Merged graphic of Navagunjara Reborn design phases: early parametric sculpting, India Art Fair model, Burning Man full proposal render, and final design with artisan-crafted elements integrated.}}
    \label{fig:design-to-build}
\end{figure}

Key takeaways from our work include:

\begin{itemize}
    \item \textbf{Successes}—A modular engineering approach that adapts to handcrafted components; effective cultural integration at scale; and deployment of open-source digital workflows~\cite{navagunjara_reborn_github,deepgis_xr_github} that can guide future STEAM-centered projects.
    \item \textbf{Challenges}—Digital tool access and training disparities, incomplete digital measurement documentation, complex transcontinental logistics, and limited post-installation assessment of community impact.
    \item \textbf{Future directions}—Bridging the digital divide with targeted digital literacy initiatives, offline-capable workflows, and resilient logistical planning. Such efforts could benefit from partnerships involving universities, small businesses, and artisan communities to enhance replicability and sustainability. The sculpture's craft panels and preserved elements are candidates for exhibition at art museums and cultural institutions, and a future all-metal rebuild—climbable and weather-hardened—is envisioned as the next iteration, extending the Navagunjara Reborn framework to permanent or semi-permanent public installation contexts. Wind-driven acoustic resonance from the aluminum superstructure, observed on-playa, opens a further design direction toward the sculpture as a passive sound instrument.
\end{itemize}

\section*{Acknowledgments}
This work is dedicated to the memory of Kamaljeet Dash. The project was executed by the Earth Innovation Hub, an Arizona non-profit organization, through a Burning Man Honoraria Arts Grant in 2025. We thank Richa Maheshwari for her role as co-artist and for coordinating artisan networks and facilitating fabrication logistics in India. We thank Anshu Arora, Badal Satpathy, Sneha Chaudhury, Maya Rachel McManus, Ayesha Dias, and Natasha Fernandes for their contributions to design motifs, crafts fabrication operations, and logistics in India. We thank Shane Till for structural welding and fabrication of critical leg elements. We thank Christina Allen for her contributions to the project planning and proposal development phase. We thank Andrew Miller, Chelsea Miller, Sam Hiatt, Alex Yankovskiy, Jeff Brochuck, Deborah Smith, Heidi Cooper-Panrucker, Lindsay Van Voorhis, and Raj Aditya for their contributions as playa build crew. We thank Sashi Wapang, Kalyan Varma, Ashwin Bhatnagar, Christopher Furst, and Saurav Kumar for their support. 

\section*{Odisha Artisans}
Rajesh Moharana (metal wireframes); Ekadashi Barik (cane-wood work); Malli Mani Nayak, Purnima Nayak, Pankajini Nayak, Arsu Marandi, Geeta Nayak (sabai neck, rooster head); Jayanti Nayak, Gowri Mahapatra, Minu Nayak, Bina Pani Patra, Basamati Nayak (sabai elephant leg); Akshaya Bariki (pattachitra painting); Jagabandhu Panika (Kotpad textile); Purnachandra Ghose (Pipli); Kotuli Sitaka (Kapdaganda textile); Adibari Sisha (Ringa textile).

\section*{Crowdfunding Contributors}
We thank our contributors: Jennifer Capricio, Kalyan Sayre, Micki Mooney, Bernd Pfrommer, Bjørt Debess Johannesen, Lindsay Van Voorhis, Chris Stevenson, Ankita Prasad, Cori Escalante, Iris Yee, Karthik Kuppa, Sai Sukesh Gorrepati, Aninda Mukherjee, Ramon Arrowsmith, James Nothnagel, Patrick Gleason, Raj Aditya, Deirdre L. Milks.

\bibliographystyle{IEEEtran}
\bibliography{references}

\end{document}